\documentclass[longauth]{aa} % for the long lists of affiliations 
\usepackage{booktabs}
\usepackage{graphicx}
\usepackage[english]{babel}
\usepackage{multirow}
\usepackage{xcolor}
\usepackage{graphicx}
\usepackage{natbib} 
\usepackage{gensymb}
\usepackage{ftnxtra}
\usepackage{fnpos}
\usepackage{txfonts}
\usepackage{soul}
\setstcolor{red}
\newcommand{\AS}[3]{$#1^{+#2}_{-#3}$}
\begin{document}

   \title{The CARMENES search for exoplanets around M dwarfs}
   \subtitle{Spectroscopic orbits of nine M-dwarf multiple systems, including two triples, two brown dwarf candidates, and one close M-dwarf--white dwarf binary}

   \author{D.~Baroch\inst{1,2} \and 
   J.~C.~Morales\inst{1,2} \and
   I.~Ribas\inst{1,2} \and
   V.\,J.\,S.~B\'ejar\inst{3,4}  \and
   S.~Reffert\inst{5} \and
   C.~Cardona~Guill\'en\inst{3,4} \and
   A.~Reiners\inst{6} \and
   J.\,A.~Caballero\inst{7} \and
   A.~Quirrenbach\inst{5} \and
   P.\,J.~Amado\inst{8} \and
   G.~Anglada-Escud\'e\inst{1,2} \and 
   J.~Colom\'e\inst{1,2} \and 
   M.~Cort\'es-Contreras\inst{7,9} \and
   S.~Dreizler\inst{6}  \and
   D.~Galad\'i-Enr\'iquez\inst{10} \and
   A.\,P.~Hatzes\inst{11} \and
   S.\,V.~Jeffers\inst{12} \and
   Th.~Henning\inst{13} \and
   E.~Herrero\inst{1,2} \and
   A.~Kaminski\inst{5} \and
   M.~K\"urster\inst{13}   \and
   M.~Lafarga\inst{1,2} \and
   N.~Lodieu\inst{3,4} \and 
   M.\,J.~L\'opez-Gonz\'alez\inst{8} \and
   D.~Montes\inst{14} \and
   E.~Pall\'e\inst{3,4} \and
   M.~Perger\inst{1,2} \and
   D.~Pollacco \inst{15} \and
   C.~Rodr\'iguez-L\'opez\inst{8} \and
   E.~Rodr\'iguez\inst{8} \and
   A.~Rosich\inst{1,2}  \and
   P.~Sch\"ofer\inst{6} \and
   A.~Schweitzer\inst{16} \and
   Y.~Shan\inst{6} \and
   L.~Tal-Or\inst{17,6} \and
   M.~Zechmeister\inst{6}
          }
   \authorrunning{D.~Baroch et al.}
   \titlerunning{Spectroscopic orbits of nine M-dwarf multiple systems}
   \institute{Institut de Ci\`encies de l'Espai (ICE, CSIC),
              Campus UAB, c/~Can Magrans s/n, E-08193 Bellaterra, Barcelona, Spain\\
              \email{baroch@ice.cat}
         \and
              Institut d'Estudis Espacials de Catalunya (IEEC),
              c/ Gran Capit\`a 2-4, E-08034 Barcelona, Spain 
        \and
              Instituto de Astrof\'isica de Canarias,
              V\'ia L\'actea s/n, E-38205 La Laguna, Tenerife, Spain
         \and
              Departamento de Astrof\'isica, Universidad de La Laguna, E-38026 La Laguna, Tenerife, Spain
        \and
              Landessternwarte, Zentrum f\"ur Astronomie der Universit\"at Heidelberg,
              K\"onigstuhl 12, D-69117 Heidelberg, Germany
         \and
              Institut f\"ur Astrophysik, Georg-August-Universit\"at,
              Friedrich-Hund-Platz 1, D-37077 G\"ottingen, Germany
        \and
              Centro de Astrobiolog\'ia (CSIC-INTA), ESAC,
              Camino Bajo del Castillo s/n, E-28692 Villanueva de la Ca\~nada, Madrid, Spain
        \and
              Instituto de Astrof\'isica de Andaluc\'ia (IAA-CSIC),
              Glorieta de la Astronom\'ia s/n, E-18008 Granada, Spain
        \and
              Spanish Virtual Observatory 
        \and
              Centro Astron\'onomico Hispano Alem\'an, Observatorio de Calar Alto, Sierra de los Filabres, E-04550 G\'ergal, Spain
        \and
              Th\"uringer Landesstenwarte Tautenburg,
              Sternwarte 5, D-07778 Tautenburg, Germany
        \and
             Max Planck Institute for Solar System Research, Justus-von-Liebig-Weg 3, D-37077 G\"ottingen, Germany
        \and
              Max-Planck-Institut f\"ur Astronomie,
              K\"onigstuhl 17, D-69117 Heidelberg, Germany
       \and
              Departamento de F\'{i}sica de la Tierra y Astrof\'{i}sica and IPARCOS-UCM (Instituto de F\'{i}sica de Part\'{i}culas y del Cosmos de la UCM), Facultad de Ciencias F\'{i}sicas, Universidad Complutense de Madrid, E-28040 Madrid, Spain
        \and
             Department of Physics, University of Warwick, Gibbet Hill Road, Coventry CV4 7AL, UK
        \and
              Hamburger Sternwarte,
              Gojenbergsweg 112, D-21029 Hamburg, Germany
        \and
              Department of Physics, Ariel University, Ariel, IL-40700, Israel
             }

   \date{Received 9 April 2021 / Accepted 29 May 2021} 

% \abstract{}{}{}{}{} 
% 5 {} token are mandatory
 
  \abstract
   {M dwarfs are ideal targets for the search of Earth-size planets in the habitable zone using the radial velocity method, attracting the attention of many ongoing surveys. As a by-product of these surveys, new multiple stellar systems are also found. This is the case also for the CARMENES survey, from which nine new double-line spectroscopic binary systems have already been announced.}
   {Throughout the five years of the survey, the accumulation of new observations has resulted in the detection of several new multiple stellar systems with long periods and low radial-velocity amplitudes. Here, we newly characterise the spectroscopic orbits and constrain the masses of eight systems and update the properties of a system that we reported earlier.}
   %their spectroscopic orbits and constrain their masses.}
   {We derive the radial velocities of the stars using two-dimensional cross correlation techniques and template matching. The measurements are modelled to determine the orbital parameters of the systems. We combine CARMENES spectroscopic observations with archival high-resolution spectra from other instruments to increase the time-span of the observations and improve our analysis. When available, we also added archival photometric, astrometric, and adaptive optics imaging data to constrain the rotation periods and absolute masses of the components.}
   {We determine the spectroscopic orbits of nine multiple systems, eight of which are presented for the first time. The sample is composed of five single-line binaries, two double-line binaries, and two triple-line spectroscopic triple systems. The companions of two of the single-line binaries, GJ\,3626 and GJ\,912, have minimum masses below the stellar boundary and, thus, could be brown dwarfs. We find a new white dwarf in a close binary orbit around the M star GJ\,207.1, located at a distance of 15.79\,pc. %is best explained as a binary system composed by an M star and a white dwarf
   From a global fit to radial velocities and astrometric measurements, we are able to determine the absolute masses of the components of GJ\,282\,C, which is one of the youngest systems with measured dynamical masses.}
   {}
 

   \maketitle

%
%-------------------------------------------------------------------

\section{Introduction} \label{sec:introduction}

%Low-mass stars are the most populous stellar constituents in our galaxy and are vital to understanding our galaxy’s structure and kinematics (Gizis et al. 2002; West et al. 2004; Bochanski et al. 2007a). Over a quarter of low-mass stars are found in binary or higher-order systems (Duchêne & Kraus 2013), making it crucial to understand how binary companions affect low-mass star formation and evolutionary processes.

%Binaries provide a window into the physical characteristics of stars not usually afforded from their single counterparts. Close, interacting systems allow us to better understand accretion properties and mass loss through winds. Wide binary systems with M dwarfs are often used to better understand the abundances and ages of low-mass stars, through a study of their co-eval companions. In addition, if these systems are eclipsing, complete orbital properties can be derived and therefore fundamental stellar parameters can be determined.

M dwarfs constitute the majority of stars in the Galaxy, in both number and mass \citep{Gizis2002}, and comprise almost 75\,\% of the stars in the solar neighbourhood \citep{Henry2006}. Over 40\,\% of M dwarfs are found in multiple systems \citep{Cortes2017,Winters2019}, making them a key ingredient to our understanding of the stellar population as a whole. Statistical studies of their properties, such as the distribution of mass ratio and orbital elements, provide crucial tests for star formation theories and evolutionary models \citep[see e.g.][]{Tohline2002,Duchene2013}. Depending on their orbital architecture, dynamical analyses of binary systems can yield individual masses and radii with a precision of a few percent independently from calibrations and stellar models \citep{Andersen1991,Torres2010,Southworth2015}. These masses and radii are of key importance for tests to stellar models, as well as to determine empirical calibrations, which can be applied to derive fundamental properties of single stars \citep{Baraffe2015,Benedict2016,Mann2019,Schweitzer2019}. The determination of precise masses and radii is particularly relevant for M-dwarf binaries, which show discrepancies with stellar model predictions \citep{Morales2010,Talor2013,Feiden2014}. 

Apart from their interest as a major galactic population, M dwarfs have also gained significant attention in the field of exoplanets. A number of surveys searching for rocky habitable planets using radial velocities (RVs) as well as the transit technique focus on M-dwarf stars, in which Earth-size planetary companions are common \citep{Dressing2013}. Moreover, compared to Sun-like stars, M dwarfs have the critical advantages of having lower mass and closer habitable zones \citep{Kasting1993,Koppaparu2013}. This means that potentially habitable planets would induce larger RV signals and be easier to detect than around earlier-type stars. As a by-product of exoplanet surveys, a large number of new stellar binary systems have also been discovered \citep[e.g.][]{Kirk2016,Baroch2018,Sperauskas2019}. 

The CARMENES survey \citep{Quirrenbach2018,Quirrenbach2020} is an example of one such exoplanet survey. It has observed more than 300 M-dwarf stars searching for exoplanets in their habitable zones, and already discovered $\sim30$ planets \citep[e.g. ][]{Ribas2018,Morales2019,Zechmeister2019,Trifonov2021}. Furthermore, CARMENES has also revealed the binary nature of nine new M-dwarf systems for which orbital elements have been computed by \citet{Baroch2018}. We announce here an additional eight new multiple systems that were not observed or could not be resolved by \cite{Baroch2018}, and also use new observations to determine the orbital parameters of the very long period SB2 system UU\,UMi, which was already analysed by \cite{Baroch2018}. These systems include two triple-line spectroscopic triple systems (ST3), two double-line spectroscopic binaries (SB2), and five single-line spectroscopic binaries (SB1), two of which have companions with masses compatible with them being substellar, and another with a white dwarf star in a close-in orbit. In Sect.\,\ref{sec:obs} we list the studied stars and describe the observations used in our analysis, including available data in public archives. Section\,\ref{sec:rvs} explains the method used to measure the RVs from observations and to obtain the orbital parameters and rotation periods. In Sect.\,\ref{sec:results} we present the results for each system. Finally, in Sect.\,\ref{sec:conclusion} we discuss the peculiarities of some of the systems and conclude our work.

\section{Sample selection and data}\label{sec:obs}

\begin{table*}[t]
\centering
\caption{Astrometric, photometric, and spectroscopic properties of the nine systems studied in this work.}
\label{tab:props}
\resizebox{\textwidth}{!}{%
\begin{tabular}{l l l l l l l l l l} 
\hline\hline
\noalign{\smallskip}
Name & Karmn & $\alpha$ (J2016)\tablefootmark{a} & $\mu_{\alpha}\cos \delta$ [mas\,a$^{-1}$]\tablefootmark{a} & $\varpi$ [mas]\tablefootmark{a} & RUWE\tablefootmark{a} & $V$ [mag]\tablefootmark{d} &   Sp.\,T\tablefootmark{b,c,d} & $M_{\star}$ [M$_{\odot}$]\tablefootmark{g}  & $v\sin i$ [km\,s$^{-1}$]\tablefootmark{i,j}  \\
 & Type  & $\delta$ (J2016)\tablefootmark{a} & $\mu_{\delta}$ [mas\,a$^{-1}$]\tablefootmark{a} & $d$ [pc]\tablefootmark{a} & $G$ [mag]\tablefootmark{a} & $J$ [mag]\tablefootmark{f} & $T_{\rm eff}$ [K]\tablefootmark{g,h} & $R_{\star}$  [R$_{\odot}$]\tablefootmark{g} &  pEW(H$\alpha$) [\AA]\tablefootmark{k}  \\
\noalign{\smallskip}
\hline
\noalign{\smallskip}
GJ\,207.1 & J05337+019 & 05:33:44.55 & $-231.54\pm0.05$ & $63.36\pm0.05$ & 2.43 & $11.50\pm0.04$ & M2.5\,V & $0.488\pm0.020$ & $9.8\pm1.5$ \\
 & SB1 &  +01:56:41.0  & $-153.99\pm0.04$ & $15.78\pm0.01$ & $10.3762\pm0.0007$ & $7.76\pm0.02$ & $3419\pm51$ &  $0.485\pm0.015$ & $-5.00\pm0.04$  \\
\noalign{\smallskip}
UCAC4 & J07001$-$190 & 07:00:07.00 &  $+145.47\pm0.03$ &  $33.53\pm0.03$ &  1.37 & $13.92\pm0.03$ &  M5.0\,V & $\cdots$ & $3.8\pm1.5$ \\
355-020729 & SB2 &  $-$20:58:34.9 &  $-89.66\pm0.03$ & $29.83\pm0.03$ & $12.3039\pm0.0007$ &  $9.03\pm0.03$ & $3100\pm50$ & $\cdots$ & $-6.99\pm0.04$  \\
\noalign{\smallskip}
GJ\,282\,C & J07361$-$031 & 07:36:07.15 & $+74.1\pm0.1$ & $70.3\pm0.1$ & 6.53 & $10.0\pm0.1$ & M1.0\,V & $0.553\pm0.021$  & $3.1\pm1.5$  \\
 & SB1 &  $-$04:53:16.6 &  $-293.1\pm0.1$ &  $14.23\pm0.03$ & $9.118\pm0.001$ & $6.79\pm0.03$ & $3894\pm51$ &  $0.547\pm0.016$ &  $-0.86\pm0.01$ \\
\noalign{\smallskip}
LP\,427-016 & J09140+196 & 09:14:03.02 & $-146.56\pm0.05$ & $43.52\pm0.05$ & 1.79 & $12.17\pm0.01$ & M3.0\,V &  $0.491\pm0.019$ & $<2$  \\
 & SB1 & +19:40:03.2 & $-160.42\pm0.04$ & $22.98\pm0.03$ & $11.101\pm0.002$ & $8.42\pm0.02$ & $3493\pm51$ &  $0.488\pm0.015$ &  $-0.1\pm0.1$ \\
 \noalign{\smallskip}
GJ\,3626 & J10504+331 & 10:50:26.07 & $+50.0\pm0.2$ & $44.2\pm0.1$ & 9.75 & $13.05\pm0.01$ & M4.0\,V & $0.403\pm0.017$ &  $<2$  \\
 & SB1 &  +33:05:54.3 & $-630.9\pm0.2$ & $22.61\pm0.06$ & $11.7962\pm0.0006$ &  $8.90\pm0.02$ & $3393\pm51$ & $0.405\pm0.012$ & $+0.09\pm0.01$ \\
\noalign{\smallskip}
UU\,UMi & J15412+759 & 15:41:20.01 & $+809\pm2$ &  $68\pm2$ &  9.18 & $12.07\pm0.05$ & M3.0\,V &  $\cdots$ &  $<3$ \\
 & SB2 & +75:59:22.9 & $-707\pm2$ & $14.6\pm0.3$ & $11.0493\pm0.0009$ & $8.26\pm0.02$ & $3430\pm51$ & $\cdots$ &  $-0.11\pm0.02$\\
\noalign{\smallskip}
GJ\,3916 & J15474$-$108 & 15:47:24.22 &  $-309.0\pm0.6$ &$62.5\pm0.3$ &  $\cdots$ & $11.26\pm0.02$ &M2.0\,V & $0.500\pm0.020$ &  $<3$ \\
 & ST3 & $-$11:06:07.0 & $-370.8\pm0.5$ & $16.00\pm0.07$ & $10.2077\pm0.0004$ & $7.58\pm0.03$ & $3528\pm51$ &  $0.497\pm0.015$ & $-0.03\pm0.02$ \\
\noalign{\smallskip}
GJ\,912 & J23556$-$061 & 23:55:39.28 & $-479.1\pm0.2$ & $56.1\pm0.1$ & 2.09  & $11.17\pm0.06$ & M2.5\,V & $0.520\pm0.020$ & $<2$ \\
 & SB1 & $-$07:51:20.8 &  $-379.01\pm0.09$ &  $17.81\pm0.03$ & $10.1368\pm0.0008$ & $7.60\pm0.02$ & $3694\pm51$ &  $0.515\pm0.015$ & $+0.07\pm0.01$ \\
\noalign{\smallskip}
GJ\,4383 & J23585+076 & 23:58:32.74 & $+77.7\pm0.4$ &  $60.9\pm0.3$ & 13.31 & $11.7\pm0.1$ & M3.0\,V &  $0.455\pm0.019$ & $<2$  \\
 & ST3 & +07:39:25.2 & $-319.8\pm0.2$ &  $16.43\pm0.07$ & $10.6312\pm0.0004$ & $7.91\pm0.02$ & $3471\pm51$ &  $0.454\pm0.015$& $+0.04\pm0.01$ \\
\noalign{\smallskip}
\hline
\end{tabular}%
}
\tablebib{ 
{\it a}: \cite{Gaia2016,Gaia2018,Gaia2020};
{\it b}: \cite{Alonso2015};
{\it c}: \cite{PMSU};
{\it d}: \cite{Lepine2013};
{\it e}: \cite{UCAC4};
{\it f}: \cite{Skrutskie2006};
{\it g}: \cite{Schweitzer2019};
{\it h}: \cite{Passegger2018};
{\it i}: \cite{Jeffers2018};
{\it j}: \cite{Reiners2018};
{\it k}: \cite{Schoefer2019}. 
}
\end{table*}
\subsection{Spectroscopic data}
\label{sec:specdata}

The CARMENES survey is spectroscopically monitoring over 300 M-dwarf stars in search for exoplanets. The visual channel of the CARMENES spectrograph covers a wavelength range from 0.52\,$\mu$m to 0.96\,$\mu$m dispersed across 61 \'echelle orders, with a measured resolving power of $R=94\,600$. A more detailed description of the CARMENES instrument was provided by % can be found in
\cite{Quirrenbach2014,Quirrenbach2016}. Reduction of raw spectra is automatically performed by the \texttt{caracal} \citep[CARMENES Reduction And Calibration,][]{Caballero2016} pipeline, which corrects for bias, flat-field, and cosmic rays, and extracts one-dimensional wavelength-calibrated spectra. High-precision RVs are routinely computed with the \texttt{serval} pipeline, which also computes activity indicators such as the chromatic index (CRX) and the differential line width (dLW), along with other spectroscopic indices \citep{Zechmeister2018}. 
%As a first check to find multiple systems within the surveyed stars, we searched for stars showing large RV excursions.
%as a primary source for selecting the multiple system candidates.
A simple search for stars showing RV excursions greater than $\sim0.5$\,km\,s$^{-1}$ revealed eight new multiple stellar system candidates, in addition to the the nine SB2s already announced by \cite{Baroch2018}. A total of 269 high-resolution spectra of these stars were obtained from January 2016 to December 2020 with the visual channel of the CARMENES spectrograph, including the UU\,UMi spectra already studied by \cite{Baroch2018}. Table~\ref{tab:props} lists the  main properties of the stars studied here. It includes their common names and CARMENES identifiers, their equatorial coordinates, distances, and proper motions, their \mbox{$G$-}, \mbox{$V$-}, and $J$-band magnitudes, and different spectroscopic indicators, all computed assuming non-binarity, and therefore they could suffer some bias. It also lists the renormalised unit weight error (RUWE) from {\it Gaia}, which is an indicator of the goodness of the fit of the astrometric observations. A RUWE %value 
$\gtrsim1.4$ may indicate that the source is not single. The astrometric properties of all systems are from the {\it Gaia} Early Data Release 3 \citep{Gaia2020}. 

We also collected additional high-resolution spectroscopic observations from public archives in order to complement our observations for long-period systems. We found a total of 104 further observations, 60 taken with the HARPS spectrograph at La Silla Observatory \citep{HARPS}, 37 with the FEROS spectrograph also at La Silla Observatory \citep{FEROS}, six spectra %were taken 
with the UVES spectrograph at Paranal Observatory \citep{UVES}, and, finally, one %was obtained using
with the CAFE spectrograph at Calar Alto Observatory \citep{CAFE}, obtained by \cite{Jeffers2018} as part of the characterisation of the CARMENES input catalogue of stars. In total, we gathered 373 high-resolution spectra for the nine systems studied in this work, increasing the time span of the observations to up to 15 years for some systems. A summary of the number of observations available from each instrument is given in Table\,\ref{tab:obs}.

\subsection{Photometric data}

In order to check for possibles eclipses or rotational variability, we collected photometric time-series from public archives of surveys such as {\em TESS} \citep{TESS}, ASAS \citep{ASAS}, NSVS \citep{NSVS}, and MEarth \citep{MEARTH}, and from not-yet public data from SWASP \citep{SWASP}. 
We refer to \cite{DiezAlonso2019} for details on these surveys.
Within the context of the CARMENES spectroscopic survey, photometric follow-up was triggered from ground-based observatories for some interesting candidates. In particular, we collected $R$-band photometric data of GJ\,207.1 at the Montsec Astronomical Observatory\footnote{\url{www.oadm.cat}} using the Telescopi {Joan Or\'o} (TJO) and the MEIA2 instrument. TJO is a fully robotic 0.8\,m Ritchey-Chr\'etien equipped with an Andor 2k\,$\times$\,2k CCD camera with a plate scale of 0.36\,arcsec\,pixel$^{-1}$. We also collected $R$-band photometric data of GJ\,282\,C from the the Sierra Nevada Observatory (OSN), Spain, using the 0.9\,m Ritchey-Chr\'etien telescope equipped with a VersArray 2k\,$\times$\,2k CCD camera, and a plate scale of 0.38\,arcsec\,pixel$^{-1}$ \citep{Rodriguez2010}. 
All CCD measurements were obtained by the method of synthetic aperture photometry without binning the CCD pixels. Each CCD frame was corrected in a standard way for bias and flat field. Different aperture sizes were also tested in order to choose the best one for our observations. A number of nearby and relatively bright stars within the frames were selected as check stars, and the best ones were used as reference stars.

To minimise the effect of flares and the systematics, photometric data were binned in equal time intervals. For the SWASP data, in which multiple observations were taken each night, we first checked for the presence of short-duration eclipses and then performed a nightly binning, except for the short-period system GJ\,207.1. We binned the {\em TESS} high-cadence data in 1\,h bins for stars with visible short-term modulations, such as GJ\,207.1, GJ\,282\,C, and UCAC4\,355-020729, while for UU\,UMi, which has a smoother light curve, we chose longer bins of 12\,h. Besides, we applied a $3\sigma$ clipping filter to all datasets. We list the final number of photometric epochs after applying this filtering, with the total number of measurements in parentheses, for each photometric survey and system in the last three columns of Table\,\ref{tab:obs}.

\begin{table*}[t]
\centering
\caption{Number of spectroscopic observations and their time-span, and photometric epochs after and before (in parentheses) filtering and binning.}
\label{tab:obs}
\begin{tabular}{l cccccc cccc} 
\hline\hline
\noalign{\smallskip}
\multirow{2}{*}{System} & \multicolumn{6}{c}{Spectroscopic data} & \multicolumn{4}{c}{Photometric data} \\
 \cmidrule(l){2-7} \cmidrule(l){8-11}
 & CARM & HARPS & FEROS & UVES & CAFE & $\Delta t$ [d] & {\em TESS} & ASAS & SWASP & Others \\
\noalign{\smallskip}
\hline
\noalign{\smallskip}
\object{GJ\,207.1} & 14 & 1 & 2 & $\cdots$ & $\cdots$ & 5010 & 1051(30645) & 452(463) & 3699(3853) & 539(560)\tablefootmark{a} \\
\object{UCAC4 355--020729} & 27 & $\cdots$ & $\cdots$ & $\cdots$ & $\cdots$ & 738 & 480(14580) & 553(572) & $\cdots$ & $\cdots$ \\
\object{GJ\,282\,C}& 43 & 38 & 3 & $\cdots$& $\cdots$ & 2464 & 542(16279) & 476(504) & 209(15238) & 1705(1715)\tablefootmark{b} \\
\object{LP\,427--016} & 22 & $\cdots$ & 2 & $\cdots$ & $\cdots$ & 2830 & $\cdots$ & 320(335) & 65(3821) & 196(199)\tablefootmark{c} \\
\object{GJ\,3626} & 42 & $\cdots$ & $\cdots$ & $\cdots$ & $\cdots$ & 2830 &  $\cdots$ & $\cdots$ & 96(5090) & $\cdots$ \\
\object{UU\,UMi} & 29 & $\cdots$ & $\cdots$ & $\cdots$ & 1 & 2674 & 329(121486) & $\cdots$ & $\cdots$ & 2696(2803)\tablefootmark{d} \\
\object{GJ\,3916} & 38 & 3 & 23 & 6 & $\cdots$ & 5609 & $\cdots$ & 457(472) & 376(73874) & $\cdots$\\
\object{GJ\,912} & 28 & 9 & 3 & $\cdots$ & $\cdots$ & 4107 & $\cdots$ & 304(311) & 256(90025) & $\cdots$ \\
\object{GJ\,4383} & 26 & 9 & 4 & $\cdots$ & $\cdots$ & 3425 & $\cdots$ & 234(241) & 239(22787) & $\cdots$ \\
\noalign{\smallskip}
\hline
\end{tabular}
\tablefoot{
\tablefoottext{a}{TJO;} \tablefoottext{b}{OSN;} \tablefoottext{c}{NSVS;} \tablefoottext{d}{MEarth.}
}
\end{table*}

\begin{table}[t]
\centering
\caption{Astrometric positions of GJ\,282\,C computed from NACO observations\tablefootmark{a}.}
\label{tab:NACO}
\begin{tabular}{l ccc} 
\hline\hline
\noalign{\smallskip}
Obs. date & $\rho$ [mas] & $\theta$ [deg] & $\Delta m$ [mag]\tablefootmark{b} \\
\noalign{\smallskip}
\hline
\noalign{\smallskip}
2015-01-22 & $516.0\pm3.0$ & $134.87\pm0.38$ & $2.400\pm0.050$ ($L_p$)\\
2016-02-02 & $455.3\pm1.6$ & $135.02\pm0.15$ & $2.560\pm0.010$ ($IB_{2.18}$)\\
2016-03-26 & $442.5\pm2.6$ & $133.77\pm0.32$ & $2.360\pm0.050$ ($L_p$)\\
\noalign{\smallskip}
\hline
\end{tabular}
\tablefoot{
\tablefoottext{a}{GJ\,282\,C is, in turn, a common proper motion companion at 57.9\,arcsec to the early K-dwarf star \object{V869 Mon} \citep[HD~61606, GJ\,282;][]{Eggen1956,Bergh1958}.} 
\tablefoottext{b}{$\Delta m$ refers to the difference in magnitude in the band indicated in parenthesis.} 
}
\end{table}

\subsection{Direct imaging data}

A bibliographical search revealed that the system GJ\,282\,C was observed and spatially resolved by \citet[][see their Fig.~10]{Kammerer2019} in an adaptive optics search for companions with the NAOS-CONICA imager \citep[NACO;][]{NACO1,NACO2}. 
NACO is an adaptive optics-assisted instrument that provides imaging, imaging polarimetry, and corono-graphy in the 1--5\,$\mu$m range. The adaptive optics system, NAOS, is equipped with both visible and infrared wavefront sensors, and contains several dichroics that split the light from the telescope between CONICA and one of the NAOS wavefront sensors. CONICA is the infrared camera and spectrometer attached to NAOS and is equipped with an Aladdin 1024\,$\times$\,1024 pixel InSb array detector. Observations were performed on 22 January 2015 and 26 March 2016 in the $L_p$ filter and on 2 February 2016 in the $IB_{2.18}$ filter. The plate scale of NACO during these observations was 27.19\,mas\,pixel$^{-1}$ for the $L_p$ filter and 13.27\,mas\,pixel$^{-1}$ at 2.18\,$\mu$m, which provided fields of view of $28\times28$\,arcsec$^2$ and $14\times14$\,arcsec$^2$, respectively. Short individual exposure times of 0.2\,s in the $L_p$ filter and 0.5\,s at 2.18\,$\mu$m were used to avoid saturation of the primary star. We downloaded these data from the European Southern Observatory archive\footnote{\url{http://archive.eso.org/cms.html}}, and reduced and analysed them consistently using routines within {\tt IRAF} \citep{Tody1986,Tody1993}. Individual images were sky-subtracted, flat-field corrected using a superflat, aligned, and combined to produce the final image. We performed aperture photometry of both stars in this combined image with {\tt daophot} to determine the magnitude difference between the primary and secondary. We determine the pixel position of the stars in each individual image using the {\tt imcentroid} routine. The average value and the standard error of the mean were adopted as the final value and its error bars. The detector positions were transformed into the sky positions using the average position angle during observations in the $L_p$ band and the plate scale and orientation from the image header. We list the astrometric positions derived from these observations in Table~\ref{tab:NACO}.

\section{Data analysis} \label{sec:rvs}
\subsection{Radial velocity measurements} \label{sec:rvs1}

After the selection of the multiple system candidates based on excursions greater than $\sim0.5$\,km\,s$^{-1}$ of the RVs provided by the CARMENES pipeline \texttt{serval}, we searched for the signature of companion stars in the spectra using \texttt{todmor} \citep{Zucker2003}, a new implementation of the two-dimensional cross-correlation method \texttt{todcor} \citep{Zucker1994} for multi-order spectra. This technique uses two different template spectra for each component, scaled according to their flux ratio. The two spectra are simultaneously Doppler-shifted, constructing a two-dimensional cross-correlation function (CCF) from which the RVs of the two components can be computed. Also, the flux ratio of the system in the wavelength range of the observations can be obtained. As templates for the calculation of the CCFs, we employed synthetic PHOENIX stellar models \citep{Husser2013}. For each system, we explored a grid of values for effective temperatures, flux ratio, and spectral line broadening (to account for rotation velocity) to look for the combination that maximised the CCF peak. Spectral orders with a signal-to-noise ratio of their CCFs below 5 or with telluric contamination were discarded beforehand.

In general, we prefer to use synthetic spectra over high signal-to-noise ratio spectra of real stars as templates because the latter are prone to systematics, due to the uncertainty of the absolute RVs of the different template stars and by residual contamination due to tellurics on the co-added spectra. However, the molecular bands of the coolest stars modelled by synthetic spectra show discrepancies with observations due to inaccurate or incomplete molecular opacities \citep{Allard2012,Passegger2020}, and for this reason, the use of real templates could reveal signals from a secondary body that otherwise, with synthetic spectra, could not be resolved. Therefore, for those stars for which the signature of the secondary component was not detected with synthetic spectra, we used high signal-to-noise co-added data from the CARMENES stars GJ\,752\,A (\object{V1428~Aql}, Karmn J19169+051N, M2.5\,V) and GJ\,1253 (\object{Wolf~1069}, Karmn J20260+585; M5.0\,V) as fiducial templates of mid and late M-dwarf stars. The main cases in which real spectra were used are in stars with a small RV difference between the components (UU\,UMi) and in those with a low flux ratio (GJ\,4383).
%The reason why we prefer to use synthetic spectra over real ones is because the use of real spectra is prone to suffer from systematics, either by different absolute RVs of the different template stars, or by residual contamination due to tellurics on the coadded spectra.
We computed the absolute RVs of the template stars by cross-correlating them with synthetic spectra.

For the GJ\,3916 and GJ\,4383 systems, the \texttt{todmor} analysis revealed three different signals in the CCF. In these cases, we first fitted the template to the component with the highest flux, setting the flux of the secondary component to zero, and subtracted it from the observed spectra. Then, we reanalysed all the residuals with \texttt{todmor} again, and extracted the RVs of the other two components. These RVs were consistent with these stars being hierarchical triple systems. However, due to their configuration, there were orbital phases in which the RVs of the primary and one of the two secondary stars almost overlapped, in which case the RV of the dimmer component could not be extracted.

Finally, for systems where the signature of the secondary component was not found in the spectra, that is, SB1s, we used the RVs provided by the CARMENES pipeline \texttt{serval} \citep{Zechmeister2018}. In this case, the spectra of the primary component is not significantly affected by the much fainter companion and \texttt{serval} provided significantly better RV precision than \texttt{todmor}. This pipeline is based on a least-squares fitting of the individual spectra against a high signal-to-noise ratio co-added spectrum constructed from all observations of the same object \citep{Anglada2012}. Therefore, several observations of the star are needed to produce the high signal-to-noise ratio template. We set this limit to five, thus allowing us to measure RVs using \texttt{serval} for CARMENES spectra and for the HARPS data of GJ\,282\,C and GJ\,912.
%,which have enough observations to construct a high signal-to-ratio template. 
A small number of FEROS and HARPS spectra are also available for four of the five analysed SB1 systems, some of them taken $\sim$10\,a before the CARMENES observations. Although isolated, they are crucial for the characterisation of long-period systems. However, two difficulties arise when trying to use these data. On the one hand, the RV offset between the different instruments needs to be computed. On the other hand, the small number of observations does not allow us to derive precise RVs using \texttt{serval}.
To solve this issue, we computed the RVs of the two FEROS spectra and the single HARPS spectrum of GJ\,207.1 with \texttt{todmor} using the one-dimensional configuration (i.e. setting the flux ratio to zero), which gives an absolute RV with respect to the synthetic spectra. In order to constrain the RV zero point of these measurements from \texttt{todmor}, we also computed the CARMENES RVs with \texttt{todmor}. We then corrected the RV offset as the difference with respect to \texttt{serval} RVs. The list of computed RVs and uncertainties for each system and instrument, together with the method used to compute them, eiher \texttt{todmor} (T) or \texttt{serval} (S), are listed in Tables\,\ref{tab:RVSB1}--\ref{tab:RVST3}. 
%in the Appendix.

\subsection{Orbital parameter determination} \label{sec:orbit}

We determined the orbital parameters by fitting an N-body Keplerian model to the RVs of all the components simultaneously. For binary systems, we fitted the six (seven for SB2 systems) parameters defining a spectroscopic Keplerian orbit: the period $P$, the semi-amplitude of the components $K_A$ (and $K_B$ for SB2), the time of periastron passage $T_0$, the eccentricity $e$, and the argument of periastron $\omega$, using $e \sin \omega$ and $e \cos \omega$ as actual parameters. Additionally, we fitted an RV offset for each instrument ($\gamma_{\rm Inst.}$), except for the corrected FEROS RVs, which was fixed as explained above. We defined the systemic velocity $\gamma$ as $\gamma_{\rm CARM}$ plus the offset between the CARMENES observations computed with \texttt{serval} and \texttt{todmor}.
%to the offset of the CARMENES RVs.
We also allowed for an adjustable RV jitter term for each instrument ($\sigma_{\rm Inst.}$), which was added in quadrature to the errors as described in \cite{Baluev2009}. We employed a hierarchical model for the triple systems (i.e. an inner binary, and a more distant star, orbiting around the centre of mass of the inner binary, without mutual interactions). We used the subscripts `AB' to refer to the wide orbit of the inner-system centre of mass around the outer and brighter component `A', and `B' to refer to the inner system, while we used `Ba' and `Bb' for the two inner components (there are no `Aa' and `Ab' components).

For GJ\,282\,C, we also integrated the three astrometric measurements and performed a simultaneous fit of the RVs, the relative astrometric positions, and a combination of the proper motions from the \textit{Hipparcos}-\textit{Gaia} Catalog of Accelerations \citep[HGCA,][]{Brandt2018} using the \texttt{orvara} software \citep{Brandt2021}. In addition to the six parameters defining the orbit of an SB1 system, the inclusion of the astrometric orbit model added three more parameters: the inclination of the system ($i$), the longitude of the ascending node ($\Omega$), and the semi-major axis of the orbit ($a$). The HGCA accelerations also provided the proper motion ($\mu_{\alpha} \cos{\delta}$ and $\mu_{\delta}$) and parallax ($\varpi$) of the barycentre.

We computed orbital parameters and uncertainties by sampling the posterior probability distribution with the \texttt{emcee} sampler \citep{Foreman2013}, an implementation of the affine-invariant ensemble sampler Markov chain Monte Carlo \citep{Goodman2010}. We sampled the posterior distribution with 250 walkers and $5\times10^3$ steps, after a burn-in of $5\times10^3$ samples, unless indicated otherwise for some specific cases. For systems with orbits larger than the time-span of the observations, we placed broad and uninformative priors in all parameters, uniformly sampling the period up to 100\,000\,d and the eccentricity to 0.9.

From the orbital parameters of the systems, we derived the projected semi-major axes, $a_{\rm A,B}\sin i$, and the minimum masses $M_{\rm A,B}\sin^3 i$ for SB2 systems, or the binary mass functions $f(M)$ for SB1 systems, which were computed as \citep{Hilditch2001}:

\begin{equation}
    a_{\rm A,B}\sin i=\frac{(1-e^2)^{1/2}}{2\pi}K_{\rm A,B}P~,
\end{equation}

\begin{equation}
    M_{\rm A,B}\sin^3 i=\frac{(1-e^2)^{3/2}}{2\pi G}(K_{\rm A}+K_{\rm B})^2K_{\rm B,A}P~,
\end{equation}

\begin{equation}
\label{eq3}
    f(M)\equiv\frac{M_{\rm B}^3 \sin^3 i}{(M_{\rm A}+M_{\rm B})^2}=\frac{P}{2\pi G}(1-e^2)^{3/2}K_{\rm A}~,
\end{equation}

\noindent where $G$ is the gravitational constant. For SB1 systems, for which we cannot directly compute the value of $M_{\rm A,B}\sin^3 i$, we computed the secondary minimum mass, $M_{\rm B, min}$, by numerically solving Eq.\,\ref{eq3} for $M_{\rm B}$ with $i=90$\,deg and assuming the primary mass listed in Table\,\ref{tab:props}.

\subsection{Rotation period determination} \label{sec:rotation}

To fully characterise the systems, we also analised the photometric data searching for variability due to stellar rotation. We computed a generalised Lomb-Scargle (GLS) periodogram \citep{Zechmeister2009} for each photometric dataset. To assess the significance of the signals found, we computed the false alarm probability (FAP) with a boostrap randomisation of the input data, and considered a signal significant if it reached a value $<0.1\%$.

For GJ\,282\,C, the high precision of the {\em TESS} photometric data revealed clear deviations from a strictly periodic sinusoid and a low-period modulation of lower amplitude. To improve the accuracy of the rotational period determination of this system, we inferred the periodicity in the light curves by using a Gaussian process regression (GP) with a sum of two quasi-periodic covariance functions \cite[see e.g.][]{Rasmussen2006,Angus2018}, therefore determining two different rotation periods, which we assigned to the individual components. Uncertainties were also computed by sampling the posterior probability with the \texttt{emcee} sampler.

\section{Results} \label{sec:results}

We report in this section on the result of the orbital fitting analysis and its interpretation for each of the nine systems separately. This includes five SB1, two SB2, and two ST3.

%\subsection{Single-line spectroscopic binaries} \label{sec:SB1}

\begin{table*}[th!]
\centering
\caption{Computed and derived parameter for the SB1 systems.}
\label{tab:paramssingle}
\begin{tabular}{l cccc} 
\hline\hline
\noalign{\smallskip}
\multirow{2}{*}{Parameter} & \multicolumn{4}{c}{System}\\
\noalign{\smallskip}
\cline{2-5}
\noalign{\smallskip}
 & GJ\,207.1 & GJ\,912 & GJ\,3626  & LP\,427-016 \\
\noalign{\smallskip}
\hline
\noalign{\smallskip}
\multicolumn{5}{c}{\textit{Orbital parameters}}\\
\noalign{\smallskip}
\hline
\noalign{\smallskip}
$P$ [d] & \AS{0.60417356}{0.00000024}{0.00000026} & \AS{5188}{58}{55} & \AS{2996}{31}{30} &  $^{>6\,400}_{<29\,000}$ \\
\noalign{\smallskip}
$T_{0}$ [BJD-2457000] & \AS{867.06173}{0.00016}{0.00015} & \AS{3\,099}{47}{44} & \AS{1\,033}{3}{3} &$\cdots$ \\
\noalign{\smallskip}
$K_A$ [km\,s$^{-1}$] & \AS{35.133}{0.032}{0.031} & \AS{1.356}{0.084}{0.074} & \AS{0.9444}{0.0051}{0.0048} & $^{>0.25}_{<1.4}$ \\
\noalign{\smallskip}
$e\sin \omega$ & 0 (fixed) & \AS{-0.4949}{0.0046}{0.0049} & \AS{-0.4172}{0.0032}{0.0033}  & $0$ (fixed) \\
\noalign{\smallskip}
$e\cos \omega$ & 0 (fixed) & \AS{-0.380}{0.021}{0.020} & \AS{-0.0785}{0.0065}{0.0066}  & $0$ (fixed) \\
\noalign{\smallskip}
$e$ & 0 (fixed) & \AS{0.6239}{0.0098}{0.0097} & \AS{0.4246}{0.0024}{0.0024}  &  $0$ (fixed) \\
\noalign{\smallskip}
$\omega$ [deg] & 0 (fixed) & \AS{232.5}{1.8}{1.6}  & \AS{259.34}{0.92}{0.94} &  $0$ (fixed) \\
\noalign{\smallskip}
$\gamma_{\rm CARM}$ [km\,s$^{-1}$] & \AS{30.994}{0.024}{0.023} & \AS{0.009}{0.027}{0.029}  & \AS{-0.6004}{0.0015}{0.0016} &  $^{>0.0}_{<0.9}$\\
\noalign{\smallskip}
$\gamma_{\rm HARPS}$ [km\,s$^{-1}$] & $\cdots$  & \AS{-0.734}{0.025}{0.027}   & $\cdots$  &  $\cdots$ \\
\noalign{\smallskip}
$\gamma$\tablefootmark{a} [km\,s$^{-1}$] & \AS{30.836}{0.024}{0.024}  & \AS{17.479}{0.031}{0.029}  & \AS{-59.870}{0.017}{0.017} & $^{>13.2}_{<14.0}$ \\
\noalign{\smallskip}
$\sigma_{\rm CARM}$ [km\,s$^{-1}$] & \AS{0.079}{0.026}{0.018} & \AS{0.00346}{0.00077}{0.00062}  & \AS{0.00201}{0.00045}{0.00041} &  $\cdots$ \\
\noalign{\smallskip}
$\sigma_{\rm HARPS}$ [km\,s$^{-1}$] & \AS{0.35}{0.38}{0.25} & \AS{0.0056}{0.0045}{0.0029}  & $\cdots$  & $\cdots$ \\
\noalign{\smallskip}
$\sigma_{\rm FEROS}$ [km\,s$^{-1}$] & \AS{0.46}{0.35}{0.32}  & \AS{0.22}{0.30}{0.16} & \AS{0.0088}{0.0098}{0.0062}  &  $\cdots$ \\
\noalign{\smallskip}
\noalign{\smallskip}
\hline
\noalign{\smallskip}
\multicolumn{5}{c}{\textit{Derived parameters}}\\
\noalign{\smallskip}
\hline
\noalign{\smallskip}
$a_{A}\sin i$ [au] & \AS{0.0019511}{0.0000018}{0.0000017} & \AS{0.505}{0.032}{0.028} & \AS{0.2355}{0.0012}{0.0011} &  $^{>0.14}_{<3.8}$\\
\noalign{\smallskip}
%$a_{A}\sin i$ [mas] & \AS{0.12415}{0.00020}{0.00020} & \AS{30.1}{2.67}{2.3} & \AS{10.393}{0.083}{0.078} & \AS{}{}{} & 4.3 to 149.7 \\
%\noalign{\smallskip}
$f(M)$ [M$_{\odot}$]& \AS{0.0027146}{0.0000074}{0.0000072} & \AS{0.00064}{0.00011}{0.00009} & \AS{0.0001941}{0.0000017}{0.0000016} &  $^{>0.000009}_{<0.008}$ \\
\noalign{\smallskip}
$M_{\rm B,min} $ [M$_{\odot}$] & \AS{0.0979}{0.0025}{0.0025} & \AS{0.0600}{0.0030}{0.0036}  & \AS{0.0331}{0.0011}{0.0011} & 0.012 \\
\noalign{\smallskip}
\hline
\noalign{\smallskip}
\multicolumn{5}{c}{\textit{Estimated parameters}}\\
\noalign{\smallskip}
\hline
\noalign{\smallskip}
$P_{\rm rot}$ [d] & \AS{0.60435}{0.00026}{0.00026} & \AS{41.8}{1.3}{1.3} & $\cdots$ & \AS{36}{3}{3}\\
\noalign{\smallskip}
$i$ [deg] & \AS{13.8}{2.2}{2.2} & $>17$ & $>19$  & $\cdots$\\
\noalign{\smallskip}
$a_A$ [au] & \AS{0.0082}{0.0015}{0.0011} & $^{>0.5}_{<3.0}$ & $^{>0.235}_{<0.656}$ &  $\cdots$\\
\noalign{\smallskip}
%$a_A$ [mas] & \AS{0.521}{0.93}{0.72} & \AS{61}{132}{25} & \AS{12.0}{7.2}{1.5} & \AS{}{}{} & $\cdots$\\
% \noalign{\smallskip}
$a$ [au] & \AS{0.01454}{0.00085}{0.00060} & $^{>4.5}_{<6.0}$ & $^{>3.08}_{<3.49}$ & $\cdots$ \\
\noalign{\smallskip}
%$a$ [mas] & \AS{0.925}{0.052}{0.039} & \AS{300}{66}{38} & \AS{137.0}{3.7}{3.0} & \AS{649}{75}{73} & 400 to 1300\\
% \noalign{\smallskip}
$M_{\rm B}$ [M$_{\odot}$] & \AS{0.63}{0.20}{0.13} & $^{>0.06}_{<0.21}$ & $^{>0.033}_{<0.11}$ & $^{>0.012}_{<0.13}$\\
\noalign{\smallskip}
$q\equiv M_{B}/M_{A}$ & \AS{1.29}{0.42}{0.26} & $^{>0.115}_{<1.0}$ & $^{>0.08}_{<0.25}$ &  $^{>0.024}_{<0.27}$\\
\noalign{\smallskip}
$K_{B}$ [km\,s$^{-1}$] & \AS{27.2}{6.9}{6.6} & $^{>1.4}_{<11.8}$ & $^{>3.78}_{<11.46}$ &  $\cdots$ \\
\noalign{\smallskip}
\hline
\end{tabular}
\tablefoot{ \tablefoottext{a}{Parameter adopted as the barycentre RV. Corresponds to $\gamma_{\rm CARM}$ plus the RV shift between the CARMENES RVs computed with \texttt{serval} and with \texttt{todmor}}.  }
\end{table*}

\subsection{Single-line spectroscopic binaries (SB1s)}
\subsubsection{\object{GJ\,207.1}}

\begin{figure}[t]
\centering
\includegraphics[width=\columnwidth]{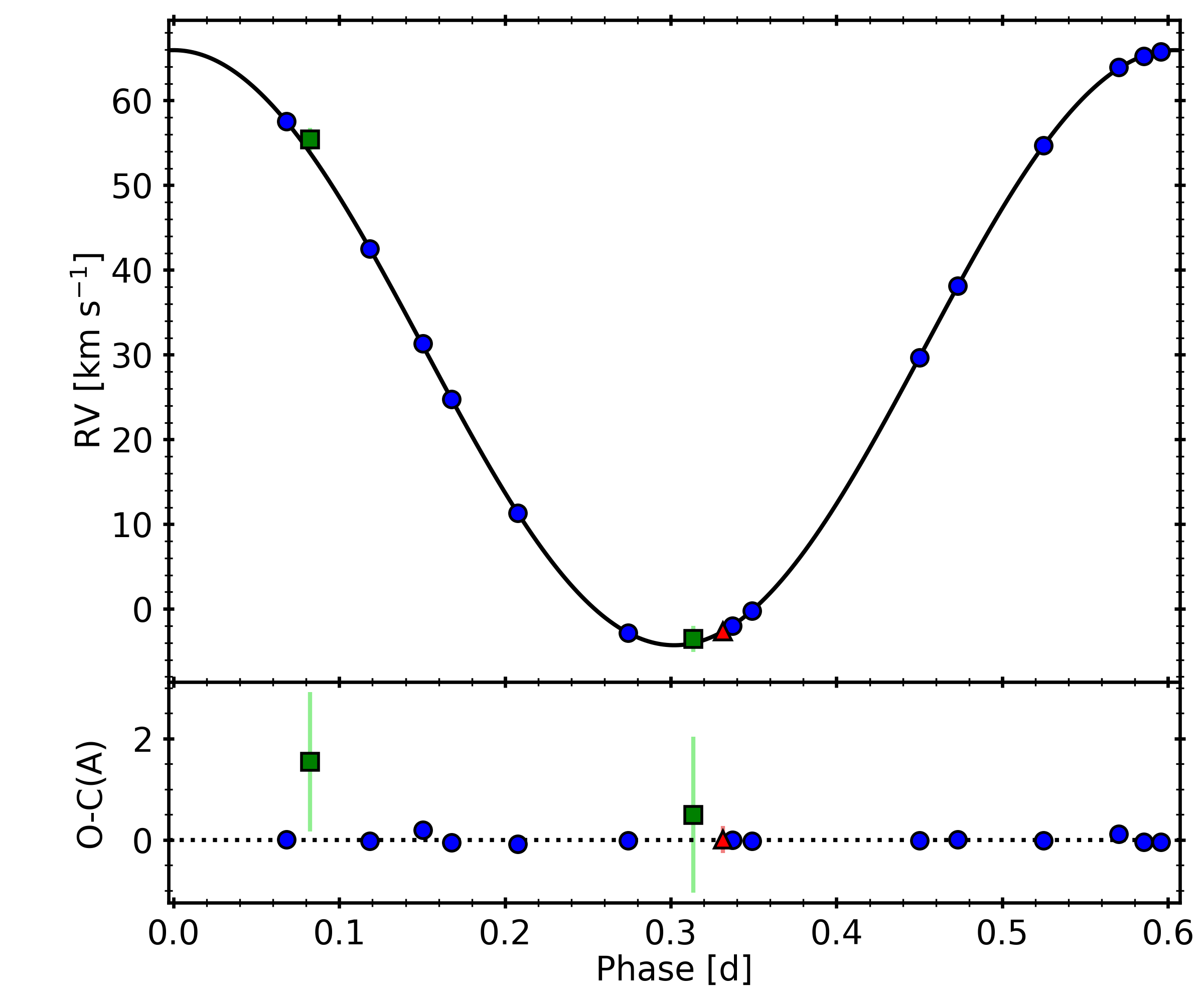}
\caption{Best orbital fit to the RVs of GJ\,207.1 from CARMENES (blue circles), HARPS (red triangle), and FEROS (green squares), phase-folded to the period of the system. The bottom panel shows the residuals from the fit.}
          \label{J05337orbit}%
\end{figure}

\begin{figure}[t]
\centering
\includegraphics[width=\columnwidth]{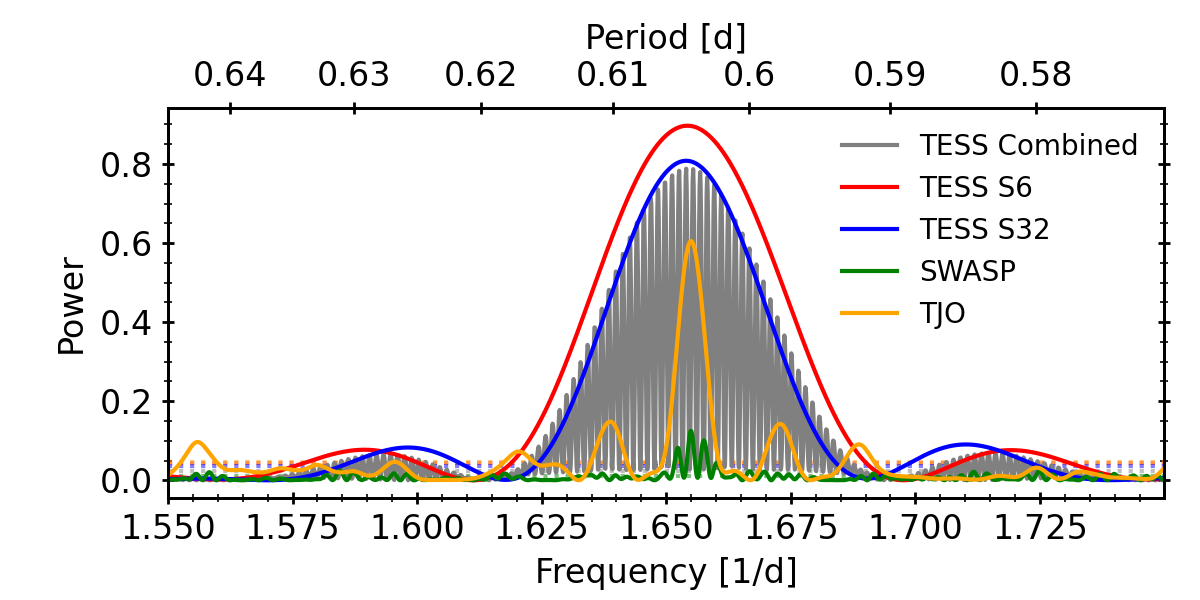}
\includegraphics[width=\columnwidth]{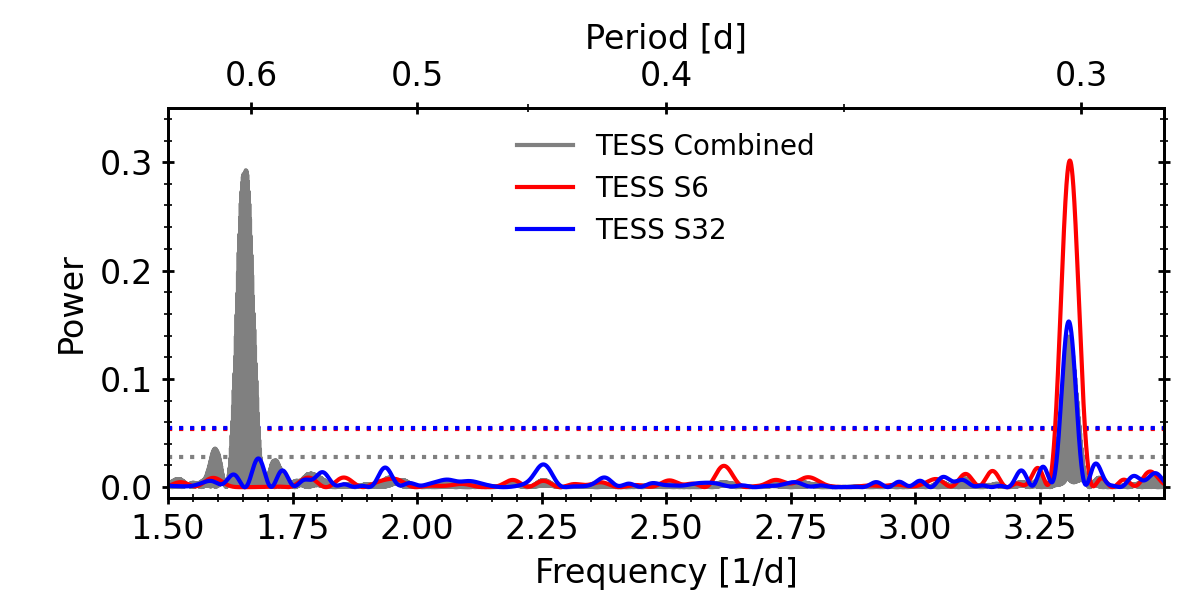}
\caption{\textit{Top:} GLS periodogram of the combined {\em TESS} data (grey) and their sectors 6 (red) and 32 (blue), SWASP (green), and TJO (orange) photometry of GJ\,207.1. \textit{Bottom:} GLS periodogram of the residuals of the combined {\em TESS} data (grey) and their sectors 6 (red) and 32 (blue) after the removal of the significant signals. The horizontal dotted lines of the corresponding colors indicate the corresponding 0.1\,\% FAP level.}
          \label{J05337GLS}%
\end{figure}

\begin{figure*}[t]
\centering
\includegraphics[width=\columnwidth]{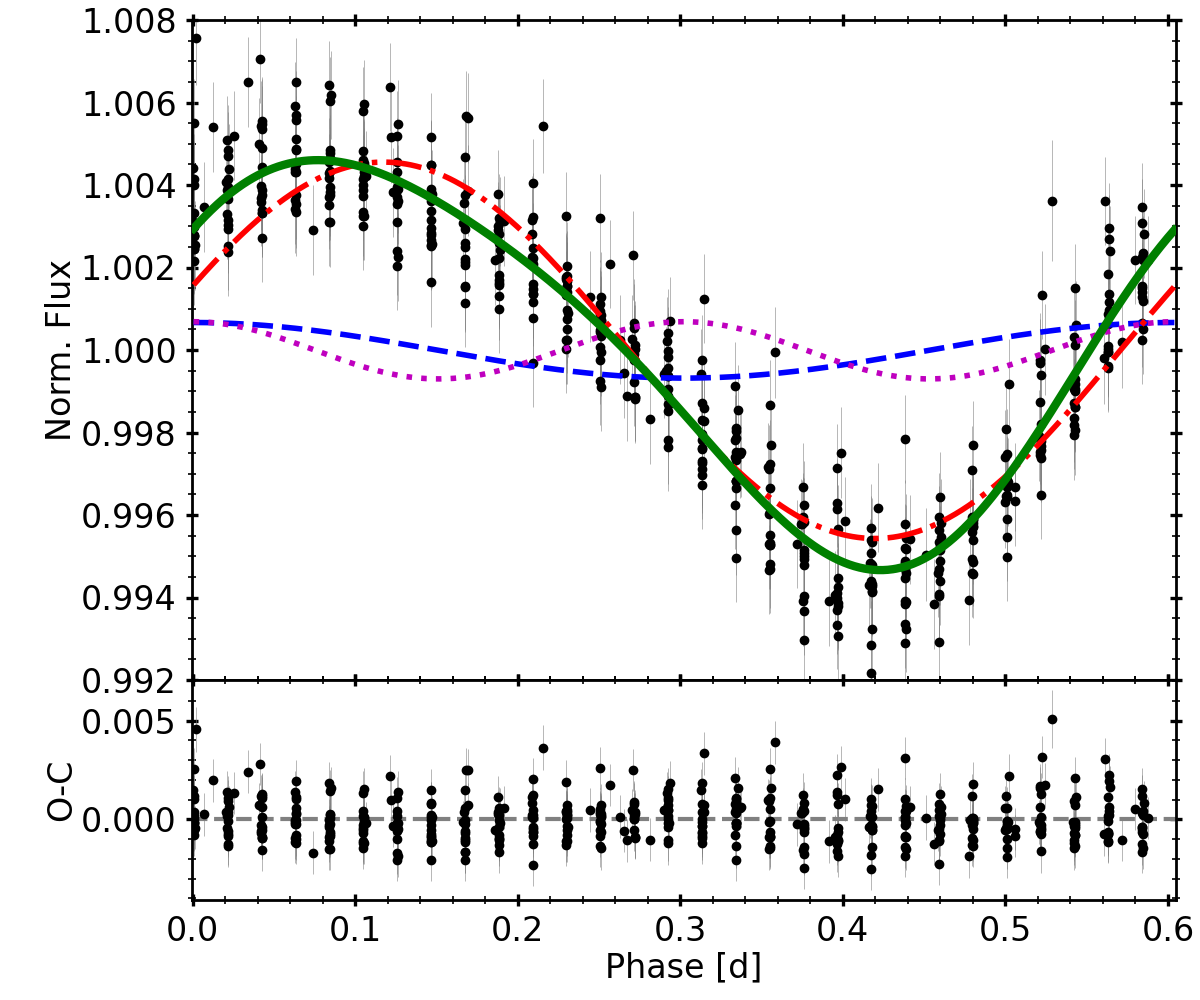}
\includegraphics[width=\columnwidth]{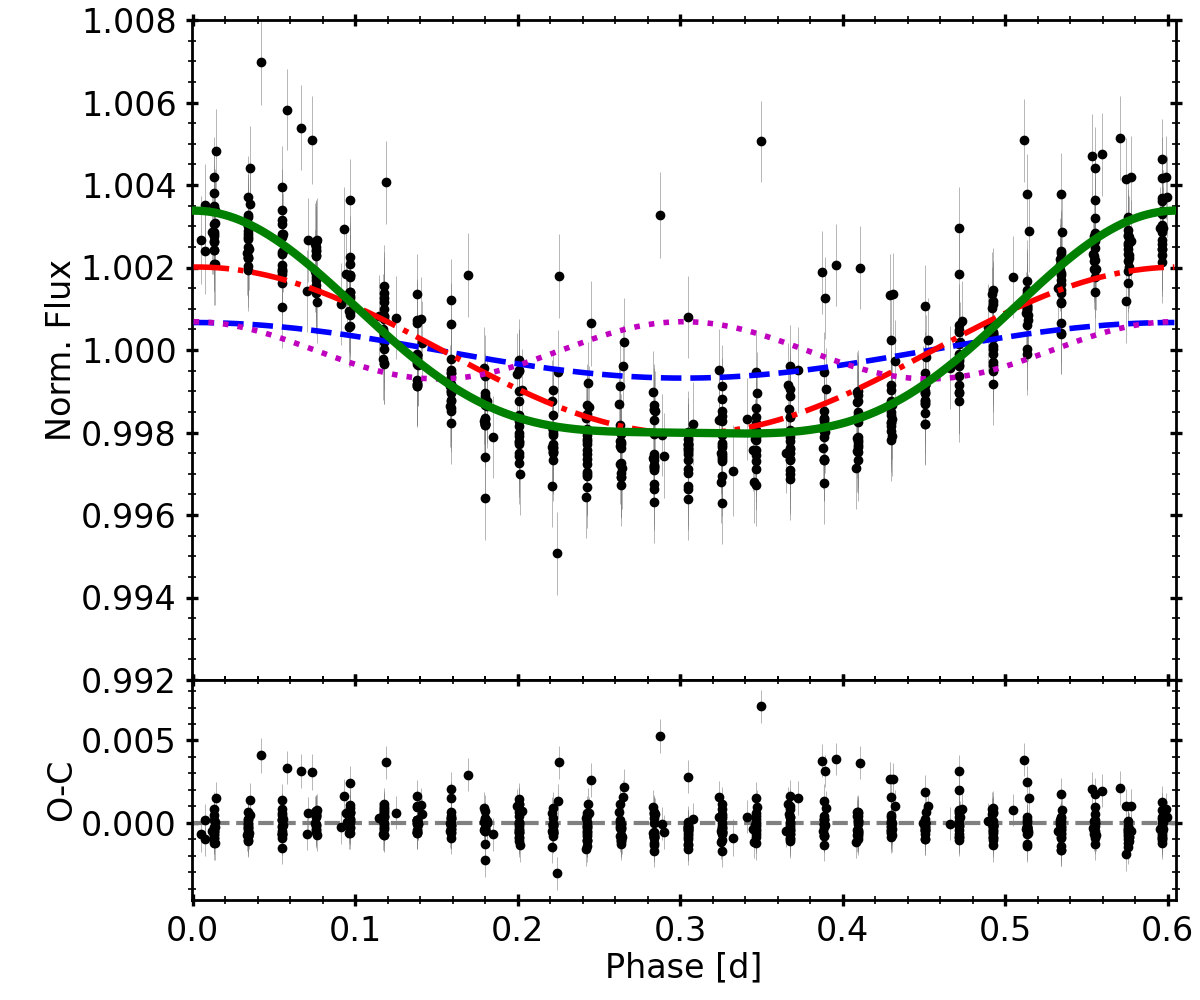}
\caption{Hourly binned photometry of GJ\,207.1 from {\em TESS} sectors 6 (\textit{left}) and 32 (\textit{right}), phase folded to the orbital period. We show the best fit, described by Eq.\,\ref{eqamp}, as a green solid line. The activity-induced variability is shown as a dash-dotted red line, while the magenta dotted line and the dashed blue line describe the ellipsoidal effect and the Doppler beaming, respectively. The bottom panels show the residuals of the best fit.}
          \label{J05337TESS}%
\end{figure*}

GJ\,207.1 is an M2.5\,V star located at $15.783\pm0.012$\,pc. 
Also known as Wachmann's Flare Star, it was first classified as a variable flaring star by \cite{Wachmann1939}, and has been repeatedly detected as an extreme-ultraviolet and soft X-ray source by wide-field surveys \citep{Pounds1993,Lampton1997,Ansdell2015}. It is listed as a member of the young disc population \citep{Montes2001}, with an estimated age of 100\,Ma \citep{Lowrance2005}. However, these estimates are based on its high level of activity and on its Galactic space-velocity components, which could be affected by binarity, therefore biasing the kinematic and age assignments. A projected rotation velocity of $v \sin i = 9.8\pm1.5$\,km\,s$^{-1}$ was reported by \cite{Jeffers2018} using CARMENES spectra. It was classified by \cite{Reiners2012} as a possible spectroscopic binary. High-resolution images ruled out companions beyond $0.2$\,arcsec with $\Delta I>1$\,mag and beyond $0.5$\,arcsec $\Delta I>3$\,mag with \citep{Jodar2013}, and planets with masses $>$10\,$M_{\rm Jup}$ beyond 12.3\,au \citep{Biller2007}, assuming an age of 100\,Ma.

A total of 14 CARMENES spectra were taken between November 2016 and November 2018. Three additional spectra are publicly available in the HARPS and FEROS archives, which were taken between 9 and 12 years before the CARMENES measurements. Although the \texttt{serval} RVs showed a large dispersion of $\sim$60\,km\,s$^{-1}$, no companion was detected with \texttt{todmor} in any of the three datasets, either using synthetic or observed template spectra. Therefore, we used the RVs provided by the CARMENES pipeline.
%For this reason, we decided to use the CARMENES \texttt{serval} RVs because they may not be significanlty affected by a very dim companion and they provide much better precision than \texttt{todmor}.
As explained in Sect\,\ref{sec:rvs1}, to put the HARPS and FEROS \texttt{todmor} measurements to the same level as the CARMENES RVs, we subtracted from them the mean difference between the CARMENES observations computed with \texttt{serval} and \texttt{todmor}. This correction, added to the RV zero point of the \texttt{serval} measurements from CARMENES ($\gamma_{\rm CARM}$) was assigned as the absolute RV of the barycentre of the system ($\gamma$).

We list the orbital parameters of the system in the second column of Table\,\ref{tab:paramssingle}, and we show in Fig.\,\ref{J05337orbit} the best fit to the RVs as a function of the orbital phase. Although the solution with non-zero eccentricity has a slightly better likelihood, it was not significantly different from a circular solution within the errors, putting a $1\sigma$ upper limit of $e=0.0017$.
%enough to compensate for the addition of two more parameters, and the obtained uncertainties are compatible with zero eccentricity.
Therefore, we decided to keep $e\sin \omega$ and $e \cos \omega$ fixed to zero. We found a sub-day orbital period of 0.60417356\,d, with an uncertainty of less than 0.1\,s, and a semi-amplitude of the RV signal of \AS{35.133}{0.032}{0.031}\,km\,s$^{-1}$. Given that the tidal circularisation timescale is small for orbital periods below 10 days, assuming a circular orbit for this system is further justified \citep{Mazeh2008}. From this solution and the estimated mass of the primary component in Table\,\ref{tab:props}, we solved Eq.\,\ref{eq3} to obtain a minimum secondary mass of $0.0979\pm0.0025$\,M$_{\odot}$, which is equivalent to a minimum mass ratio of 0.2.

%\textbf{The short orbital period of this system may cause significant RV shifts during the exposure time of the observations, which we estimate to reach up to 7.6\,km\,s$^{-1}$, depending on the orbital phase at which the observation was taken and assuming the exposure time of 30\,min. This shift produces a broadening of the spectral lines that causes an overestimation of the true $v \sin i$ from the spectral analysis. However, we can estimate the effect of this RV shift over the measured $v \sin i$ taking into account that the variance of the observed line profile is the addition of the variances of the kernels defining the true line profile and the RV shift. The kernel of both the observed and true line profiles can be defined, to first order, by a Wigner semicircle distribution \citep{Gray2005} with radius equal to the projected rotational velocity, and variance $(v\sin i)^2/4$, while the RV shift effect is defined by a square function with length $l$ equal to the shift and variance $l^2/12$. Therefore, the true projected rotational velocity, $v_{\rm T} \sin i$, is defined by $v_{\rm T} \sin i = 2\sqrt{(v \sin i)^2/4-l^2/12}$. This yields a minimum value of $v_{\rm T} \sin i=8.8$\,km\,s$^{-1}$ for $l=7.6$\,km\,s$^{-1}$. Since this value is within the uncertainty of the measured $v \sin i = 9.8\pm1.5$\,km\,s$^{-1}$ even assuming the maximum RV shift, we omit this correction for the subsequent calculations}.

We analysed the available photometric data listed in Table\,\ref{tab:obs} to look for significant signals in the GLS periodograms. We found no significant periods in the ASAS photometry. However, as shown in the top panel of Fig.\,\ref{J05337GLS}, we found very significant signals at $0.6042\pm0.0012$, $0.6043\pm0.0003$, and $0.60460\pm0.00013$\,d in the TJO, SWASP, and {\em TESS} photometry, respectively, very similar to the orbital period of the system. We also found
the same significant period at $0.605\pm0.007$\,d when analysing the {\em TESS} data separately for sectors 6 and 32. After removal of the main signal, the residuals of the combined data still show an accumulation of significant peaks around 0.6\,d. 
This accumulation could be an indication that the signal, while maintaining the same period, slightly changed its phase between the two sectors, producing an interference pattern in the periodogram. The periodograms of the residuals of the individual sectors, however, do not show any significant signal around 0.6\,d, but they do at the first harmonic at about 0.3\,d, as can be seen in the bottom panel of Fig.\,\ref{J05337GLS}.

Tidal evolution theory predicts that the synchronisation and alignment timescales should be two or three orders of magnitude shorter than the circularisation timescale \citep{Zahn1989,Mazeh2008}. Given that the orbit of GJ\,207.1 is already circularised, and that the orbital and rotation periods are virtually identical, one may expect that spin-orbit alignment should have already been achieved. If that is the case, we can estimate the orbital inclination by combining the measured $v \sin i$ (9.8\,km\,$^{-1}$, Table\,\ref{tab:props}) of the star together with the measured rotation period and the stellar radius. This yields a significantly pole-on orbital inclination of $i=$\AS{13.8}{2.2}{2.2}\,deg, which results in a companion mass of \AS{0.63}{0.20}{0.13}\,M$_{\odot}$ with an RV semi-amplitude of \AS{27.2}{6.9}{6.6}\,km\,s$^{-1}$.

The measured $v \sin i$ of GJ\,207.1 may be biased due to its short orbital period. Indeed, we estimate that the RV shifts during the exposure time of the spectroscopic observations, which varies between 8 and 30\,min, can reach up to 7.6\,km\,s$^{-1}$, depending on the orbital phase at which the observation was taken. 
This shift produces a broadening of the spectral lines that causes an overestimation of the true $v \sin i$ from the spectral analysis. However, we can estimate the effect of this RV shift over the measured $v \sin i$ taking into account that the variance of the observed line profile is the addition of the variances of the kernels defining the true line profile and the RV shift. The kernel of both the observed and true line profiles can be defined, to first order, by a Wigner semicircle distribution \citep{Gray2005} with radius equal to the projected rotational velocity and variance $(v\sin i)^2/4$, while the RV shift effect is defined by a rectangular function with length $l$ equal to the shift and variance $l^2/12$. Therefore, the true projected rotational velocity, $v_{\rm T} \sin i$, is defined by $v_{\rm T} \sin i = 2\sqrt{(v \sin i)^2/4-l^2/12}$. Even in the worst case scenario in which the RV shift caused by the exposure time is $l=7.6$\,km\,s$^{-1}$, this equation yields a minimum value of $v_{\rm T} \sin i=8.8$\,km\,s$^{-1}$ which is well within the uncertainty of the measured $v \sin i = 9.8\pm1.5$\,km\,s$^{-1}$.

In general, the light curves of short-period binary systems present photometric modulations induced by the presence of a close companion. These include the reflection modulation and the beaming effect, which modulate the stellar brightness with orbital periodicity, and the ellipsoidal variation effect, which causes cyclic modulations at half the orbital period \citep{Zucker2007,Mazeh2008,Faigler2011}. These modulations can be approximated by sinusoidal functions, with harmonic periods and phase differences well defined with respect to the time of conjunction, and with constant amplitudes. These effects should be readily visible in the {\em TESS} photometry, but this appears to be dominated by activity-induced variations. This hypothesis is further favoured by the phase change observed between the two {\em TESS} sectors.

The theoretically predicted amplitudes of the Doppler beaming and ellipsoidal and reflection effects, are $\sim670$, $\sim780$, and $\sim0.7$\,ppm, as computed from Eq.\,2 in \citet{Zucker2007} and Eqs.\,2 and 3 in \cite{Faigler2011}. While we can neglect the reflection effect, the amplitudes of the Doppler beaming and ellipsoidal effect should be visible in the {\em TESS} photometry shown in Fig.\,\ref{J05337TESS} and, therefore, we can model them. To do so, we assumed the activity-induced variability to be a sinusoidal modulation at the orbital period ($P=0.60417356$\,d, Table\,\ref{tab:paramssingle}) and different phases and amplitudes between the two sectors. To model the Doppler beaming and ellipsoidal effect, we assumed two cosine functions with periods $P$ (for the Doppler beaming) and $P/2$ (for the ellipsoidal effect), with a reference time fixed to the maximum of the radial velocities and with the same amplitude in both sectors. The full model that we adopted to describe the variability in each {\em TESS} sector $j$ is defined by:

\begin{equation}
    A_j=C+a_j\cos{\left(\frac{2\pi}{P}T+\delta_j\right)} + a_b\cos{\left(\frac{2\pi}{P}T\right)} + a_e\cos{\left(\frac{4\pi}{P}T\right)},
    \label{eqamp}
\end{equation}

\noindent where $a_j$ and $\delta_j$ are the amplitude and phase of the activity-induced modulations in each sector, $a_b$ and $a_e$ are the amplitudes of the Doppler beaming and the ellipsoidal effect, and $T$ is defined as $t-t_{\rm ref}$, where $t_{\rm ref}$ is the time of maximum radial velocity. In addition to the orbital period, we also fixed the amplitude of the Doppler beaming modulation, since it depends on the well constrained RV semi-amplitude $K_{\rm A}$, while we left as free parameters $C$, $\delta_i$, $a_i$, and $a_e$. We computed the parameters and uncertainties by sampling the posterior probability distribution using \texttt{emcee}. %\citep{Foreman2013}. 
The best-model fit is shown in Fig\,\ref{J05337TESS}. We obtain amplitudes of $a_6=4557\pm67$\,ppm and $a_{32}=2023\pm60$\,ppm for the activity-induced variability of {\em TESS} sectors 6 and 32, respectively, with a shift between them of $1.19\pm0.03$\,rad. We obtained an amplitude of the ellipsoidal effect of $a_e=694\pm47$\,ppm, slightly lower than the predicted value of $780$\,ppm. From Eq.\,2 in \cite{Faigler2011}, and using the computed $a_e$ and the estimated inclination, we derive a secondary mass of about 0.56\,M$_{\odot}$, which is in good agreement with the mass derived from RVs, within uncertainties. Since the photometric amplitude of the ellipsoidal effect is larger for inclinations closer to 90\,degrees, it is possible to put constraints on the inclination of the system independently of our former assumption about spin-orbit alignment. To do so, we estimated the amplitude $a_e$ at which the model is no longer compatible with the observed photometry. By setting a log-likelihood decrease 
%$\Delta \ln{\mathcal{L}}$ 
of 15 with respect to the best-fit, we obtained the maximum $a_e$ amplitude at $\sim 1000$\,ppm. This upper limit rules out orbital inclinations higher than 20\,deg and, therefore, excludes companions with masses below 0.37\,M$_{\odot}$. For such a massive companion, the spectral lines of both components of the system should be easily detected using \texttt{todmor} if it were a normal main-sequence M dwarf.

%For companion masses larger than 0.37\,M$_{\odot}$, the spectral lines of both components of the system should be easily detected using \texttt{todmor} if it was a normal main-sequence M dwarf. The fact that they are not
All the evidences points towards an under-luminous companion to GJ\,207.1A. Interestingly, the absolute mass of the unseen component derived from the RVs under the assumption of zero obliquity of \AS{0.63}{0.20}{0.13}\,M$_{\odot}$ lies in the middle of the white-dwarf mass distribution \citep{Catalan2008}. Therefore, we conclude that the secondary component of GJ\,207.1 is most likely a white dwarf. We estimate that the progenitor star could have had a mass of \AS{2.1}{1.6}{1.4}\,M$_{\odot}$, as computed from initial-to-final mass relationships \citep[cf.][]{Catalan2008}. The time required for an object of this mass range to become a white dwarf is at least \AS{0.59}{22}{0.45}\,Ga \citep{Marigo2001}. However, %we note here that 
the absolute masses and radii used to determine the mass of the white-dwarf progenitor should be taken with caution. This binary system has probably suffered episodes of mass transfer during the red giant phase of the secondary, and the M dwarf could have truncated the evolution of the white dwarf precursor when it was ascending the giant branch. This truncation would produce a white dwarf with a lower mass than if it were isolated \citep{Iben1993,Rebassa2019}, and may have changed the properties of both stars \citep{Rebassa2011}.

\cite{Bar2017} conducted a spectroscopic search for white dwarfs around nearby M dwarfs. For GJ\,207.1, they set an upper limit of 8000\,K for the temperature of any existing white dwarf in the system, estimated from the non-detection of any flux contribution from a companion in an X-SHOOTER \citep{Vernet2011} spectrum down to 300\,nm. From the cooling sequences of hydrogen white dwarfs \citep{Chabrier2000}, we estimate that it takes at least 1\,Ga for a 0.6\,M$_{\odot}$ white dwarf to cool down to a temperature of $\sim$8000\,K, and even more for more massive white dwarfs. All this evidence points towards an age older than 1\,Ga for this system, in contradiction with the young age proposed in previous works \citep{Lowrance2005}.

To investigate this disagreement, we recomputed the Galactic space-velocity components determined by \cite{Montes2001} using the absolute RV of the system determined in this work and the newly available \emph{Gaia} proper motions and parallax. These new parameters yield space velocity components $U$, $V$, and $W$ of $-$21.4, $-$10.5, and $-$28.6 km\,s$^{-1}$, respectively \citep{Johnson1987}, which place the system in the thin disc population \citep{Bensby2003}, and outside the young disc population limits \citep{Leggett1992}. Other stellar youth indicators for GJ\,207.1, such as its UV and X-ray overluminosity, relatively strong H$\alpha$ emisson, and flaring activity can be explained by the fast rotation of the M-dwarf component caused by the spin-orbit synchronisation at the short orbital period of the binary. The characteristic age of the thin disc population is younger than $\sim8$\,Ga \citep{Fuhrmann1998}, which is compatible with the cooling age of the putative white dwarf. The upper limit of $\sim8$\,Ga on the age, set by the kinematics of the system, and the lower limit of $\sim1$\,Ga, set by the cooling time of the white dwarf, mean that the main sequence phase of the progenitor star should not have lasted more than $\sim7$\,Ga. Only stars with masses $\gtrsim0.9$\,M$_{\odot}$ would reach the white dwarf stage within that time \citep{Marigo2001} and would yield a final mass $\gtrsim0.52$\,M$_{\odot}$ \citep{Catalan2008}. 

% Finally, we cannot exclude an alternative scenario explaining the non-detection of the secondary component of GJ\,207.1. It is possible that spin-orbit alignment is not fulfilled. In this case, the orbital inclination could be different, and hence the mass of the secondary component. From our experience with the other binary systems discovered with CARMENES observations, we estimate that the signal of the secondary component could be detected if the mass ratio is above $\sim$0.3. This threshold on the mass ratio imposes an inclination of the orbit below $\sim40$\,deg. Assuming a flat distribution in $\cos i$, the probability of such inclinations is $\sim$75\%. Nevertheless, circularization and synchronization is achieved very fast in short-period binaries \citep{Mazeh2008}, so a third object would be needed to explain the spin-orbit misalignment of this system.

\subsubsection{GJ\,912}

\begin{figure}[t]
\centering
\includegraphics[width=\columnwidth]{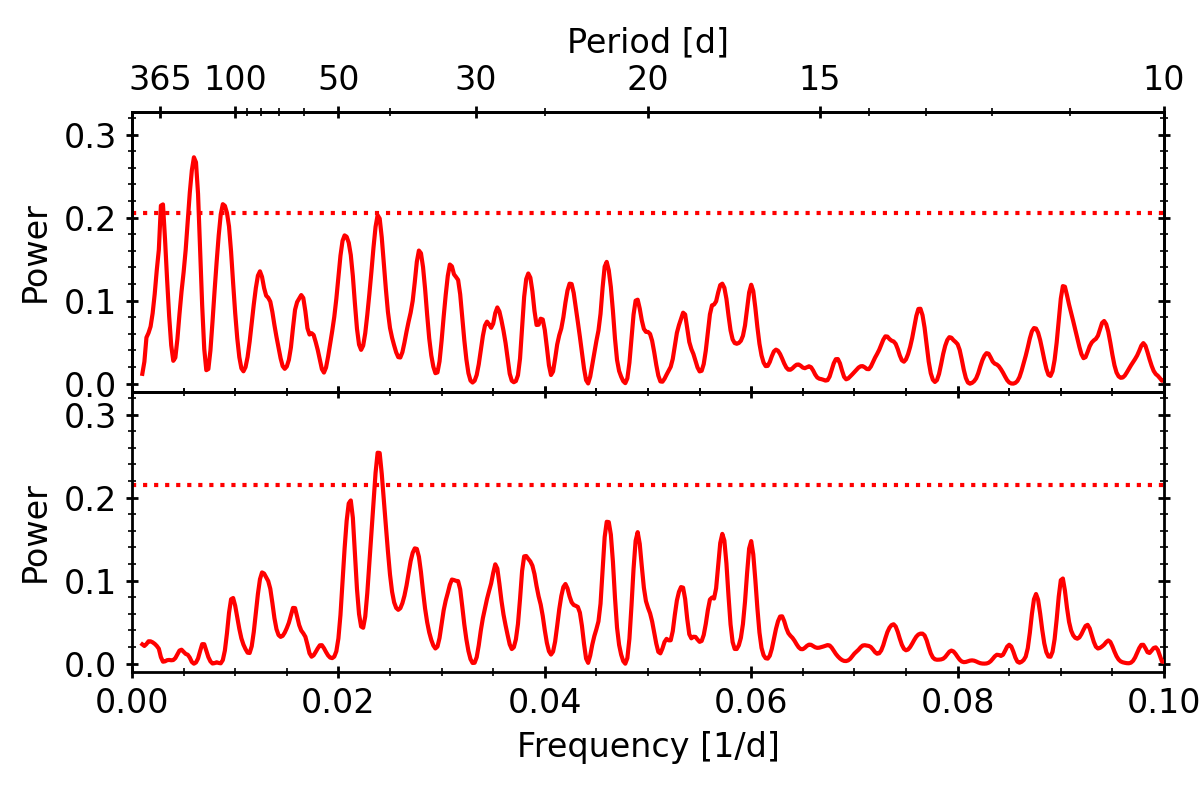}
\caption{GLS periodogram of the SWASP photometry of GJ\,912 (\textit{top}) and of the residuals after the removal of the significant signal (\textit{bottom}). The horizontal dotted lines indicate the corresponding 0.1\,\% FAP level.}
          \label{J23556GLS}%
\end{figure}

\begin{figure}[t]
\centering
\includegraphics[width=\columnwidth]{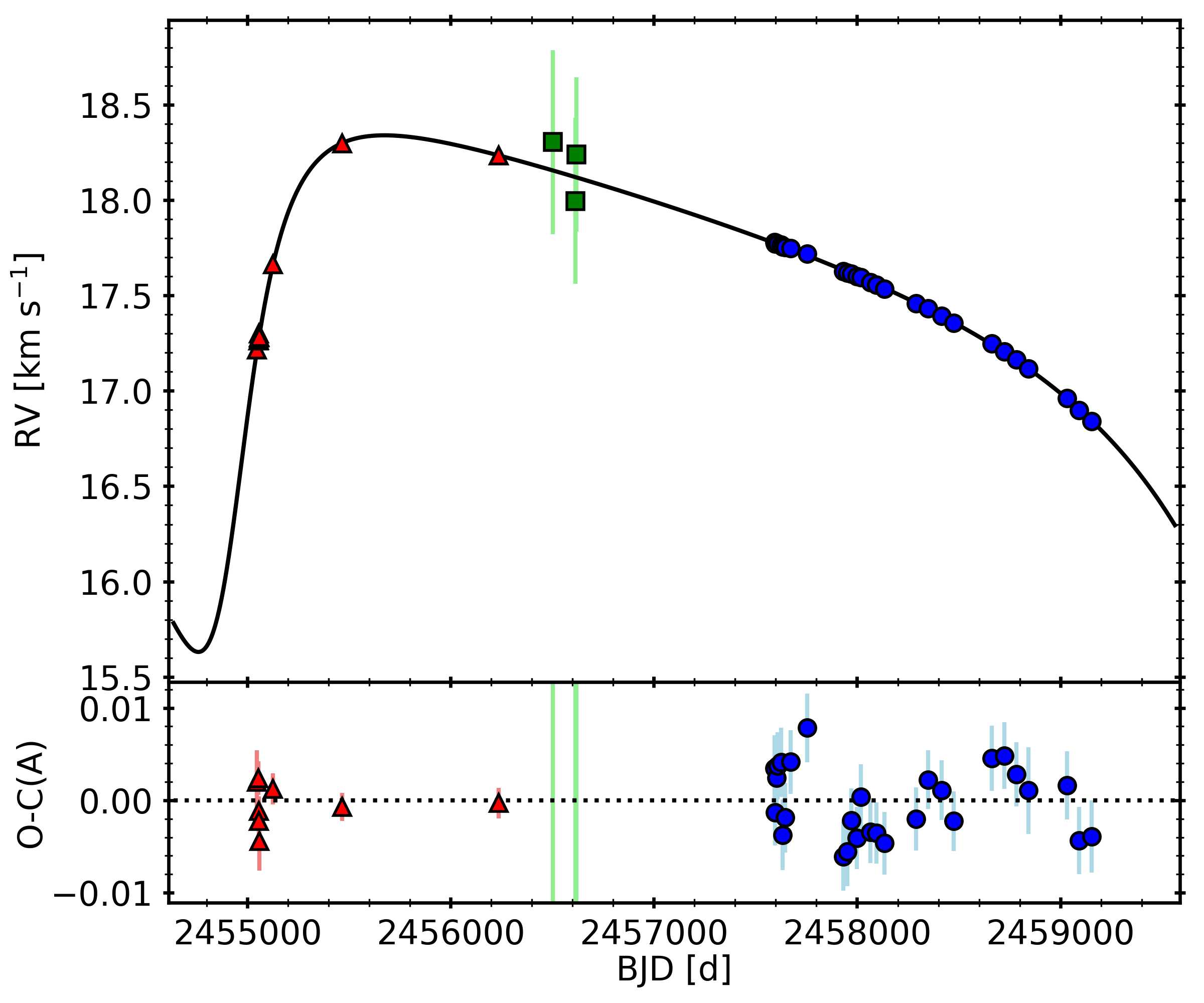}
\caption{Best orbital fit to the RVs of GJ\,912 from CARMENES (blue circles), HARPS (red triangles), and FEROS (green squares), as a function of time. Residuals from the best fit are depicted in the bottom panel. Those corresponding to the FEROS observations are outside the Y-axis scale. 
}
          \label{J23556orbit}%
\end{figure}

\begin{figure}[t]
\centering
\includegraphics[width=\columnwidth]{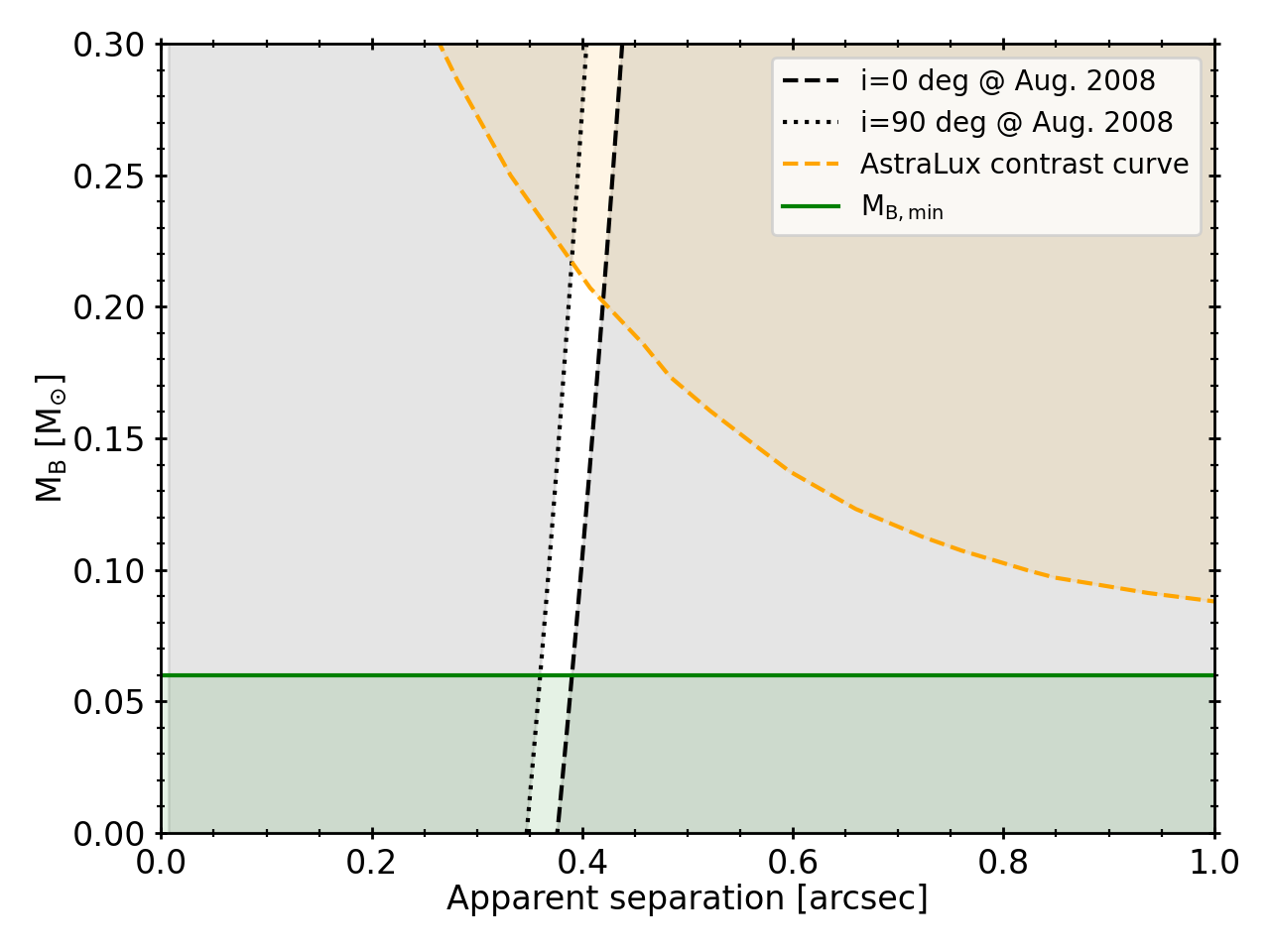}
\caption{Constraint to the mass of the secondary star in GJ\,912 as a function of the apparent angular separation of the two components. The orange dashed line represents the contrast curve of the AstraLux observation by \cite{Jodar2013}. The green solid line indicates the minimum mass limit computed from our RVs. The dashed and dotted black lines are constraints from RV to the angular separation at the time of the observation, assuming inclinations between 0\,deg (dashed) and 90\,deg (dotted). All shaded regions mark mass values that are not compatible with the observations.}\label{J23556CC}%
\end{figure}

GJ\,912 is a high-proper motion M2.5\,V star located at a distance of $17.81\pm0.03$\,pc. From its galactocentric space-velocity computed by \cite{Hawley1996}, this system could be a member of the thin disc population, with an estimated age younger than 8\,Ga \citep{Fuhrmann1998}. No companions were found at angular separations between 0.1\,arcsec and 3.5\,arcsec with AstraLux \citep{Jodar2013}, or at wider separations  using low-resolution imaging \citep{Winters2019}. The sensitivity of these surveys at small separations depended, however, on the flux ratio of the components. The rotation period estimated from the $R'_{\rm HK}$ index of the star, $\log (R'_{\rm HK})=-5.01\pm0.19$, is of the order of 42\,d \citep{Astudillo2017}. 

Twenty-eight measurements for GJ\,912 were obtained with CARMENES between July 2016 and November 2020. Also, nine archival observations from HARPS and three from FEROS taken between August 2009 and November 2013 are available. An inspection of all the spectra with \texttt{todmor} revealed no secondary peaks in the CCF, indicating either an extreme flux ratio or a small relative RV between the two components. As for GJ\,207.1, we used the difference between the CARMENES RVs computed with \texttt{serval} and with \texttt{todmor} to correct the offset of the FEROS \texttt{todmor} RVs.%, which could not be obtained with \texttt{serval}.

The orbital parameters resulting from the MCMC analysis are shown in the third column of Table\,\ref{tab:paramssingle}. The best-fit solution corresponds to a period of \AS{5188}{58}{55}\,d (about 14.2\,a), with an RV semi-amplitude of \AS{1.356}{0.084}{0.074}\,km\,s$^{-1}$. The orbit is highly eccentric, $e$ = \AS{0.6239}{0.0098}{0.0097}. The RVs and the best-fit solution are shown in Fig.\,\ref{J23556orbit} as a function of the orbital phase. Using the derived mass function $f(M)$ and the estimated mass of the primary component (0.520\,M$_{\odot}$, Table\,\ref{tab:props}), we determine a minimum companion mass of \AS{0.0600}{0.0030}{0.0036}\,M$_{\odot}$, which is compatible with a brown dwarf \citep[i.e. mass lower than 0.072\,M$_{\odot}$,][]{Chabrier2000b} for inclinations above $\sim$50\,deg. Assuming an inclination probability distribution uniform in $\cos i$ (which implies a companion mass function favouring lower-mass companions), this corresponds to a probability of $\sim$63\,\%. An inspection of the RV residuals of the best fit yielded no significant signals.

We also analysed the available photometry of this object to search for signals of its rotation period. We binned the available SWASP data in nightly bins in order to reduce the scatter. Furthermore, we discarded observations made after BJD = $2\,456\,000$, since they have systematical nightly deviations of more than three magnitudes, and the median of the errors is ten times higher than that of the previous data. This reduced the number of individual epochs in the light curve from 90\,025 to only 256. The GLS periodogram shows a significant peak at 165\,d. However, the long period was similar to the time-span of each season of observation and, after its removal, a significant peak at 41.8\,d arises, as shown in Fig.\,\ref{J23556GLS}. Therefore, we attributed the $41.8\pm1.3$\,d signal to the rotation period of the star, which is consistent with the period estimated by \cite{Astudillo2017}. The error of the period was estimated from the width of the peak at half its maximum. No significant signals were found in the ASAS data. Based on the rotation period of this system, we can safely assume that this system is older than the objects of NGC\,6811, with an estimated age of 1\,Ga \citep{Curtis2019,Curtis2019b}.

Finally, a constraint to the inclination of the system and, therefore, of the maximum mass of the companion, can be set by using 3.1\,a of astrometric data from {\em Hipparcos} \citep{Hipparcos,Leeuwen2007}. Besides the five standard astrometric parameters, we fitted an astrometric orbit to the {\em Hipparcos} abscissa residuals, from which the inclination and ascending node could be determined via a least squares minimisation \cite[see][for more details on the method]{Reffert2011} for GJ\,912. An astrometic orbit was detected with 94\,\% probability, and the inclination was constrained to be greater than 5.5\,deg at 1$\sigma$. An even more constraining limit can be set from the non-detection of a companion in the AstraLux observation by \cite{Jodar2013}, made in August 2008. Assuming the 5\,Ga stellar evolutionary model from \citet[][hereafter BHAC15]{Baraffe2015} and the estimated primary mass ($0.520\pm0.02$\,M$_{\odot}$, Table\,\ref{tab:paramssingle}), we converted the brightness contrast curve shown in Fig.\,3 by \cite{Jodar2013} into mass detection limits, and show it in Fig.\,\ref{J23556CC} as a function of the angular separation. We also show in Fig\,\ref{J23556CC} the minimum mass allowed by the radial velocities, and the range of possible angular separations between the components at the date of observation, given any inclination between 0 and 90\,deg. We can therefore determine that any companion with a mass above 0.21\,M$_{\odot}$ should be detected in the AstraLux observation, assuming that the companion is a main-sequence star. This limits the inclination to values higher than $\sim$17\,deg, which pushes the probability of the secondary being a brown dwarf up to $\sim$66\,\%.

\subsubsection{GJ\,3626}

\begin{figure}[t]
\centering
\includegraphics[width=\columnwidth]{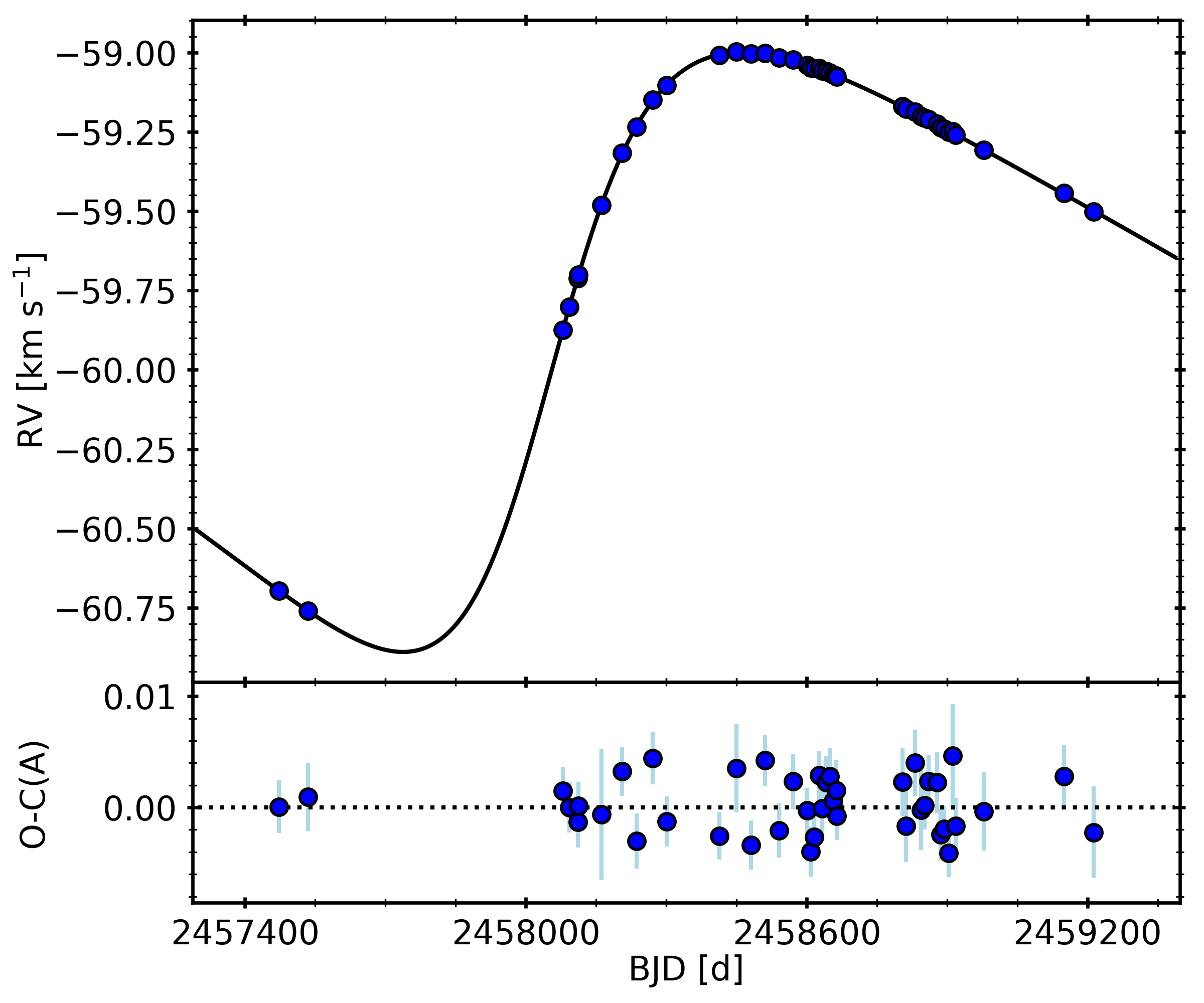}
\caption{Best orbital fit to the RVs of GJ\,3626 from CARMENES as a function of time. The residuals from the fit are shown in the bottom panel.}
          \label{J10504orbit}%
\end{figure}

\begin{figure}[t]
\centering
\includegraphics[width=\columnwidth]{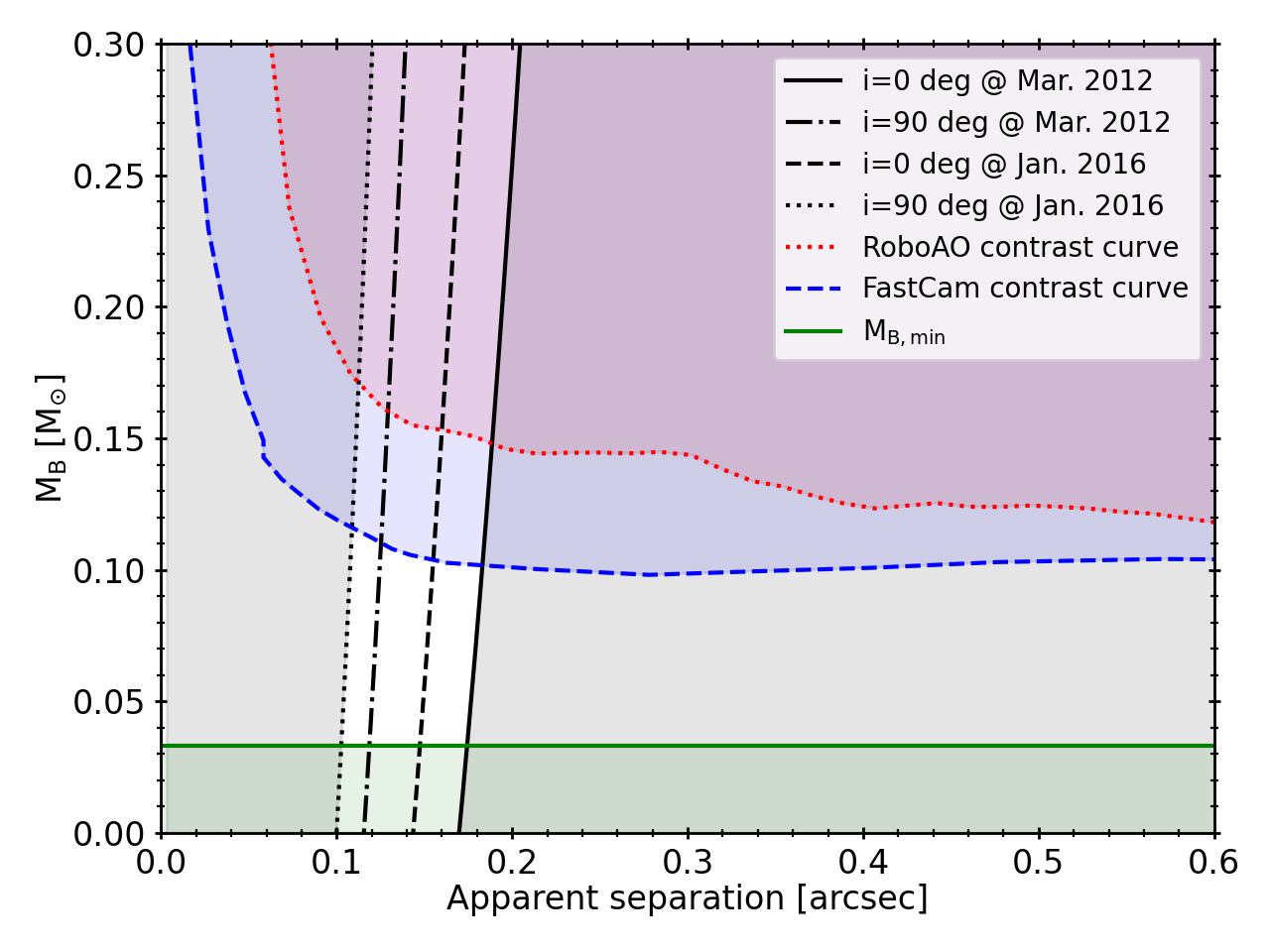}
\caption{Constraints on the mass of the secondary star in GJ\,3626 as a function of the apparent angular separation of the two components. The red dotted and blue dashed lines represent the contrast curves of the Robo-AO and FastCam observations, respectively. The green solid line indicates the minimum mass limit computed from RVs. The solid and dot-dashed, and the dashed and dotted lines are constraints from RVs to the angular separation at the time of the FastCam and Robo-AO observations, respectively, for orbital inclinations between 0 and 90\,deg. All the shaded regions mark mass values that are not compatible with the observations.}\label{J10504CC}%
\end{figure}

Located at a distance of $22.61\pm0.06$\,pc, GJ\,3626 is an M4.0\,V high-proper motion star. It has a low activity level, with an H$\alpha$ pseudo-equivalent width pEW(H$\alpha$) = +0.092$\pm$0.014\,{\AA} \citep{Schoefer2019}, and an upper limit to the projected rotational velocity of 2\,km\,s$^{-1}$ \citep{Reiners2018}. No rotation period has been reported for this star. %It is probably a member of the thin disc population, deduced from its Galactocentric space-velocities computed in \textbf{Cortes in prep}, for which \citep{Fuhrmann1998} estimate an age not older than 8\,Ga.
In the context of the sample selection of the CARMENES input catalogue, \cite{Cortes2017} searched for low-mass companions to this star using the high-resolution lucky imaging instrument FastCam. From these observations, performed in March 2012, they excluded the presence of companions with $\Delta I<4$\,mag between 0.2 and 2\,arcsec, and with $\Delta I<5$\,mag beyond 2\,arcsec. 
Also, observations taken for the Robo-AO adaptative optics M-dwarf multiplicity survey \citep{Lamman2020}, performed in January 2016, put limits to companions with $\Delta i'<2$\,mag beyond 0.1\,arcsec, $\Delta i'<3$\,mag beyond 0.5\,arcsec, and $\Delta i'<4$\,mag beyond 1\,arcsec. 

We observed GJ\,3626 with CARMENES between March 2016 and November 2020, obtaining 42 spectra. We searched for the signature of a secondary component in the spectra using \texttt{todmor}, but no additional peaks in the CCF were detected. Therefore, we used the high-precision RVs computed with \texttt{serval} to determine the orbital parameters of the system as an SB1 %a single-line binary.
However, to measure the absolute centre-of-mass RV, we used the RVs computed with \texttt{todmor} using synthetic spectra. 

We show in Fig.\,\ref{J10504orbit} the best-fitting orbital solution, and the parameters are listed in Table\,\ref{tab:paramssingle}. The solution corresponds to a rather eccentric ($e=$ \AS{0.4257}{0.0029}{0.0028}) orbit with a period of \AS{3009}{37}{35}\,d (about 8.24\,a) and a semi-amplitude of \AS{0.9431}{0.0053}{0.0053}\,km\,s$^{-1}$. From the mass function computed from the fits and the primary component mass ($0.403\pm0.017$\,M$_{\odot}$; Table\,\ref{tab:props}), we calculate a minimum mass of $0.0331^{+0.0010}_{-0.0011}$\,M$_{\odot}$ for the secondary component, well below the brown dwarf mass limit. We sampled the posterior distribution of minimum masses for 10$^5$ random values for the inclination, assuming a uniform distribution in $\cos i$, and estimated a true secondary mass of \AS{0.039}{0.025}{0.005}\,M$_{\odot}$ and a 87\,\% chance of it being a brown dwarf.

As an additional constraint, and as we have done for GJ\,912, we used the Robo-AO \citep{Lamman2020} and the FastCam \citep{Cortes2017} contrast curves. Using the 5\,Ga BHAC15 stellar evolutionary model and the estimated primary mass, %($0.40\pm0.02$\,M$_{\odot}$, Table\,\ref{tab:props}), 
we again converted the brightness contrast curves into mass detection limit, which are shown in Fig.\,\ref{J10504CC} as a function of the orbital separation. We also illustrate in Fig.\,\ref{J10504CC} the possible range of apparent angular separations of the two components, given any inclination between 0 and 90\,deg, at the time of the FastCam and RoboAO observations. It can be seen from this figure that a companion with a mass above 0.11\,M$_{\odot}$ would have been detected in the FastCam observations. Using this limit as an upper bound to the companion mass, we set a lower bound to the inclination of the system to 19\,deg, pushing the probability of the secondary component being a brown dwarf up to 93\,\%.
 
We also analysed the available photometry from SWASP to look for signals induced by rotation. We first removed outliers using a 3\,$\sigma$ clipping procedure and binned the light curve into nightly bins, resulting in 96 epochs between May 2004 and June 2007. The GLS periodogram shows a significant peak at 199\,d. However, the photometry is divided in two different subsets of less than 200 days, with $\sim$1\,000 days in between, which makes the interpretation of this signal difficult. Therefore, it is unclear whether this signal may be due to rotation. As a further check, we analysed the residuals from the fits to the RV curve with GLS. The periodogram shows a peak at 72.7\,d just below the 10\,\% FAP level, much higher than our adopted 0.1\% FAP threshold. 
%The signal has an amplitude of 2.4\,m\,s$^{-1}$. 
We also analysed the CARMENES activity indicators in order to check if this signal is caused by stellar activity. However, none of the indicators have any significant peak at this period or at any other frequency. 
%Hence, we cannot rule out the possibility that the 72.6 days signal is caused by a 8.3\,M$_{\oplus}$ planet orbiting the primary star. More data of this system to determine the rotation period and improve the RV signal significance is needed to confirm this hypothesis.

\subsubsection{LP\,427-016}

\begin{figure}[t]
\centering
\includegraphics[width=\columnwidth]{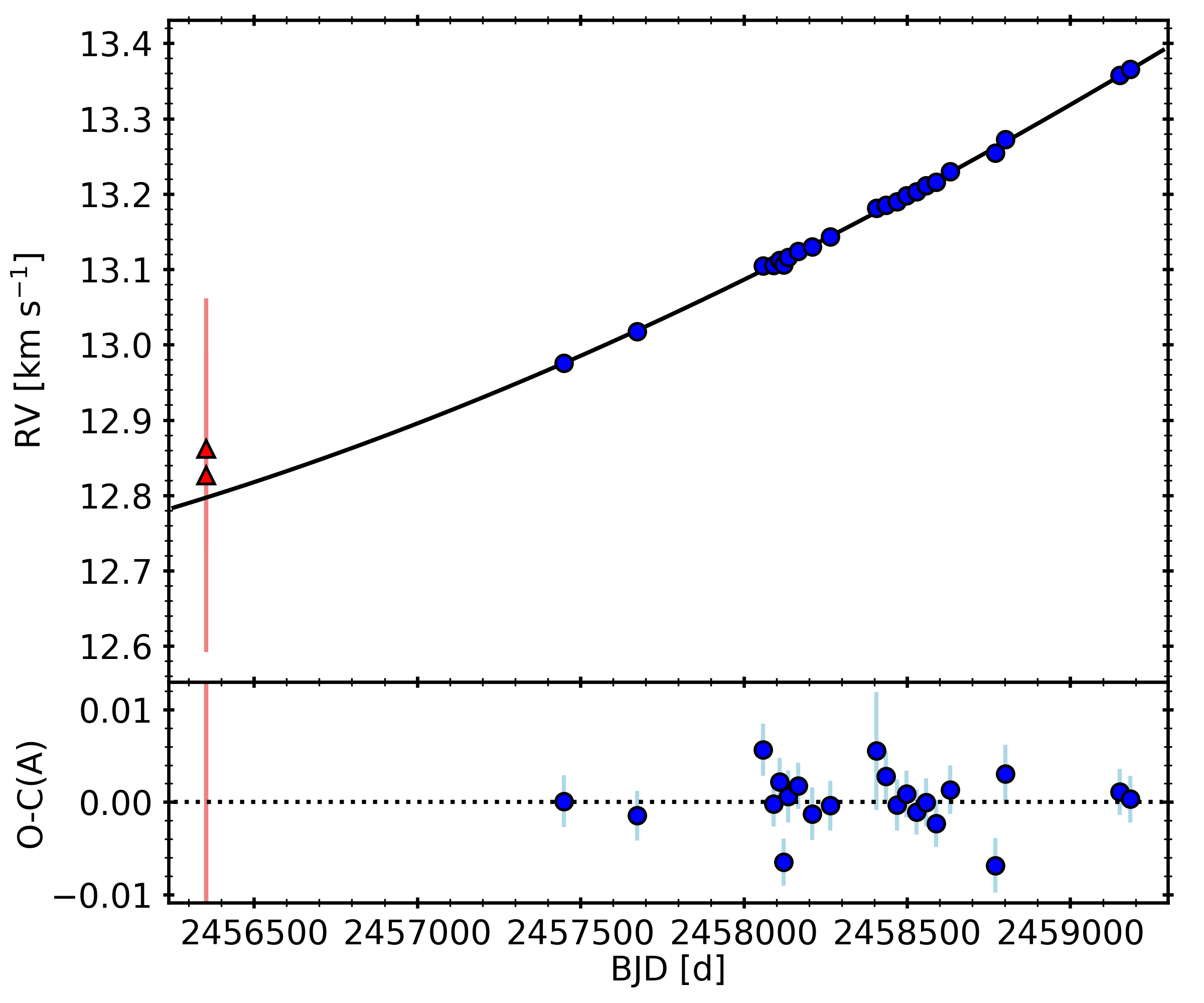}
\caption{Best fit to the RVs of LP\,427-016 taken with CARMENES (blue circles) and FEROS (red triangle), assuming a circular orbit. The bottom panel shows the residuals from the fit. 
Those corresponding to FEROS observations are outside the Y-axis scale.
%For clarity, we left the residuals corresponding to FEROS observations outside the y axis scale, which is defined to illustrate CARMENES RV residuals.
}
\label{J09140orbit}%
\end{figure}

\begin{figure}[t]
\centering
\includegraphics[width=\columnwidth]{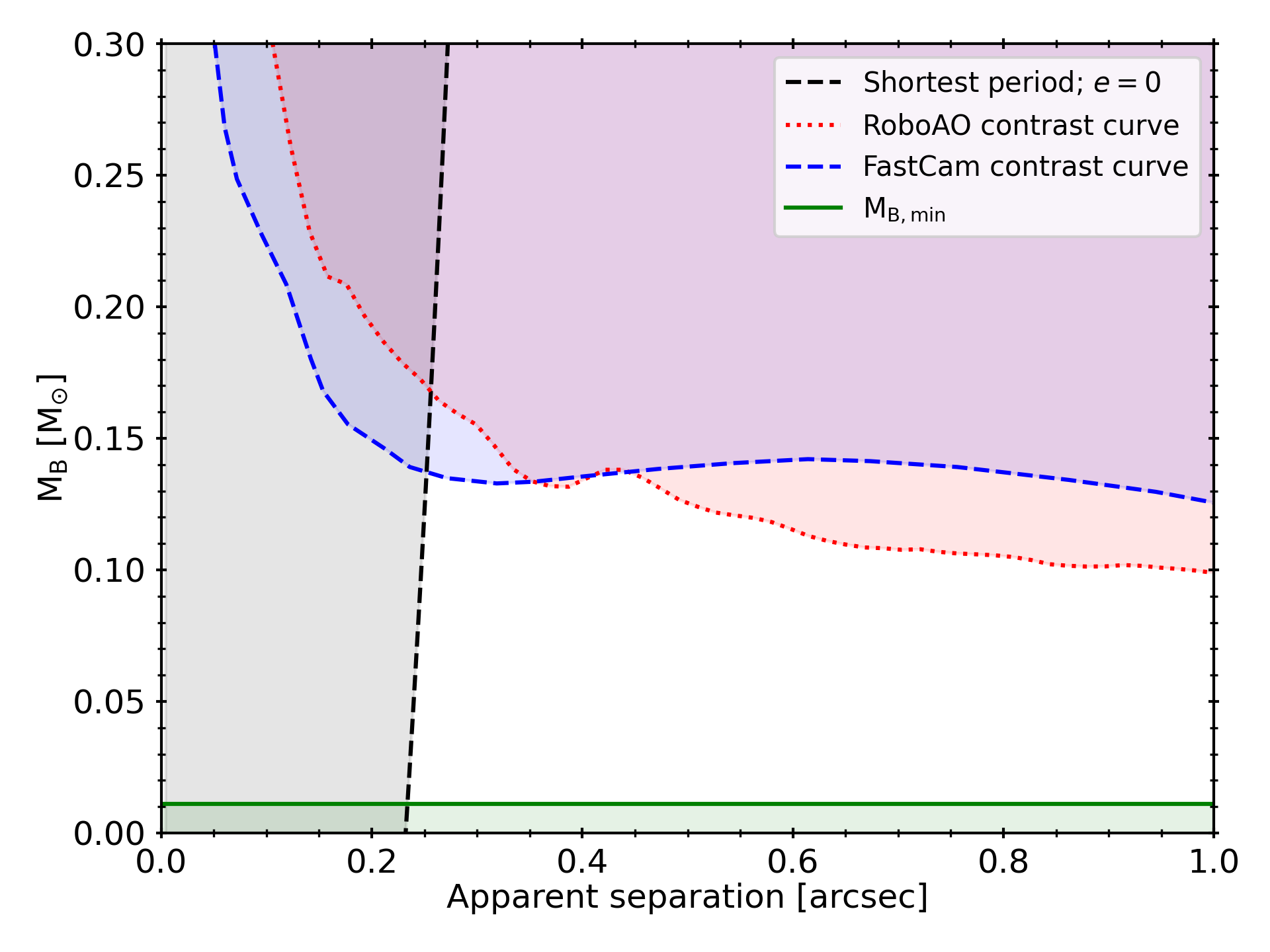}
\caption{Constraints on the mass of the secondary star in LP\,427-016 as a function of the apparent separation of the two components. The red dotted and blue dashed lines depict the contrast curves of the RoboAO and FastCam observations, respectively. The green line indicates the minimum mass limit computed from RVs, of 0.011\,M$_{\odot}$. The black dashed line marks the apparent separation for the lower limits to the period, assuming a circular orbit and $i=90$\,deg. All the shaded regions mark solutions that are not consistent with the observations.}\label{J09140CC}%
\end{figure}

\begin{figure}[t]
\centering
\includegraphics[width=\columnwidth]{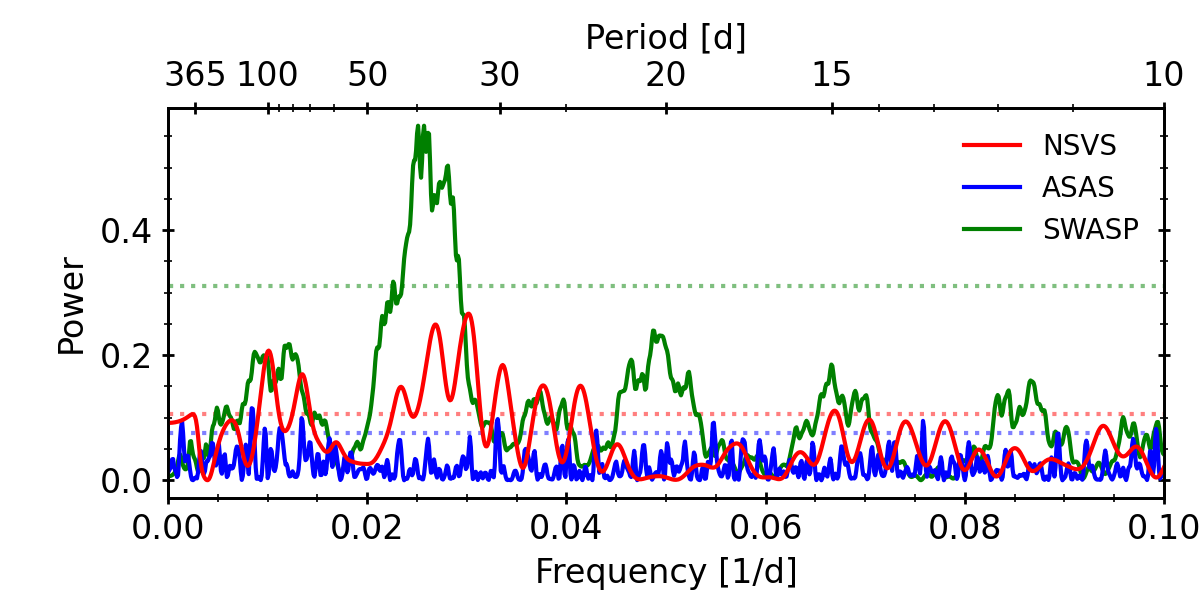}
\caption{GLS periodogram of the NSVS (red), ASAS (blue), and SWASP (green) photometry of LP\,427--016. The horizontal dotted lines of each color indicate the corresponding 0.1\,\% FAP level.}
          \label{J09140GLS}%
\end{figure}

LP\,427-016 is an M3.0\,V star located at a distance of $22.98\pm0.03$\,pc. From its galactocentric space velocity, it was classified as a member of the Local Association kinematic group \citep{Cortes2016}, with an estimated age between 10 and 150\,Ma \citep{Bell2015}. The H$\alpha$ emission shows low levels of chromospheric emission, pEW(H$\alpha$) = $-0.10\pm$0.01\,\AA, and the spectroscopic observations are consistent with slow rotation, $v\sin i<$2\,km\,s$^{-1}$ \citep{Schoefer2019,Jeffers2018}. Furthermore, \cite{DiezAlonso2019} reported a rotation period of $89.9\pm2.0$\,d using ASAS photometry, although with a poor FAP of 1.8\,\%.
Considering its hypothetical young age, the star would be expected to be magnetically active \citep{Terndrup2000,West2008,Feng2018} and to have a shorter rotation period \citep{Barnes2003,Curtis2019}. However, this is not supported by the measured activity indicators or the rotation period of this star, pointing towards an older age for this system than suggested by its kinematic assignation \citep[see][for an assessment of the contamination by old stars in purely kinematic searches of members of nearby young moving groups]{Lopez2019}.

No close companions with $\Delta I<3$\,mag at an angular separation greater than 0.2\,arcsec were found in observations from the FastCam high-resolution lucky imager \citep{Cortes2017}. In addition, observations from the Robo-AO adaptative optics M dwarf multiplicity survey \citep{Lamman2020} ruled out companions with $\Delta i'<1$\,mag beyond 0.1\,arcsec, and $\Delta i'<5$\,mag beyond 1\,arcsec.

This object was spectroscopically observed with CARMENES for a total of 22 times, between March 2016 and November 2020. Two FEROS observations, obtained two years before those of CARMENES, are also available. The RVs for this system are listed in Table\,\ref{tab:RVSB1}, and shown in Fig.\,\ref{J09140orbit}. The CARMENES RVs show an unambiguous long-term trend indicating binarity, but no companion was detected with \texttt{todmor}. In order to move the FEROS RVs to the same offset level as the CARMENES ones, we again used the difference between the CARMENES RVs computed using \texttt{todmor} and \texttt{serval}. An attempt to determine the best orbital fit resulted in the MCMCs diverging towards large periods, as expected.
The best circular orbit fit, shown in Fig.\,\ref{J09140orbit}, corresponds to a period of $\sim35\,500$\,d (97\,a), although the circular model is compatible ($\Delta \ln \mathcal{L} <15$) with orbital periods between $\sim6\,400$\,d (17.5\,a) and $\sim100\,000$\,d (274\,a). We therefore adopt $6\,400$\,d as the shortest possible period for a circular orbit, which yields a semi-amplitude of 0.25\,km\,s$^{-1}$ and a minimum secondary mass of 0.012\,$M_{\odot}$.

More stringent bounds can be derived from the available observations with high-resolution imaging. Fig.\,\ref{J09140CC} shows the contrast curves obtained from the FastCam and Robo-AO high-resolution observations. We used the BHAC15 5\,Ga evolutionary models to estimate the mass limit of the secondary component from the magnitude difference of the contrast curves and the mass of the primary component as for GJ\,912. We also plot the minimum mass corresponding to the minimum allowed period derived from the RVs and the lower limit to the apparent separation (all assuming $e=0$) computed with Kepler's Third Law and the lower bound to the period. From this figure, we conclude that any companion with a mass higher than $\sim$0.13\,$M_{\odot}$ would have been detected in high-resolution imaging observations. From this value, we obtain an upper bound to the period of $\sim29\,000$\,d (79\,a). We suggest that the secondary component may have a mass in the range 0.011--0.13\,M$_{\odot}$, at an angular separation between 0.5\,arcsec and 1.3\,arcsec. We list in Table\,\ref{tab:paramssingle} all the limits to the orbital parameters resulting from this analysis.

%We note, however, that a
Allowing for non-zero eccentricity would affect the lower limit of the period, which could be as small as the time-span of the data, $\sim3\,000$\,d. Extreme values of the eccentricity would also allow very small apparent projected separations, increasing the upper limit to the mass of the companion.

LP\,427-016 was photometrically monitored by the ASAS, NSVS, and SWASP surveys. As shown in Fig.\,\ref{J09140GLS}, the GLS periodogram shows significant signals at periods of $33.2\pm1.3$\,d and $38.9\pm4.8$\,d for the NSVS and SWASP data, respectively, which are compatible within their errors. The second highest peak in the periodogram of the NSVS data, also significant and related to $33.2\pm1.3$\,d by the window function, is at a period of $37.2\pm1.3$\,d, coincident with the signal found in the SWASP data. In the ASAS data, we find excess power at a period of 89.9\,d as claimed by \cite{DiezAlonso2019}, but with a significance much lower than our 0.1\,\% FAP threshold. Given that the time-span of the SWASP data is five times longer than that of NSVS, although less precise, we attribute the $38.9\pm4.8$\,d signal to the rotation period of the more massive star. As in the case of GJ\,912, based on the rotation period of this system, we can safely assume that it is older than the objects of NGC\,6811, with an estimated age of 1\,Ga \citep{Curtis2019,Curtis2019b} and, therefore, rule out its membership to the Local Association kinematic group as suggested by its space velocity, which, incidentally, could be biased by the binarity of the system.

\subsubsection{GJ\,282\,C}

\begin{table}[t]
\centering
\caption{Orbital parameters for the astrometric binary GJ\,282\,C.}
\label{tab:J07361}
\begin{tabular}{lc} 
\hline\hline
\noalign{\smallskip}
Parameter & System\\
\noalign{\smallskip}
\hline
\noalign{\smallskip}
\multicolumn{2}{c}{\textit{Orbital parameters}}\\
\noalign{\smallskip}
\hline
\noalign{\smallskip}

$P$ [d] & \AS{6591}{136}{177} \\
\noalign{\smallskip}
$T_0$ [BJD] & \AS{2460004}{90}{91} \\
\noalign{\smallskip}
$K_{Ca}$ [km\,s$^{-1}$] & \AS{2.66}{0.12}{0.12} \\
\noalign{\smallskip}
$e\sin \omega$ & \AS{-0.0109}{0.0054}{0.0057}\\
\noalign{\smallskip}
$e\cos \omega$ & \AS{-0.213}{0.010}{0.010}\\
\noalign{\smallskip}
$e$ & \AS{0.213}{0.010}{0.010}\\
\noalign{\smallskip}
$\omega$ [deg] & \AS{183.0}{1.5}{1.4}\\
\noalign{\smallskip}
$\Omega$ [deg] & \AS{136.95}{0.46}{0.46}\\
\noalign{\smallskip}
$a$ [au]  & \AS{6.224}{0.081}{0.080}\\
\noalign{\smallskip}
$i$ [deg]  & \AS{93.96}{0.55}{0.55}\\
\noalign{\smallskip}
$\gamma_{\rm CARM}$ [km\,s$^{-1}$]  & \AS{+0.89}{0.11}{0.12}\\
\noalign{\smallskip}
$\gamma_{\rm HARPS}$ [km\,s$^{-1}$] & \AS{-0.20}{0.11}{0.12}\\
\noalign{\smallskip}
$\gamma$ [km\,s$^{-1}$]  & \AS{-18.35}{0.11}{0.12}\\
\noalign{\smallskip}
$\sigma_{\rm ALL}$\tablefootmark{a} [m\,s$^{-1}$]  & \AS{10.52}{0.80}{0.71}\\

\noalign{\smallskip}
\hline
\noalign{\smallskip}
\multicolumn{2}{c}{\textit{Derived parameters}}\\
\noalign{\smallskip}
\hline
\noalign{\smallskip}

$M_{Ca}$ [M$_{\odot}$] & \AS{0.554}{0.058}{0.049}\\
\noalign{\smallskip}
$M_{Cb}$ [M$_{\odot}$] & \AS{0.1881}{0.0048}{0.0047}\\
\noalign{\smallskip}
$q$  & \AS{0.340}{0.034}{0.034}\\
\noalign{\smallskip}
$K_{Cb}$ [km\,s$^{-1}$] & \AS{7.83}{0.49}{0.41}\\
\noalign{\smallskip}
$\mu_{\alpha} \cos{\delta}$ [mas\,a$^{-1}$] & \AS{+60.72}{0.38}{0.29}\\
\noalign{\smallskip}
$\mu_{\delta}$ [mas\,a$^{-1}$] & \AS{-278.61}{0.32}{0.44}\\
\noalign{\smallskip}
$\varpi$ [mas] & \AS{71.053}{0.027}{0.027}\\

\noalign{\smallskip}
\hline
\noalign{\smallskip}
\multicolumn{2}{c}{\textit{Estimated parameters}}\\
\noalign{\smallskip}
\hline
\noalign{\smallskip}

$P_{\rm rot,Ca}$ [d] & $12.15\pm0.02$\\
\noalign{\smallskip}
$P_{\rm rot,Cb}$ [d] & $0.3257\pm0.0002$\\

\hline
\end{tabular}
\tablefoot{ \tablefoottext{a}{\texttt{orvara} code only fits a single jitter term to all the RV sets.}  }
\end{table}

\begin{figure*}[t]
\centering
\includegraphics[width=\columnwidth]{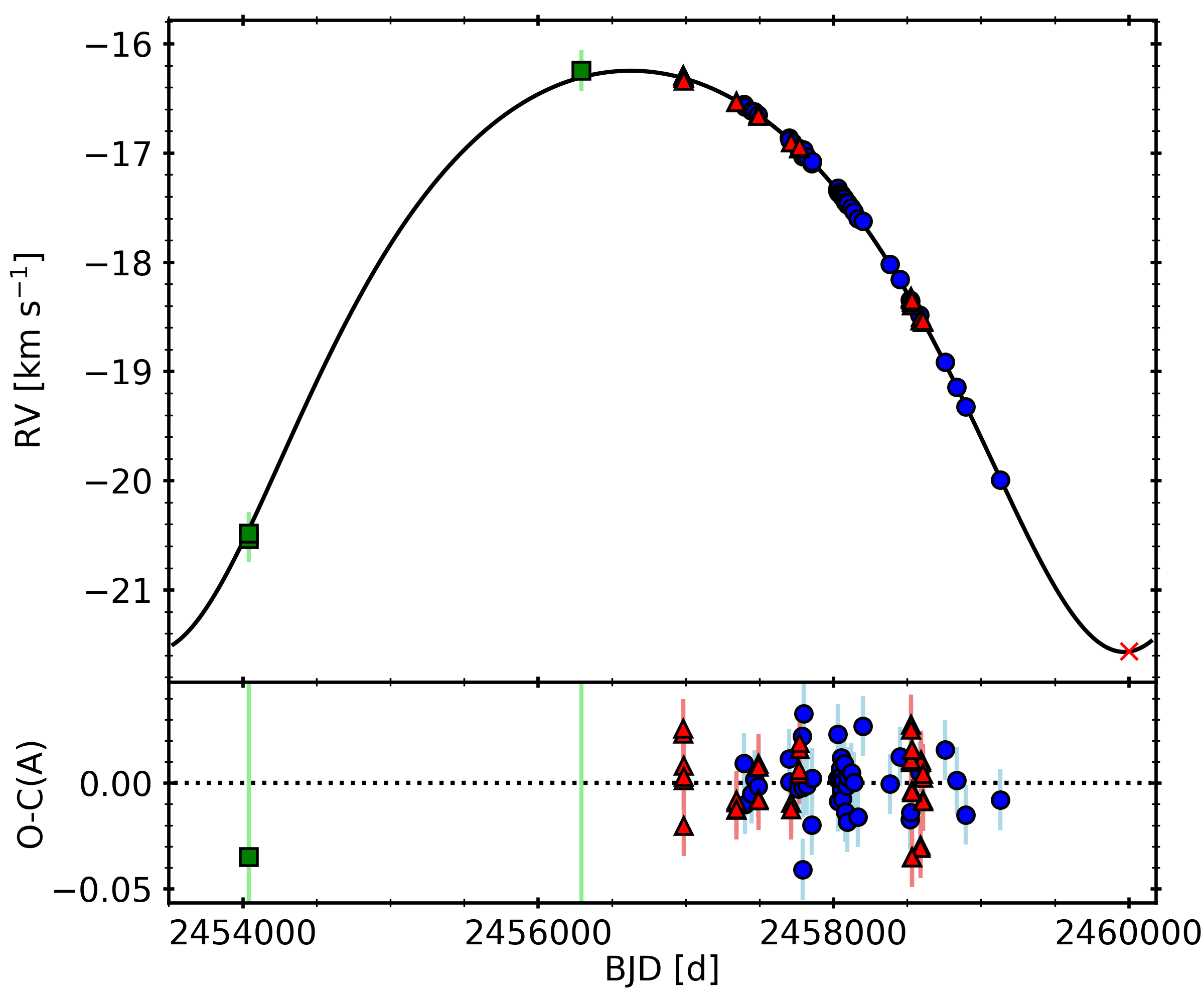}
\includegraphics[width=\columnwidth]{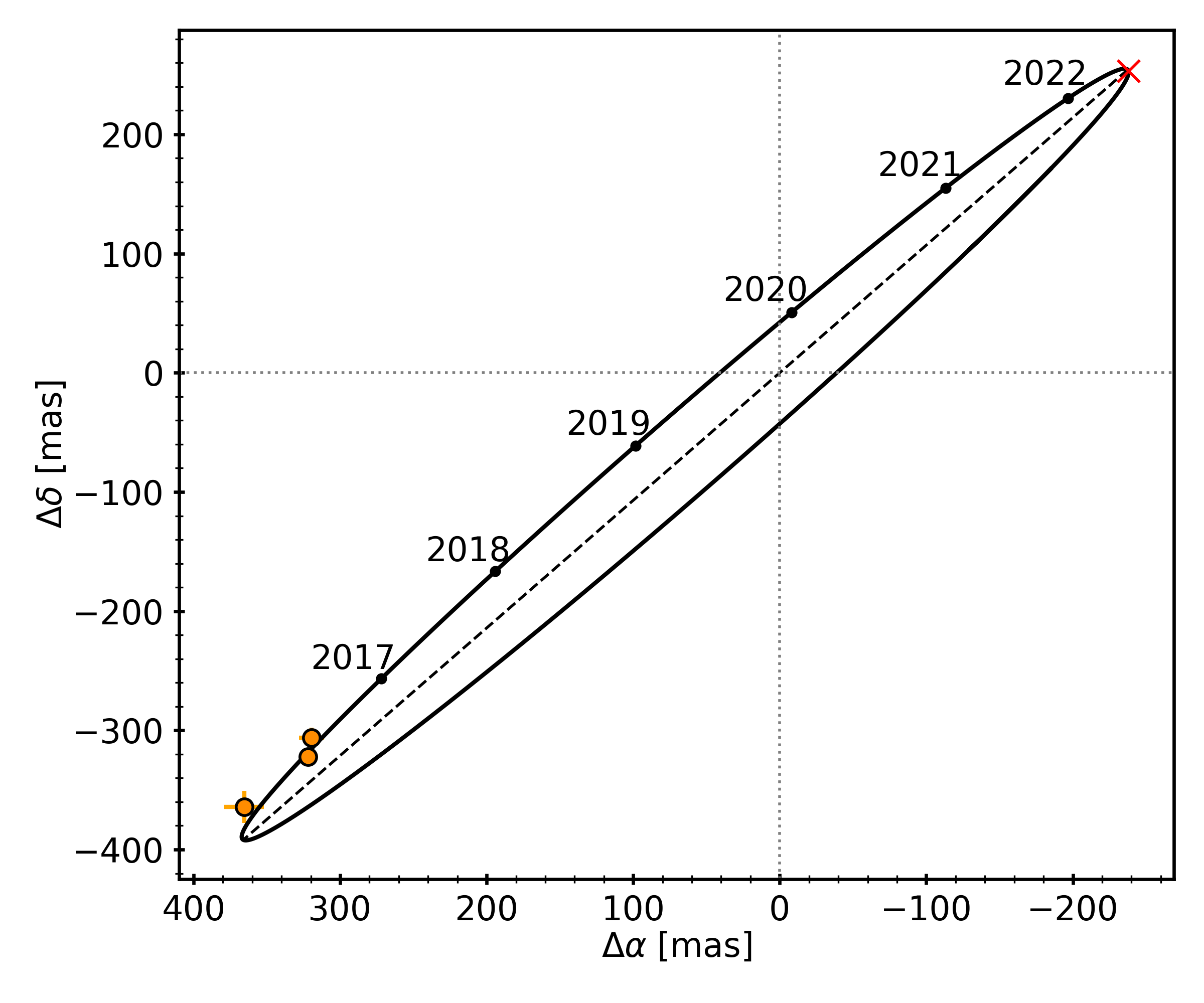}
\caption{Best joint orbital fit to the spectroscopic data and differential astrometry of GJ\,282\,C. \textit{Left}: RVs of GJ\,282\,C from CARMENES (blue cicles), HARPS (red triangles) and FEROS (green squares), as a function of time. The residuals from the fit are shown in the bottom panel, except for FEROS, which are outside the Y-axis scale. \textit{Right:} differential astrometry between GJ\,282\,Ca and GJ\,282\,Cb. Orange circles show the astrometric relative position measured from NACO observations. The red cross marks the position of the periastron, while the dashed line indicates the line of nodes.  }\label{J07361orbit}%
\end{figure*}

\begin{figure}[t]
\centering
\includegraphics[width=\columnwidth]{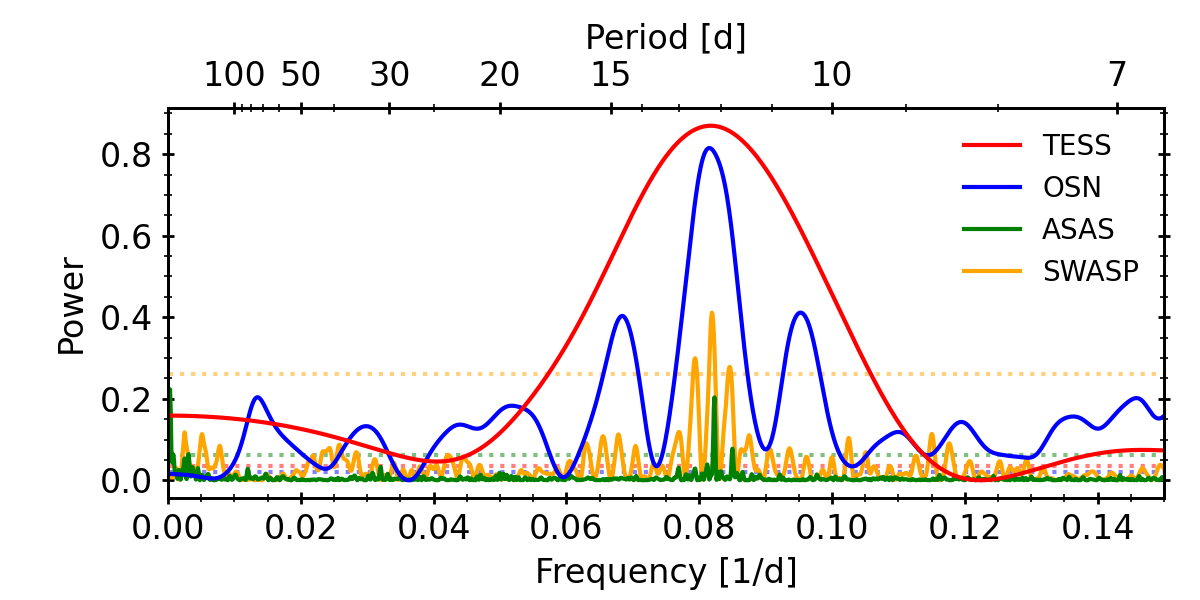}
\caption{GLS periodogram of the {\em TESS} (red), OSN (blue), ASAS (green), and SWASP (orange) photometry of GJ\,282\,C. The horizontal dotted lines of each color indicate the corresponding 0.1\,\% FAP level.}
\label{J07361GLS}%
\end{figure}

\begin{figure}[t]
\centering
\includegraphics[width=\columnwidth]{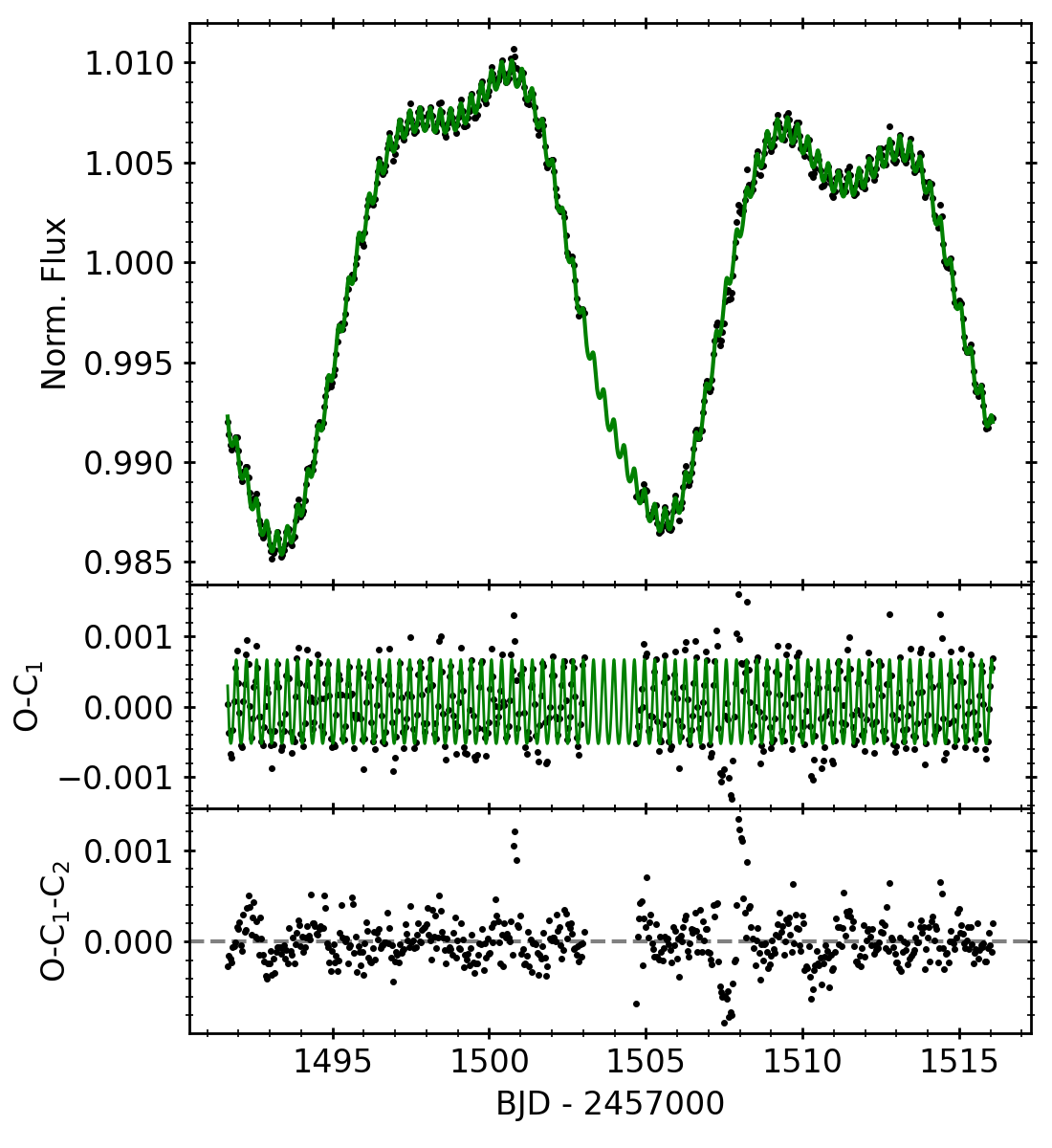}
\caption{\textit{Top}: {\em TESS} photometry of GJ\,282\,C with an hourly binning and the GP fit using as covariance function the sum of two quasi-periodic kernels, in green. \textit{Middle:} residuals and best short-period GP model after subtracting the long-period model. \textit{Bottom:} Residuals of the {\em TESS} data.}
\label{J07361TESS}%
\end{figure}

GJ\,282\,C is an M1.0\,V star located at a distance of $14.23\pm0.03$\,pc. It is active, with pEW(H$\alpha$) = $-0.86\pm0.01$\,{\AA} \citep{Schoefer2019}, and also a bright X-ray source as observed by the {\em ROSAT} satellite \citep{Voges1999,Kiraga2012}. \cite{DiezAlonso2019} reported a rotational period of $12.2\pm0.1$\,d, based on ground-based photometry from ASAS. The projected rotational velocity is $3.1\pm1.5$\,km\,s$^{-1}$ \citep{Reiners2018}. \cite{Cortes2017} observed this object with the FastCam high-resolution lucky imager, but no companions were detected.
\cite{Poveda2009} determined that GJ\,282\,C is bound to the GJ\,282\,AB system, a pair of K3\,V and K7\,V stars \citep{PMSU,Montes2018}, based on their small projected and radial separations, 55\,300 and 25\,000\,au, respectively, their similar proper motions and RVs, and their identification as X-ray sources in the {\em ROSAT} catalogue. Nevertheless, \cite{Poveda2009} could not attribute the small discrepancies observed between the values of the proper motions and the RVs to the A--B orbit. From the kinematics of GJ\,282\,A, and therefore, also GJ\,282\,C, \cite{Tabernero2017} classified this system as a member of the Ursa Major moving group, with an estimated age of about 300\,Ma \citep{Soderblom1993}. 
 
In the recent years, several works have published claims of the detection of a secondary companion to GJ\,282\,Ca (dubbed GJ\,282\,Cb), although none of them provides a robust determination of the orbital parameters and masses. In order to determine the presence of perturbing objects in nearby stars, \cite{Kervella2019} searched for proper motion anomalies between the long-term proper motion vector, computed from the difference in the astrometric position between {\em Hipparcos} and {\it Gaia} and the proper motion measurements from the two catalogues. From this study, \cite{Kervella2019} obtained a lower limit to the secondary mass, normalised at an orbital radius of 1\,au, of $35.4^{+9.9}_{-3.8}$\,M$_{\rm Jup}$\,au$^{-1/2}$. Furthermore, from high-contrast imaging observations made with NACO at the VLT, \cite{Kammerer2019} detected a companion at $441.5\pm0.02$\,mas, with an $L_p$-band contrast of $2.619\pm0.005$\,mag, which they classified as a candidate stellar-mass companion. Assuming a primary mass of 0.553\,M$_{\odot}$, as listed in Table\,\ref{tab:props}, and using the 300\,Ma BHAC15 models, this contrast would correspond to a $\sim$0.16\,M$_{\odot}$ companion. Finally, \cite{Grandjean2020} reported the detection of a low-mass stellar companion from the analysis of the HARPS RV data, in which they found a long-term trend with an amplitude of 700\,m\,s$^{-1}$ and also a short-term variation with an amplitude of 50\,m\,s$^{-1}$. Although they attributed the short-term variation to magnetic activity, they modelled the data with two Keplerian terms, obtaining a long period of $8500\pm1900$\,d and an eccentricity of $0.55\pm0.04$, corresponding to a companion with a minimum mass of $29^{+18}_{-16}$\,$M_{\rm Jup}$, and a period of 22\,d for the short-term signal. %We note here that 
However, this short period is not consistent with the rotation period of 12.2\,d reported from photometry \citep{DiezAlonso2019}, which would be the expected variability timescale for activity-related signals. 
 
We observed GJ\,282\,C with CARMENES between January 2016 and November 2020, gathering a total of 42 spectra. Additionally, we retrieved 38 and three spectra from the HARPS and FEROS archives, respectively. A preliminary analysis of the RVs obtained with \texttt{todmor} revealsed a clear long-period modulation with an amplitude of $\sim4$\,km\,s$^{-1}$. No signature of the companion was found in the spectra. As for the other systems, we used the difference between the CARMENES RVs derived with \texttt{todmor} and the RVs from \texttt{serval}, to correct the offset for the FEROS RVs, which could not be computed with \texttt{serval}. %We provide in Table\,\ref{tab:RVSB1} in the Appendix the list of RVs used in the analysis.
Besides the RVs, we also analised three high-contrast images acquired with NACO in January 2015, February 2016, and March 2016 to derive the angular separations and position angles of the two components, which we list in Table\,\ref{tab:NACO}, together with the measured magnitude differences.

We simultaneously fitted the RVs and the differential astrometry using the code \texttt{orvara}, which also uses the acceleration on the plane of the sky from the Hipparcos-{\it Gaia} Catalog of Accelerations. To avoid systematic effects due to the different filters used in the astrometric observations, we scaled the errors of the separation and position angle listed in Table\,\ref{tab:NACO} by a constant value of 3.7 and 4.5, respectively, so that the reduced $\chi^2$ statistic of the astrometric fit was equal to one. The orbital parameters resulting from the MCMC analysis are listed in Table\,\ref{tab:J07361}, and the best-fit to the RVs and to the relative astrometric orbit are shown in Fig.\,\ref{J07361orbit}. The best-fitting model has a period of \AS{6591}{136}{177}\,d with an eccentricity of \AS{0.213}{0.010}{0.010}, and an almost edge-on orbit with $i=$\AS{93.96}{0.55}{0.55}\,deg. From the computed orbital parameters, we derive absolute dynamical masses of \AS{0.554}{0.058}{0.049}\,M$_{\odot}$ and \AS{0.1881}{0.0048}{0.0047}\,M$_{\odot}$ for the primary and secondary components, respectively. The mass of the primary is in agreement with the mass derived by \cite{Schweitzer2019}, of $0.553\pm0.02$\,M$_{\odot}$, from empirical mass-luminosity calibrations. In addition, the luminosity contrast corresponding to these two masses in the $L$- and $K$-band, estimated from the 300\,Ma BHAC15 models, are 2.45 and 2.56, respectively, which are very similar to those computed from the NACO observations, listed in Table\,\ref{tab:NACO}, although their filters are not exactly matching those available from the models. The estimated contrast in the $I$-band is 3.0, which lies right at the detection limit of the FastCam high-resolution imaging observations made in March 2014 by \cite{Cortes2017}, with a predicted angular separation of 500\,mas. We find an additional significant signal at 12.1\,d in the RV residuals of the best-fit, which is in agreement with the rotation period found by \cite{DiezAlonso2019}. No more significant signals are found in the data.

We analised the available photometry, listed in Table\,\ref{tab:obs}, to search for rotational signals. As shown in the GLS periodograms in Fig.\,\ref{J07361GLS}, we found significant periods at $12.15\pm0.03$\,d, $12.21\pm0.09$\,d, $12.27\pm0.6$\,d, and $12.2\pm2.5$\,d in the ASAS, SWASP, OSN, and {\em TESS} data, respectively, confirming the period found by \cite{DiezAlonso2019} using the ASAS photometry only. 
We also found short-period modulations of small amplitude in the {\em TESS} data, which could be produced by the companion. To compute their period, we fitted the {\em TESS} light curve with a GP, using a sum of two quasi-periodic kernels as a covariance function. We show the {\em TESS} photometry with the best-fit GP model in Fig.\,\ref{J07361TESS}. 
We obtained periods of $12.41\pm0.10$\,d and $0.32574\pm0.00018$\,d for the large- and small-amplitude modulations, respectively. 
The estimated contribution of the companion to the total flux in the {\em TESS} band, based on the luminosity contrast measured by direct imaging, is around 3\,\%. 
The short-period modulations, with an amplitude of 0.12\,\% as can be seen in the middle panel of Fig.\,\ref{J07361TESS}, could thus be explained by 4\,\% modulations on the brightness of the companion.
Another possibility is that the modulation is produced by a background star located within the {\em TESS} photometric mask around this object (Gaia EDR3 3060789275061364480). This star has a {\em TESS}-band magnitude of $T$ = 12.0\,mag and it is, thus, brighter than GJ\,282\,Cb, which has an estimated $T$ of $\sim12.6$\,mag. The background star contributes with 7\,\% of the global flux, which means that variations of 1.7\,\% in its brightness would be required to produce the observed modulation. With an effective temperature for the background star of 6000\,K estimated from \textit{Gaia} data, such variations would be highly unusual \citep{McQuillan2012}. Given the young age and strong activity of GJ\,282\,C, we suggest that the short-period photometric signal is plausibly produced by the low-mass companion rather than the background star. We attribute the 12.2\,d signal to the rotation of the primary component. In this case, we adopt the period from the ASAS photometry because the light curve covers a large number of cycles, thus leading to a tighter constraint. 

Finally, using the systemic radial and astrometric properties derived from our analysis, listed in Table\,\ref{tab:J07361}, we recalculated the galactocentric velocity, obtaining $25.1$, $-2.5$, and $-7.8$\,km\,s$^{-1}$ for the $U$, $V$, and $W$ components, respectively. These values are very similar to those of GJ\,282\,A listed by \cite{Tabernero2017} and compatible with members of the Ursa Major moving group \citep{Montes2001}, which supports the physical association of the quadruple system.
Moreover, the pEW(H$\alpha$) of GJ\,282\,C is similar to or grater in absolute value than the values for members of the Hyades \citep[$ \sim $650--800\,Ma;][and references therein]{Douglas2019}, measured by \cite{Terndrup2000} and \cite{Fang2018}.
Likewise, the X-ray luminosity $\log{L_\mathrm{X}} = 29.29\pm 0.17$\,erg\,s$^{-1}$ \citep{Voges1999} is also compatible with members of the Hyades \citep{Nunez2016}, even if we assume that the Ca and Cb components have similar levels of X-ray emission. 
The rotational periods of GJ\,282\,Ca and Cb are longer than those measured by \citet{Rebull2016} for members of the Pleaides \citep[$ \sim $110--130\,Ma;][]{Stauffer1998,Dahm2015} but shorter than for stars of Praesepe \citep[$\sim $600--750\,Ma;][and references therein]{Douglas2019} as measured by \cite{Douglas2016}. All this information supports the membership of GJ\,282\,C in the Ursa Major moving group that, considering the age of this association, makes this one of the youngest multiple stellar systems with known ages and dynamical masses \citep{Lodieu2020}. Furthermore, the masses are comparable with field objects of similar spectral type \citep[][and references therein]{Benedict2016}, which indicates that that these low-mass stars in the mass range of 0.2--0.5\,M$_{\odot}$ have already arrived on the main sequence at the age of the moving group \citep{Zapatero2014}.

%\subsection{SB2} \label{sec:SB2}

\begin{table}[t]
\centering
\caption{Computed and derived parameters for the SB2 systems UU\,UMi and UCAC4\,355--020729.}
\label{tab:paramsdouble}
\begin{tabular}{lcc} 
\hline\hline
\noalign{\smallskip}
\multirow{2}{*}{Parameter} & \multicolumn{2}{c}{System}\\
\cline{2-3}
\noalign{\smallskip}
 & UCAC4\,355--020729 & UU\,UMi  \\
\noalign{\smallskip}
\hline
\noalign{\smallskip}
\multicolumn{3}{c}{\textit{Orbital parameters}}\\
\noalign{\smallskip}
\hline
\noalign{\smallskip}
$P$ [d] & \AS{6.56025}{0.00030}{0.00029} & \AS{5240}{410}{290}\\
\noalign{\smallskip}
$T_{0}$ [BJD] & \AS{2458002.367}{0.054}{0.054} & \AS{2455220}{160}{220}\\
\noalign{\smallskip}
$K_A$ [km\,s$^{-1}$] & \AS{2.157}{0.012}{0.012} & \AS{3.82}{0.44}{0.43}\\
\noalign{\smallskip}
$K_B$ [km\,s$^{-1}$] & \AS{2.510}{0.019}{0.019} & \AS{6.42}{0.73}{0.71}\\
\noalign{\smallskip}
$e\sin \omega$  & \AS{0.0226}{0.0060}{0.0060} & \AS{0.056}{0.019}{0.021} \\
\noalign{\smallskip}
$e\cos \omega$  & \AS{-0.1055}{0.0057}{0.0058} & \AS{-0.304}{0.086}{0.068} \\
\noalign{\smallskip}
$e$ & \AS{0.1080}{0.0058}{0.0057}  & \AS{0.310}{0.069}{0.087} \\
\noalign{\smallskip}
$\omega$ [deg]  & \AS{167.9}{3.2}{3.2} & \AS{169.5}{3.2}{3.7} \\
\noalign{\smallskip}
$\gamma$ [km\,s$^{-1}$] & \AS{13.9593}{0.0088}{0.0091} & \AS{-41.28}{0.09}{0.10} \\
\noalign{\smallskip}
$\sigma$ [km\,s$^{-1}$] & \AS{0.010}{0.011}{0.007} & \AS{0.0067}{0.0070}{0.0047} \\
\noalign{\smallskip}
\hline
\noalign{\smallskip}
\multicolumn{3}{c}{\textit{Derived parameters}}\\
\noalign{\smallskip}
\hline
\noalign{\smallskip}
$q\equiv M_{B}/M_{A}$ & \AS{0.8593}{0.0082}{0.0080}  & \AS{0.593}{0.040}{0.035}\\
\noalign{\smallskip}
$a\sin{i}$ [au] & \AS{0.0027974}{0.0000073}{0.0000072} & \AS{4.70}{0.10}{0.09}\\
\noalign{\smallskip}
%$a\sin{i}$ [mas] & \AS{0.09624}{0.00069}{0.00070} & \AS{328}{10}{10}\\
%\noalign{\smallskip}
$M_A\sin^3{i}$ [M$_{\odot}$] & \AS{0.00003650}{0.00000061}{0.00000061} & \AS{0.315}{0.061}{0.056}\\
\noalign{\smallskip}
$M_B\sin^3{i}$ [M$_{\odot}$] & \AS{0.00003136}{0.00000043}{0.00000043} & \AS{0.187}{0.037}{0.034}\\
\noalign{\smallskip}
\hline
\end{tabular}
\end{table}

\subsection{Double-line spectroscopic binaries (SB2s)}
\subsubsection{UCAC4\,355--020729}

\begin{figure}[t]
\centering
\includegraphics[width=\columnwidth]{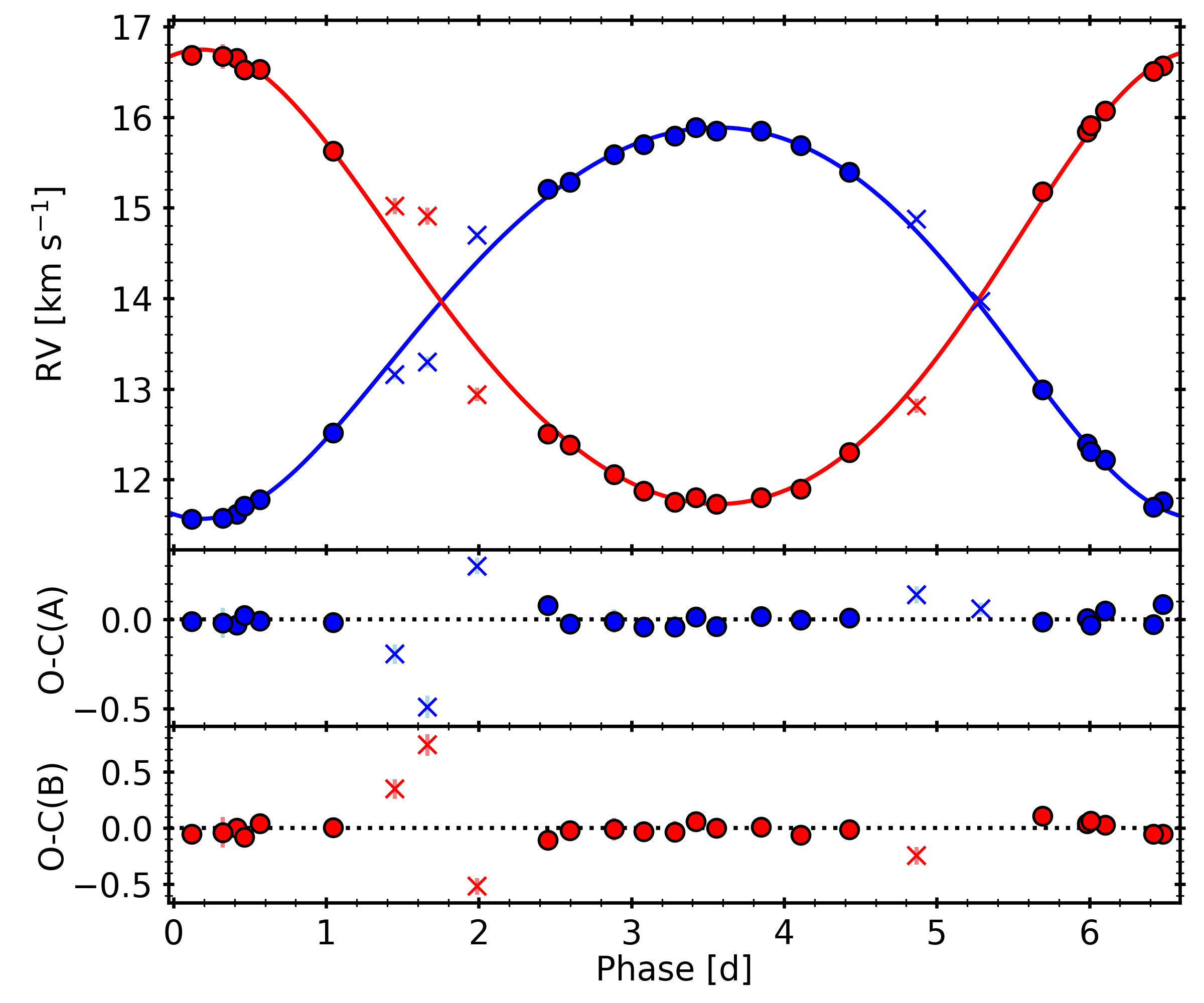}
\caption{\texttt{todmor} RV curves of UCAC4\,355--020729 as a function of the orbital phase, along with their best-fitting models. Blue and red circles correspond to the primary and secondary components, respectively, while the crosses indicate data points not used in the fit. The residuals from the fit are shown in the bottom panels.}
\label{J07001orbit}%
\end{figure}

\begin{figure}[t]
\centering
\includegraphics[width=\columnwidth]{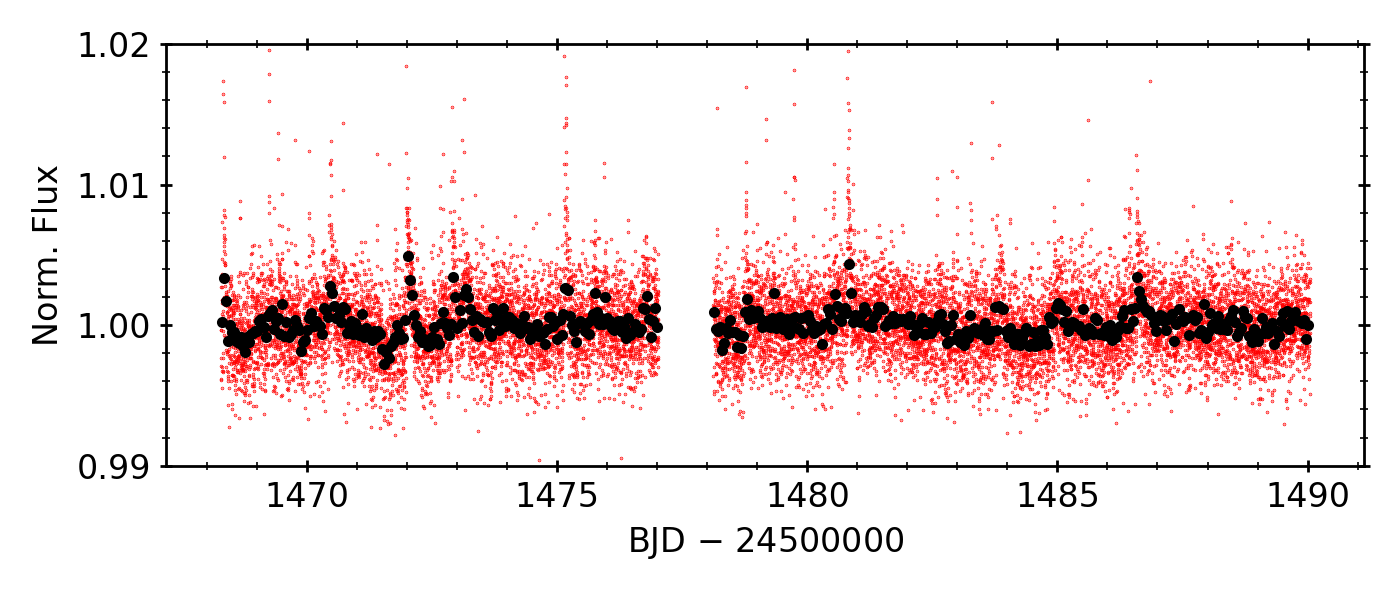}
\includegraphics[width=\columnwidth]{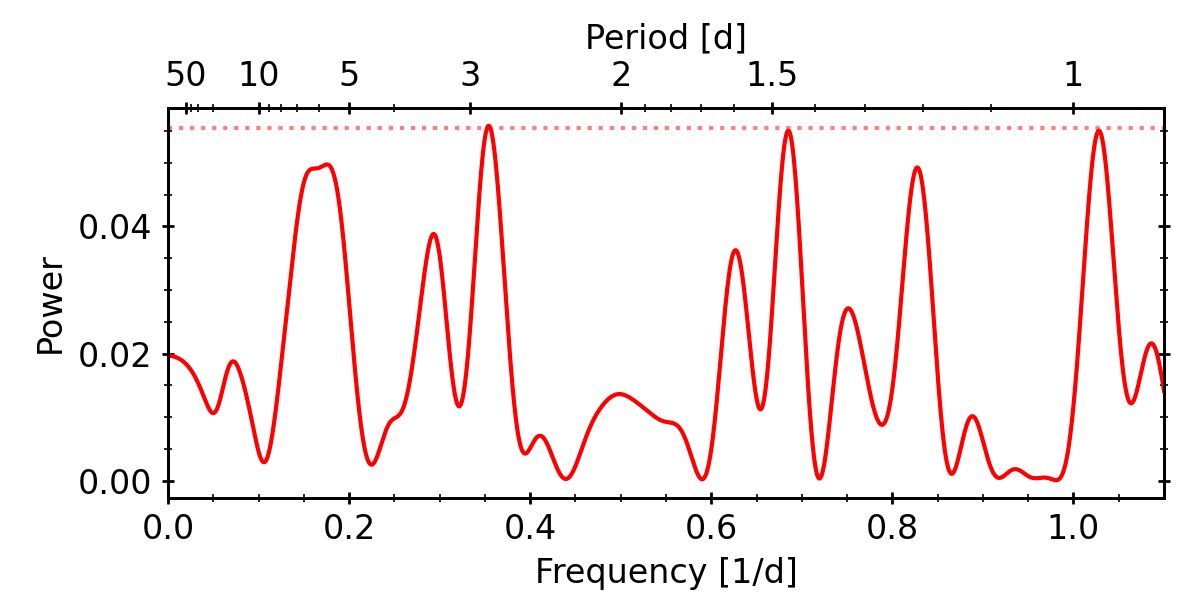}
\caption{\textit{Top panel: }{\em TESS} photometry of UCAC4\,355--020729 (red) and its 1 hour binning (black). \textit{Bottom panel:} GLS periodogram of the hourly binned {\em TESS} photometry. The horizontal dotted line indicates the 0.1\% FAP level.}
          \label{J07001TESS}%
\end{figure}

UCAC4\,355--020729 is an M5.0\,V star located at a distance of $29.83\pm0.03$\,pc. Catalogued as a bright X-ray source in the {\em ROSAT} bright source catalogue \citep{Voges1999,Haakonsen2009}, it is an active star, with pEW(H$\alpha$) = $-6.99\pm0.04$\,\AA, despite the relatively low value of its projected rotational velocity of $3.1\pm1.5$\,km\,s$^{-1}$.

We observed this system with CARMENES between November 2016 and November 2018, obtaining a total of 27 spectra. Although we could detect two signals in all spectra using \texttt{todmor}, the resulting RVs near the conjunctions were probably affected by systematics produced by blended lines due to the small RV difference between the two signals. We therefore removed these RVs from our orbital analysis.
Table\,\ref{tab:paramsdouble} lists the orbital parameters corresponding to the best-fitting solution to the RVs, while in Fig.\,\ref{J07001orbit} we show the RVs phase folded to the best orbital period. The best-fitting solution yields a slightly eccentric orbit, with $e=0.1123\pm0.0054$, and an orbital period of $P=6.56045\pm0.00028$\,d. The semi-amplitudes of the RVs yield a mass ratio of \AS{0.8603}{0.0082}{0.0081}, and minimum masses of $3650\pm61\times10^{-8}$\,M$_{\odot}$ and $3136\pm43\times10^{-8}$\,M$_{\odot}$
%$3.650\pm0.061\times10^{-5}$ and $3.136\pm0.043\times10^{-5}$\,M$_{\odot}$ 
for the primary and secondary components, respectively.

The very low minimum masses obtained could be an indication of an almost pole-on orientation of the orbit. Following the method described by \cite{Baroch2018}, from the $V$-band magnitude of this system and the derived mass ratio listed in Table\,\ref{tab:paramsdouble}, and using the empirical mass-luminosity relation by \cite{Mann2019}, we estimated the individual masses of the two components of the system, $M_{\rm A}=0.24\pm0.02$ and $M_{\rm B}=0.22\pm0.02$\,M$_{\odot}$. Using the obtained minimum masses, the orbital inclination should be lower than 3.0\,deg. This low inclination would be consistent with the low value of the measured projected rotational velocity. This argument is even more convincing considering that this value is probably overestimated because of the blending of the spectral lines.

Finally, we analysed the photometry and activity indicators in search for stellar rotation signals, although the low inclination of the system is likely to greatly reduce the amplitude of a possible modulation. No significant signal was found in any of the activity indicators, aside from signals in the CRX and dLW that can be attributed to the blending of the two component spectra. The light curve from {\em TESS} (top panel of Fig.\,\ref{J07001TESS}) reveals several flare events during the 27\,d of continuous monitoring, as expected from the high activity of the star. After removing the flares with a sigma clipping and a rebinning of 1\,h, the GLS periodogram (bottom panel of Fig\,\ref{J07001TESS}) shows three signals just above the 0.1\,\% FAP level, at periods of 0.97, 1.46, and 2.82\,d, which are not related to the 6.56\,d by any alias or harmonic. There are also no significant signals in the ASAS photometry.

\subsubsection{UU\,UMi}

\begin{figure}[t]
\centering
\includegraphics[width=\columnwidth]{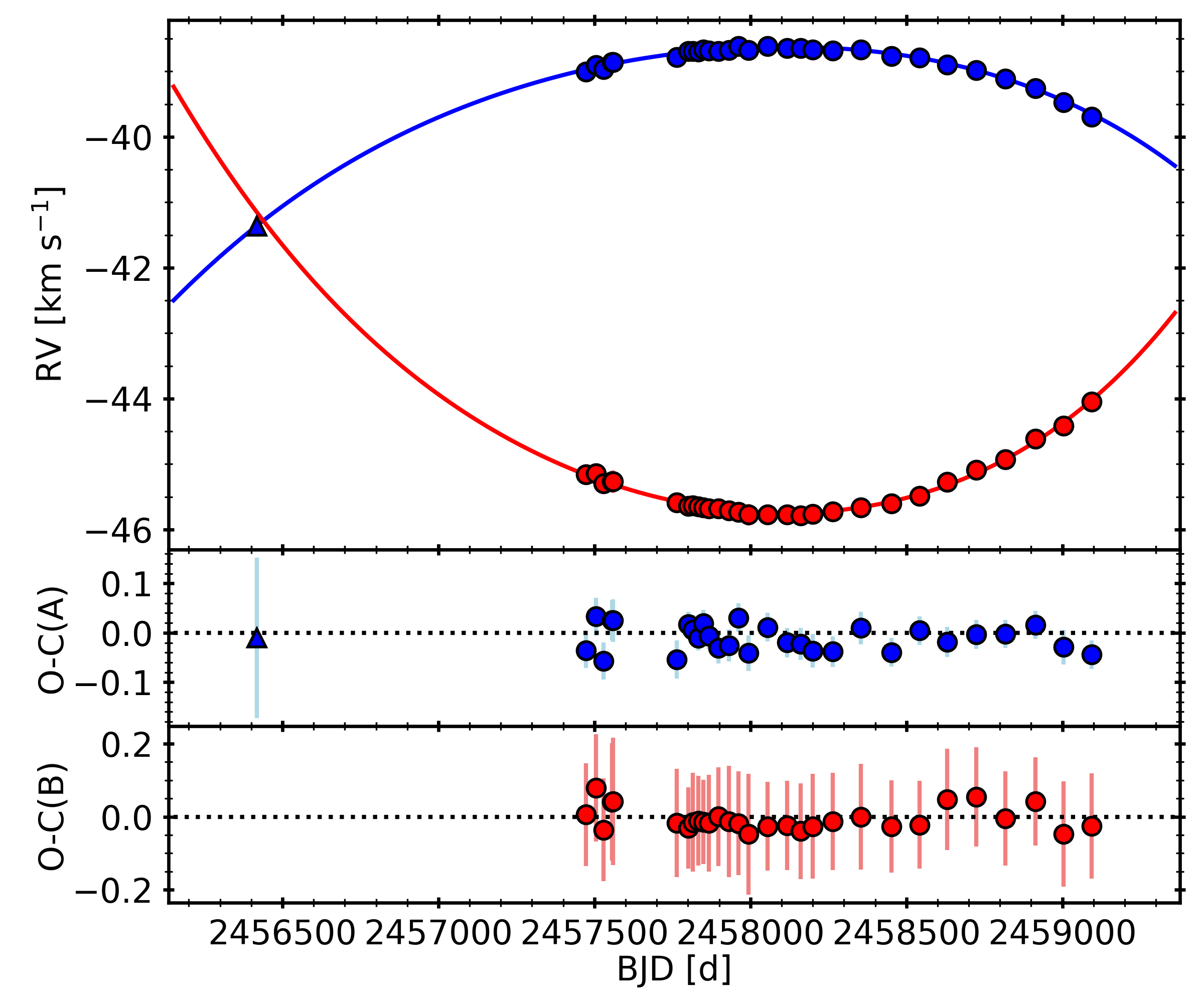}
\caption{Best fit to the \texttt{todmor} RVs of UU\,UMi as a function of time. Blue and red colors correspond to the primary and secondary components, respectively. Circular and triangular symbols correspond to the RVs from CARMENES and CAFE, respectively. The residuals from the best fit are shown in the bottom panels.}
          \label{J15412orbit}%
\end{figure}

\begin{figure}[t]
\centering
\includegraphics[width=\columnwidth]{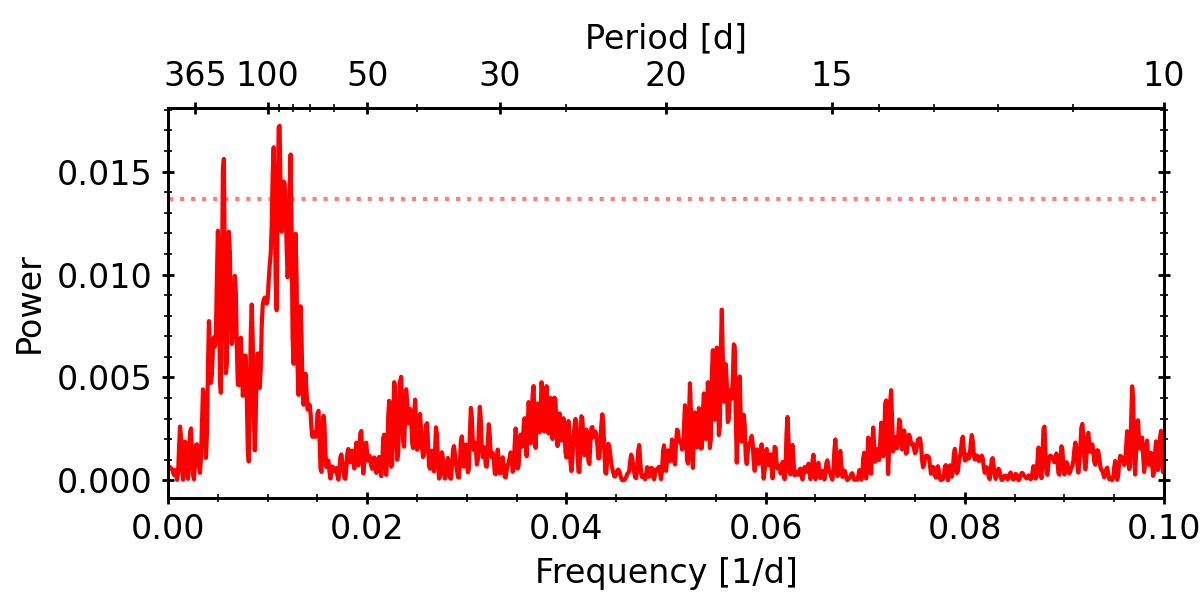}
\caption{GLS periodogram of the MEarth photometry of UU\,UMi. The horizontal dotted line indicates the corresponding 0.1\% FAP level.}
\label{J15412GLS}%
\end{figure}

Catalogued as the high proper-motion star Ross\,1057 \citep{Ross39}, UU\,UMi is an M3.0\,V star located at a distance of $14.6\pm0.3$\,pc. It is an inactive system, with pEW(H$\alpha$) $=-0.10\pm0.10$\,{\AA} and a measured projected rotational velocity below 3\,km\,s$^{-1}$. \cite{Newton2016} reported a probable rotation period of 91\,d from MEarth photometry, while \cite{DiezAlonso2019} found $83.4\pm10.4$\,d using the same dataset. %\cite{Jodar2013} could not detect companions with $\Delta I<1.5$ at more than 250\,mas in archival high-resolution lucky imaging observations using FastCam. \cite{Passegger2018} mass of $0.34\pm0.03$ from the mass-luminosity empirical relation by \cite{Benedict2016}. \citep{Gaidos2014}, mass and radii of 0.36, estimated from luminosity. 
A first indication of its binarity was suggested by \cite{Lippincott1981}, who found evidence of variable proper motion in the residuals of plate series taken between 1970 and 1979, showing a perturbation with a period of 10--15\,a and an amplitude of 0.04\,arcsec. 
\cite{Heintz1993} expanded the study including plates up to 1990, and found UU\,UMi to be an unresolved binary with a large astrometric variation amplitude and a companion mass of about 0.10\,M$_{\odot}$. The authors derived a circular orbit with a period of 14.5\,a, an inclination of $i=92$\,deg, an amplitude of $a$ = 0.0745\,arcsec, and $\Omega=21$\,deg. 
%\cite{Kervella2019}: As for GJ\,282\,C, from Hipparcos-Gaia, found secondary of 52.56 Mj/au-1/2.

The companion was detected spectroscopically in our previous study of binary systems discovered with the CARMENES instrument \citep{Baroch2018}, making this an SB2 system. We reported a mass ratio of $0.57\pm0.06$. However, from the modest fraction of the orbit covered by the data, the authors were not able to discern between a circular and an eccentric orbit and, therefore, the eccentricity was fixed to zero. Here, we update this analysis with eight additional CARMENES observations and one archival measurement taken with the CAFE spectrograph in May 2013. Although only the primary star can be resolved in this latter case, the almost equal RV of both spectra is a useful information to constrain the systemic velocity and the time of conjunction. Table\,\ref{tab:RVSB2} lists the full set of RVs used in this analysis for UU\,UMi.

From a Keplerian fit to the RVs of this system, which is shown in Fig.\,\ref{J15412orbit}, we obtained a period of \AS{5240}{410}{290}\,d with an eccentricity of \AS{0.310}{0.069}{0.087} and a mass ratio of \AS{0.593}{0.040}{0.035}. The eccentric orbit is significantly favoured over a circular one, with a Bayesian information criterion (BIC) difference of 14 \cite[a strong evidence against the higher BIC value is obtained for $\Delta$BIC $>$ 6 --][]{Claeskens2008,Lillo2014}. Interestingly, the fitted period of \AS{14.3}{1.1}{0.8}\,a is in agreement with the value of 14.5\,a found using astrometric observations by \cite{Heintz1993}. Using the almost edge-on inclination derived by \citet{Heintz1993}, $i=92$\,deg, we derived absolute masses of $0.32\pm0.06$\,M${_\odot}$ and $0.19\pm0.04$\,M${_\odot}$ for the primary and secondary components, respectively. %We note here that
Uncertainties in the mass may be underestimated due to the fact that the inclination had no reported error bar.

Besides, we inspected the activity indices computed with \texttt{serval} to look for signs of stellar rotation, but without success. We also analysed {\em TESS} and ASAS photometry, equally unsuccesfully. In a GLS periodogram of the seven years of MEarth photometry, which we show in Fig.\,\ref{J15412GLS}, we found a signal below the 0.1\,\% FAP level at a period of $90\pm10$\,d with an amplitude of 1\,mmag, which was already reported by \cite{Newton2016} and \cite{DiezAlonso2019}. 
This signal could be a product of the stellar rotation of one of the components and indicate an age of the system older than 1\,Ga \citep{Curtis2019}. Finally, we also reanalysed the {\em Hipparcos} photometry \citep{Hipparcos} from which this object was catalogued as variable \citep{Kazarovets1999,Adelman2001}. We found that all the points leading to the classification of UU\,UMi as a variable star are flagged as bad quality measurements. 
 
 %the available photometry for UU\,UMi in search for variability. We found a signifi searched all available photometry and the \texttt{serval} activity indicators for signs of stellar rotation. We find a significant signal at $\sim90$\,d, also found by \cite{Newton2016} and \cite{Alonso2015}. However, this signal is very significant in the nightly-average of all M-dwarf also provided in the photometric files, which correlates heavily with measured humidity and air temperature \citep{MEARTH}, and therefore we deem this signal as arising from systematic trends in the lightcurve. No significant signals are found in neither the {\em TESS} photometry nor the activity indices. We also reanalised the Hipparcos photometry \citep{Hipparcos} from which this target was catalogued as variable \citep{Kazarovets1999,Adelman2001}, and find that all the points leading to the classification as a variable star are flagged as bad quality measurements.

%\subsection{ST3} \label{sec:ST3}

\begin{table}[t]
\centering
\caption{Computed and derived parameters for the ST3 systems GJ\,3916 and GJ\,4383.}
\label{tab:paramstriple}
\begin{tabular}{lcc} 
\hline\hline
\noalign{\smallskip}
\multirow{2}{*}{Parameter} & \multicolumn{2}{c}{System}\\
\cline{2-3}
\noalign{\smallskip}
 & GJ\,3916 & GJ\,4383 \\
\noalign{\smallskip}
\hline
\noalign{\smallskip}
\multicolumn{3}{c}{\textit{Orbital parameters}}\\
\noalign{\smallskip}
\hline
\noalign{\smallskip}
$P_{AB}$ [d] & \AS{3028}{23}{23} & \AS{4634}{17}{17}\\
\noalign{\smallskip}
$T_{\rm 0,AB}$ [BJD] & \AS{2456972}{39}{38}& \AS{2458700}{39}{36}\\
\noalign{\smallskip}
$K_A$ [km\,s$^{-1}$] & \AS{7.43}{0.37}{0.32}& \AS{4.206}{0.037}{0.036}\\
\noalign{\smallskip}
$K_B$ [km\,s$^{-1}$] & \AS{7.07}{0.35}{0.32} & \AS{7.66}{0.11}{0.11}\\
\noalign{\smallskip}
$e_{AB}\sin \omega_{AB}$ & \AS{0.318}{0.023}{0.023} & \AS{0.0176}{0.0061}{0.0066}\\
\noalign{\smallskip}
$e_{AB}\cos \omega_{AB}$ & \AS{0.328}{0.031}{0.030}& \AS{-0.1052}{0.0050}{0.0048}\\
\noalign{\smallskip}
$e_{AB}$ & \AS{0.458}{0.026}{0.026}& \AS{0.1068}{0.0049}{0.0049}\\
\noalign{\smallskip}
$\omega_{AB}$ [deg] & \AS{44.1}{3.6}{3.6}& \AS{170.4}{3.6}{3.3}\\
\noalign{\smallskip}
$P_{B}$ [d] & \AS{132.959}{0.013}{0.014} &\AS{216.07}{0.13}{0.14}\\
\noalign{\smallskip}
$T_{\rm 0,B}$ [BJD] & \AS{2458153.2}{1.1}{1.1} & \AS{2458122.5}{1.1}{1.1}\\
\noalign{\smallskip}
$K_{Ba}$ [km\,s$^{-1}$] & \AS{13.56}{0.10}{0.10}& \AS{11.26}{0.17}{0.17} \\
\noalign{\smallskip}
$K_{Bb}$ [km\,s$^{-1}$] & \AS{15.14}{0.11}{0.10} & \AS{12.02}{0.20}{0.19} \\
\noalign{\smallskip}
$e_{B}\sin \omega_{B}$ & \AS{-0.0793}{0.0055}{0.0055} & \AS{0.277}{0.012}{0.012} \\
\noalign{\smallskip}
$e_{B}\cos \omega_{B}$ & \AS{-0.0767}{0.0051}{0.0052}& \AS{0.059}{0.011}{0.011} \\
\noalign{\smallskip}
$e_{B}$ & \AS{0.1104}{0.0049}{0.0046}& \AS{0.283}{0.012}{0.011}\\
\noalign{\smallskip}
$\omega_{B}$ [deg] & \AS{46.0}{2.9}{3.0}& \AS{77.9}{2.2}{2.3} \\
\noalign{\smallskip}
$\gamma$ [km\,s$^{-1}$] & \AS{2.962}{0.077}{0.065}& \AS{-11.409}{0.029}{0.027}\\
\noalign{\smallskip}
$\sigma_{CARM}$ [km\,s$^{-1}$] & \AS{0.397}{0.041}{0.038}& \AS{0.0231}{0.0089}{0.0087}\\
\noalign{\smallskip}
$\sigma_{FEROS}$ [km\,s$^{-1}$] & \AS{0.76}{0.10}{0.09}& \AS{0.13}{0.19}{0.09}\\
\noalign{\smallskip}
$\sigma_{HARPS}$ [km\,s$^{-1}$] & \AS{0.44}{0.32}{0.22}& \AS{0.033}{0.039}{0.023}\\
\noalign{\smallskip}
$\sigma_{UVES}$ [km\,s$^{-1}$] & \AS{0.53}{0.31}{0.22} & $\cdots$ \\
\noalign{\smallskip}
\hline
\noalign{\smallskip}
\multicolumn{3}{c}{\textit{Derived parameters}}\\
\noalign{\smallskip}
\hline
\noalign{\smallskip}
$q_{AB}\equiv M_{B}/M_{A}$ & \AS{1.049}{0.051}{0.040} & \AS{0.549}{0.010}{0.010}\\
\noalign{\smallskip}
$a_{AB}\sin{i_{AB}}$ [au] & \AS{3.59}{0.10}{0.09}& \AS{5.025}{0.051}{0.048}\\
\noalign{\smallskip}
$M_A\sin^3{i_{AB}}$ [M$_{\odot}$]& \AS{0.328}{0.033}{0.028}& \AS{0.509}{0.017}{0.016}\\
\noalign{\smallskip}
$M_B\sin^3{i_{AB}}$ [M$_{\odot}$] & \AS{0.344}{0.035}{0.028}& \AS{0.2795}{0.0061}{0.0058}\\
\noalign{\smallskip}
$q_{B}\equiv M_{Bb}/M_{Ba} $ & \AS{0.8952}{0.0094}{0.0089} & \AS{0.937}{0.021}{0.020}\\
\noalign{\smallskip}
$a_{B}\sin{i_{B}}$ [au] &\AS{0.3487}{0.0018}{0.0018} & \AS{0.4436}{0.0047}{0.0050}\\
\noalign{\smallskip}
$M_{Ba}\sin^3{i_{B}}$ [M$_{\odot}$] & \AS{0.1511}{0.0025}{0.0025} & \AS{0.1206}{0.0040}{0.0040}\\
\noalign{\smallskip}
$M_{Bb}\sin^3{i_{B}}$ [M$_{\odot}$] &  \AS{0.1353}{0.0030}{0.0030} & \AS{0.1130}{0.0048}{0.0049}\\
\noalign{\smallskip}
$\sin{i_{AB}}/\sin{i_{B}}$ &\AS{1.063}{0.036}{0.031} & \AS{1.062}{0.017}{0.015}\\
\noalign{\smallskip}
\hline
\end{tabular}
\end{table}

\subsection{Triple-line spectroscopic triples (ST3s)}
\subsubsection{GJ 3916}

\begin{figure}[p]
\centering
\includegraphics[width=0.96\columnwidth]{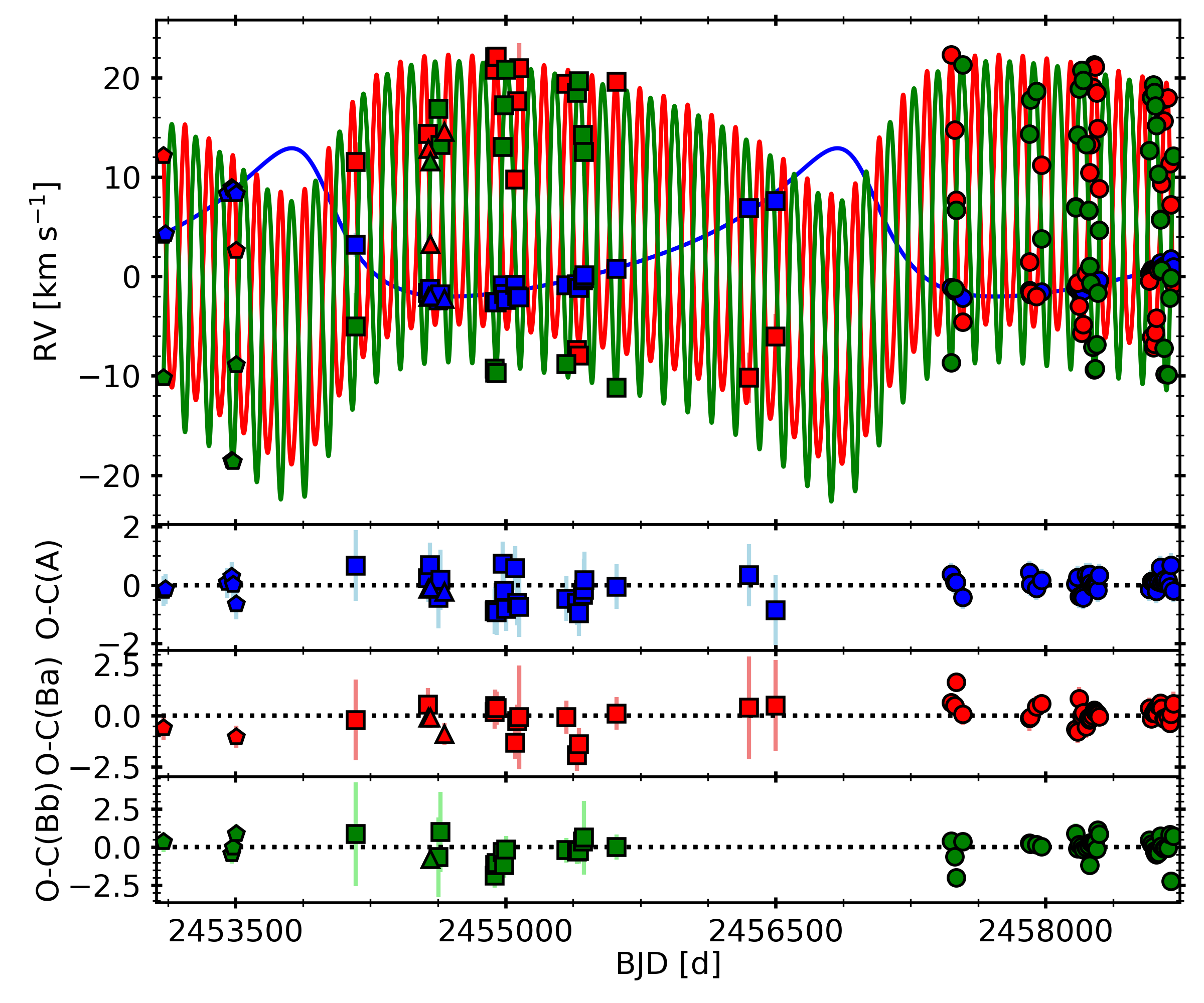}
\includegraphics[width=0.96\columnwidth]{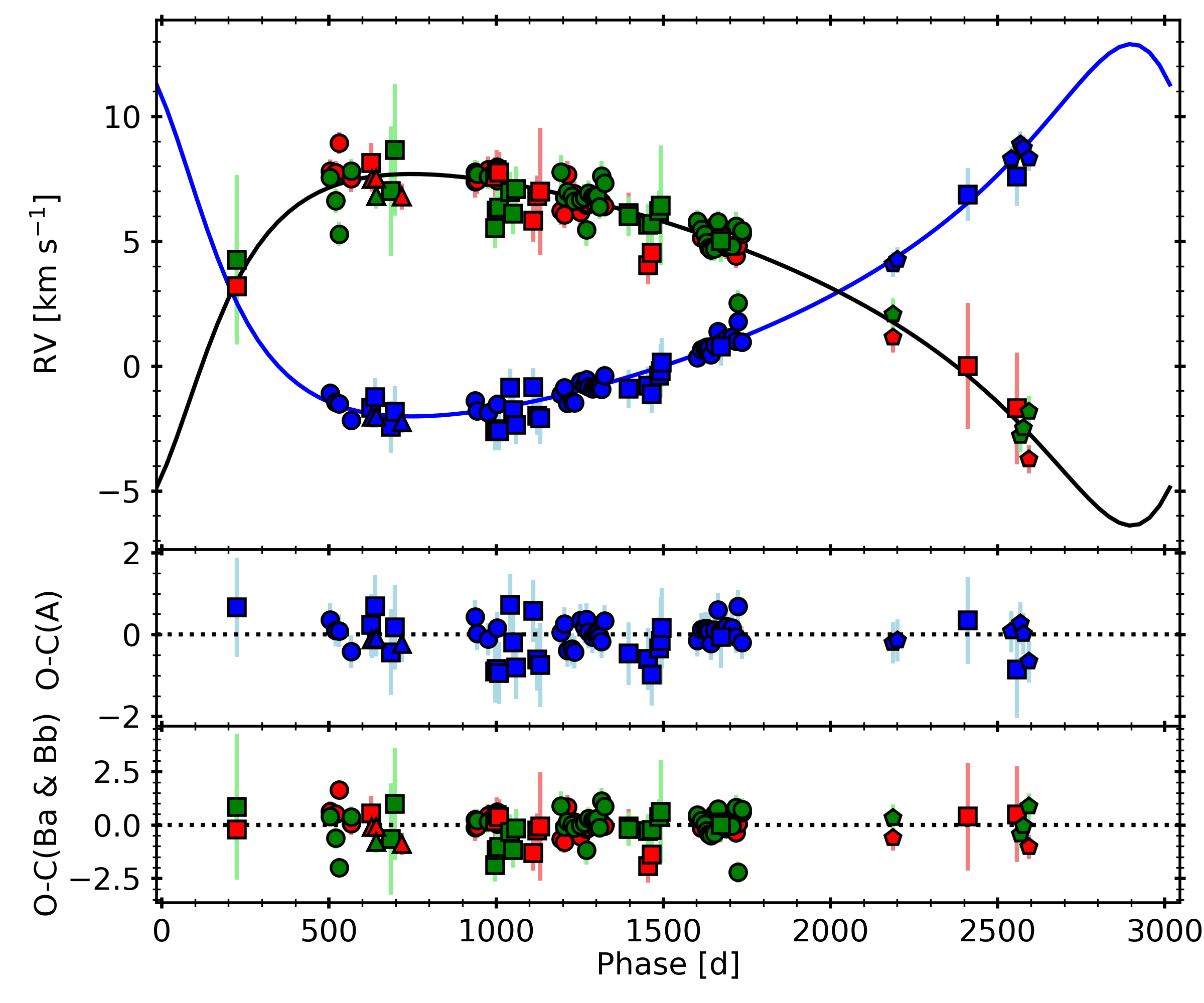}
\includegraphics[width=0.96\columnwidth]{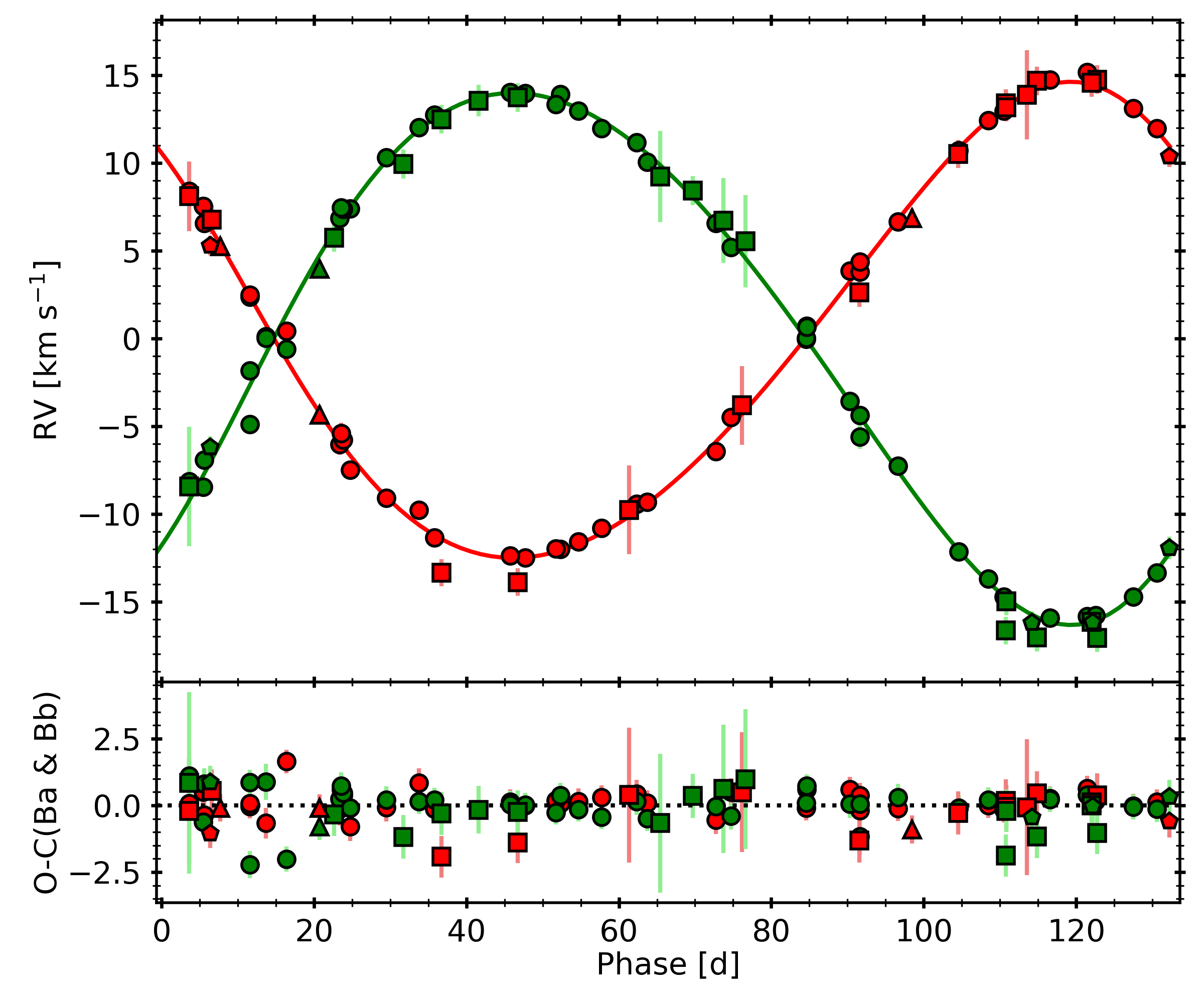}
\caption{{\em Top panel}: Fit to the RVs of the three components of GJ\,3916 as a function of time. 
{\em Middle panel}: Same fit but phase folded to the period of the outer orbit and with the inner orbit subtracted. 
{\em Bottom panel}: Fit to the inner orbit of the components Ba and Bb with outer orbit subtracted. 
Blue, red, and green symbols indicate the RVs of components A, Ba, and Bb, respectively. 
Measurements from CARMENES, FEROS, UVES, and HARPS are plotted as circular, squared, pentagonal, and triangular symbols, respectively.}
          \label{J15474orbit}%
\end{figure}

\begin{figure}[t]
\centering
\includegraphics[width=\columnwidth]{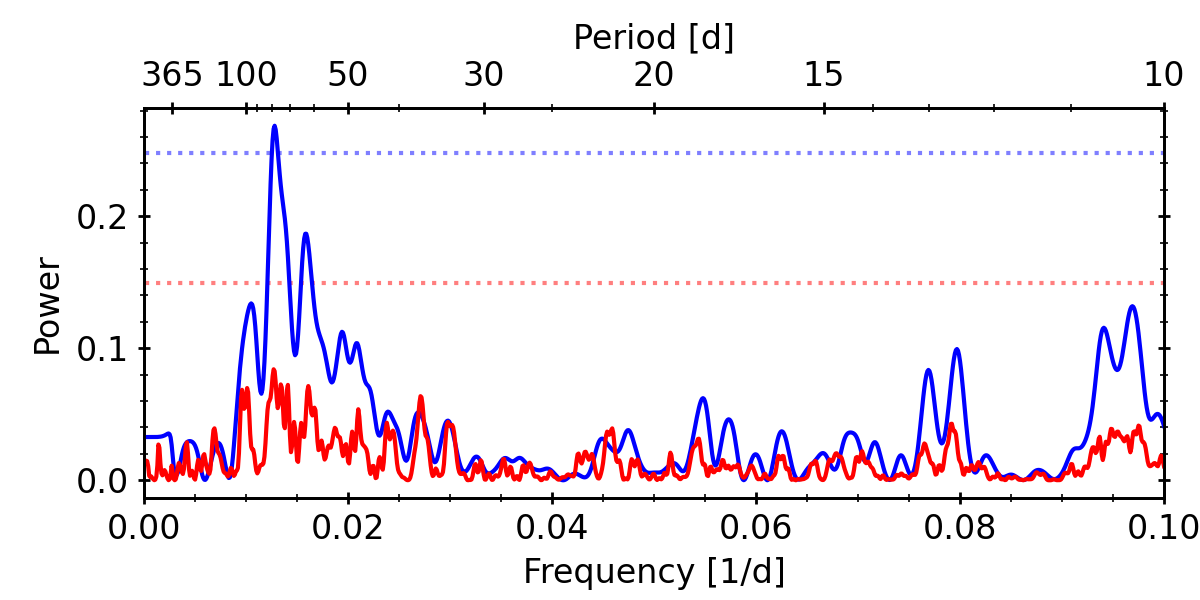}
\caption{GLS periodogram of the SWASP photometry of GJ\,3916, using the entire dataset (red) and the first two epochs (blue). The horizontal dotted lines of each color indicate the corresponding 0.1\% FAP level.}
          \label{J15474GLS}%
\end{figure}

GJ\,3916 is catalogued as an M2.0\,V high-proper motion star \citep{Luyten1955}, located at a distance of $16.00\pm0.07$\,pc. It is a quiet star, with pEW(H$\alpha$) = $-0.03\pm0.02$\,{\AA} and a projected rotational velocity below 3\,km\,s$^{-1}$. Using empirical relationships, \cite{Astudillo2017} estimated a rotation period of 78\,d from the emission strength of the Ca~{\sc ii} H\&K lines. 

The first indication of the binarity of this system was deduced from {\em Hipparcos} astrometric data %\citep{Hipparcos}, 
through the detection of an accelerated proper motion \citep{Makarov2005,Frankowski2007}. More recently, \cite{Kervella2019} reported a difference in the proper motion between {\em Hipparcos} and {\em Gaia}, from which they obtained a lower limit to the secondary mass, normalised to an orbital separation of 1\,au, of $144^{+40}_{-15}$\,M$_{\rm Jup}$\,au$^{-1/2}$. The presence of a companion to GJ\,3916 was spectroscopically confirmed by \cite{Zechmeister2009}, who reported a 4.1\,km\,s$^{-1}$\,a$^{-1}$ trend in six high-precision RVs obtained as part of the planet search programme with VLT+UVES \citep{UVES}. Moreover, \cite{Houdebine2015} detected the signal of the secondary companion by cross-correlation in three high-resolution spectra taken with HARPS. They also obtained projected rotational velocities of $1.24\pm0.14$ and $3.51\pm0.05$\,km\,s$^{-1}$ for the primary and secondary signals, respectively. 

We collected 32 spectra of GJ\,3916 between April 2016 and August 2019 with CARMENES. We also used 23 spectra taken between March 2007 and July 2013 with the FEROS spectrograph, along with the UVES and HARPS spectra mentioned above. An inspection of the CARMENES spectra with \texttt{todmor}, using synthetic spectra, revealed a clear signal from a tertiary component in the system, in addition to the two expected signals.  We proceeded as explained in Sect.\,\ref{sec:rvs1} and extracted the RVs of the three signals in the spectra. We repeated the same process for the observations from UVES, FEROS, and HARPS, where the third signal was not so evident as in the CARMENES spectra. From the available 32 archival spectra, we could only detect all three signals in 11 spectra, two signals in 17 spectra, and only the signal from the brightest companion in 4. This effect was mostly due to the unfavourable orbital position of some of the components and the lower exposure time of the HARPS, UVES, and FEROS observations with respect to those from CARMENES. We list the RVs in Table\,\ref{tab:RVST3}. %The last column in this table indicates the instrument used (U, F, H, C for UVES, FEROS, HARPS, and CARMENES, respectively).

The RVs suggest a hierarchical triple system configuration, with the most massive star, A, tracing out a long-period orbit around the common barycentre with the two other stars, Ba and Bb, which in turn are orbiting each other with a shorter orbital period. The best fit to the hierarchical Keplerian model used to describe this system is shown in Fig.\,\ref{J15474orbit}. 
The top panel depicts the RVs of the three components. 
In the middle panel, we show the long-period orbit of components A around the Ba+Bb system after subtracting the RVs of the short-period orbit. % from their RVs. 
In the bottom panel, we show the orbital motion of Ba and Bb after subtracting the orbital motion of their centre of mass around component A. 
The fitted parameters are reported in Table\,\ref{tab:paramstriple}. The outer orbit has a period of $3028\pm23$\,d with an eccentricity of $0.458\pm0.026$, while the inner orbit has a period of \AS{132.959}{0.013}{0.014}\,d and an eccentricity of \AS{0.1104}{0.0049}{0.0047}. The minimum masses are \AS{0.328}{0.033}{0.028}, \AS{0.1511}{0.0025}{0.0025}, and \AS{0.1353}{0.0031}{0.0030}\,M$_{\odot}$ for the components A, Ba, and Bb, respectively.
%We show in the top panel in Fig.\,\ref{J15474orbit} the resulting RV best fit of GJ\,3916. 

Additionally, as we did for GJ\,912, we constrained the inclination of the system by using the astrometric observations from {\em Hipparcos}. %\citep{Hipparcos,Leeuwen2007}.
Following \cite{Reffert2011} and using the orbital solution in Table\,\ref{tab:paramstriple}, we fitted an astrometric orbit to the {\em Hipparcos} abscissa residuals, and determined an inclination of $i_{AB}=99\pm8$\,deg and an ascending node of $\Omega=304\pm8$\,deg at 1\,$\sigma$ level. %It should be noted that the 
The astrometric orbit fitted here corresponds to the orbit of the photocentre, not of component A. As a result, the true apparent orbit would  depend on the flux ratio between the most and least massive (i.e. brightest and faintest) %and the lighter 
components, although this effect only affects the amplitude of the apparent orbit, and not the derived inclination. From the inclination of the outer orbit and the relation between $\sin i_{AB}$ and $\sin i_B$ we measured a difference between the inner and outer inclinations of $\Delta i=13\pm7$\,deg, consistent with coplanarity within $2\sigma$. 
We also used %can also use 
the estimated outer inclination, the minimum masses, and the mass ratios to assign absolute individual masses of $0.352\pm0.047$\,M$_{\odot}$, $0.195\pm0.026$\,M$_{\odot}$, and $0.174\pm0.024$\,M$_{\odot}$ for the A, Ba, and Bb components, respectively, which yields a mass ratio between the three components very close to 2:1:1.

We finally used the available photometry from ASAS and SWASP to search for signals of the rotation period of the components. No significant signals in the ASAS photometry could be detected as already reported by \cite{DiezAlonso2019}. 
Although not significant, the highest peak in the periodogram of the SWASP data, shown in Fig.\,\ref{J15474GLS}, is at $\sim78$\,d, identical to the one estimated by \cite{Astudillo2017} from the emission of the Ca~{\sc ii} H\&K lines. Using only the first two epochs of the SWASP data, which have half the dispersion and uncertainty of the rest, the peak at $78.1\pm8.6$\,d becomes significant and, therefore, we attributed it to the rotation period of star A, given the low expected flux of the other components.

\subsubsection{GJ 4383}

\begin{figure}[t]
\centering
\includegraphics[width=0.96\columnwidth]{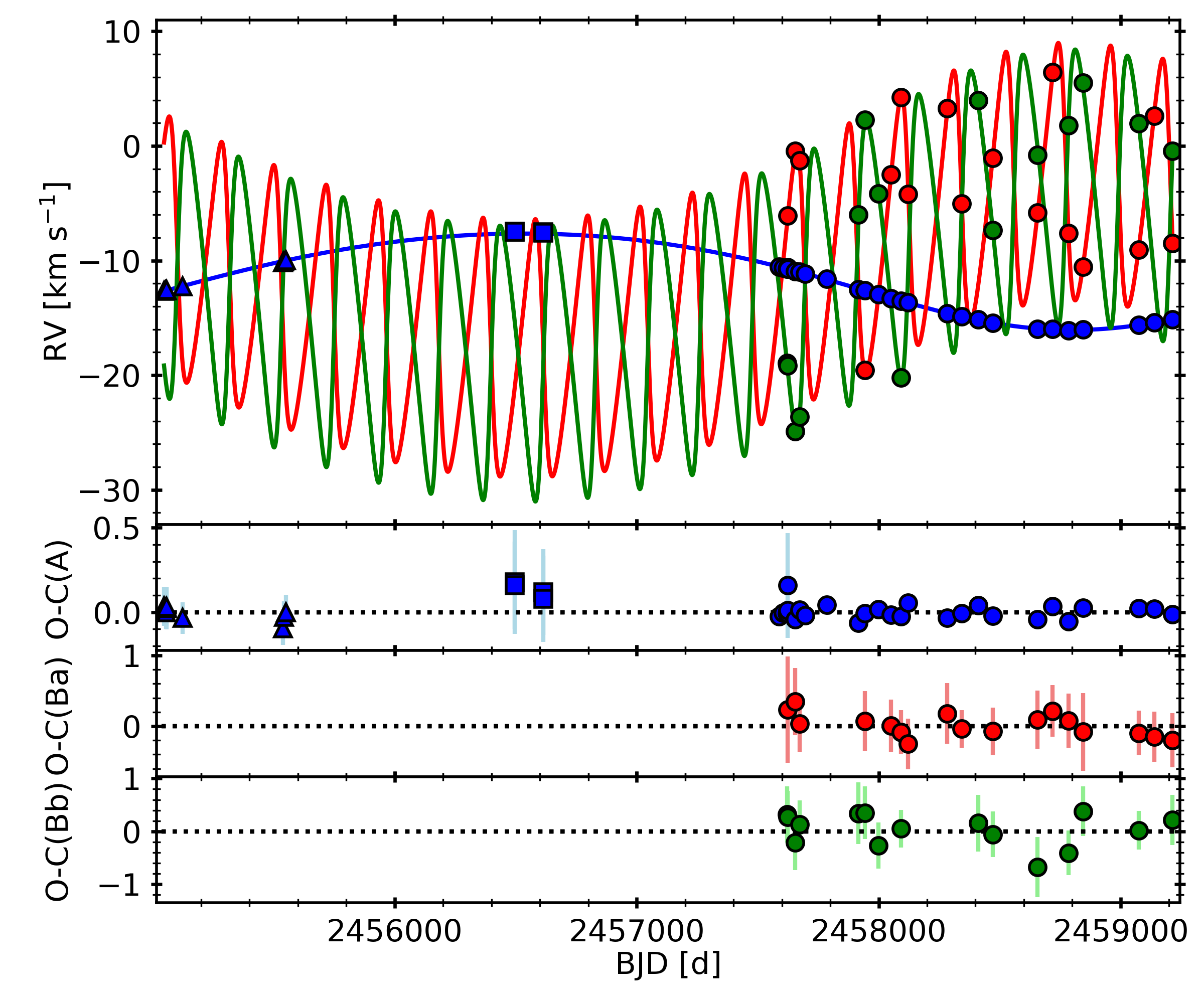}
\includegraphics[width=0.96\columnwidth]{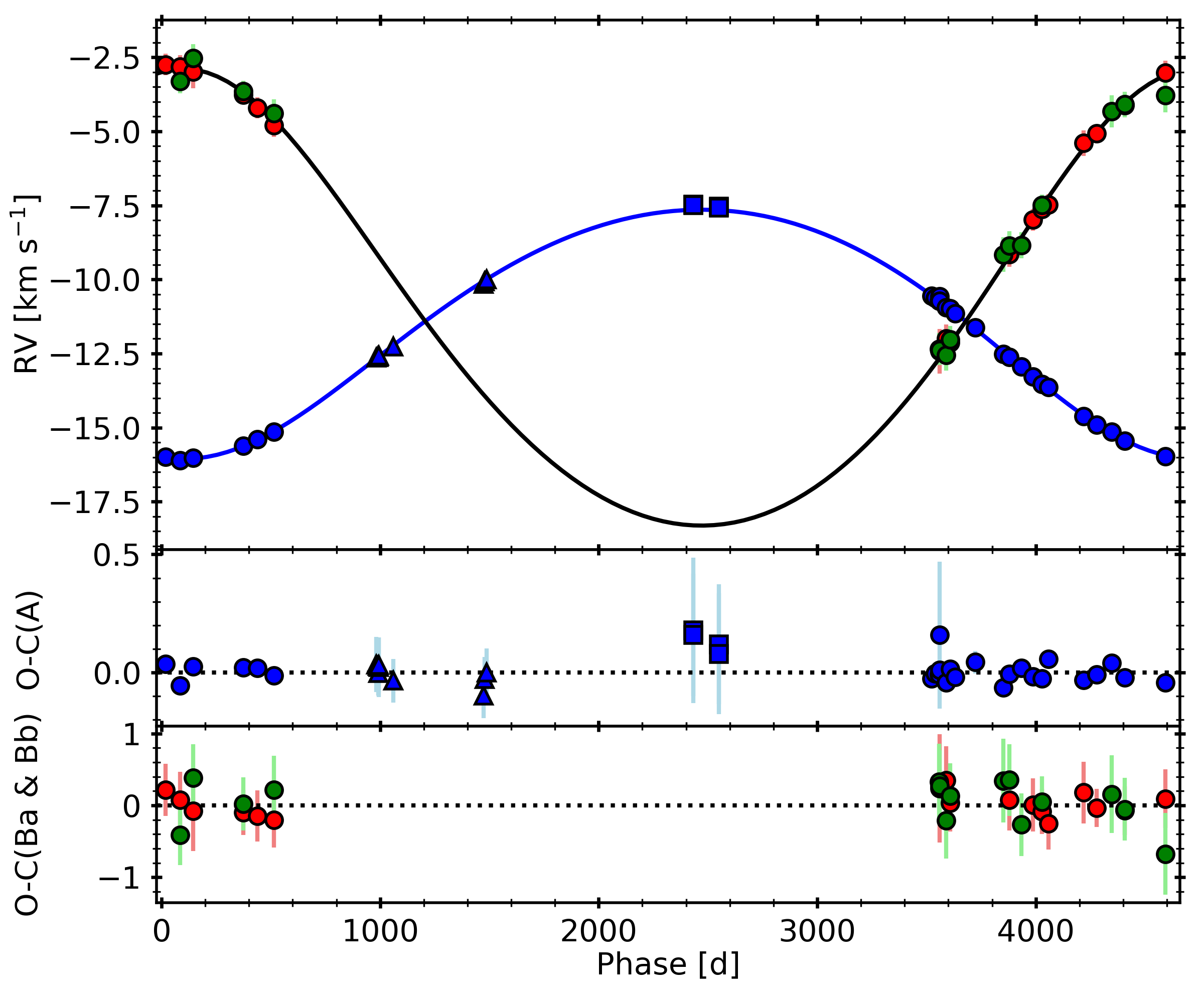}
\includegraphics[width=0.96\columnwidth]{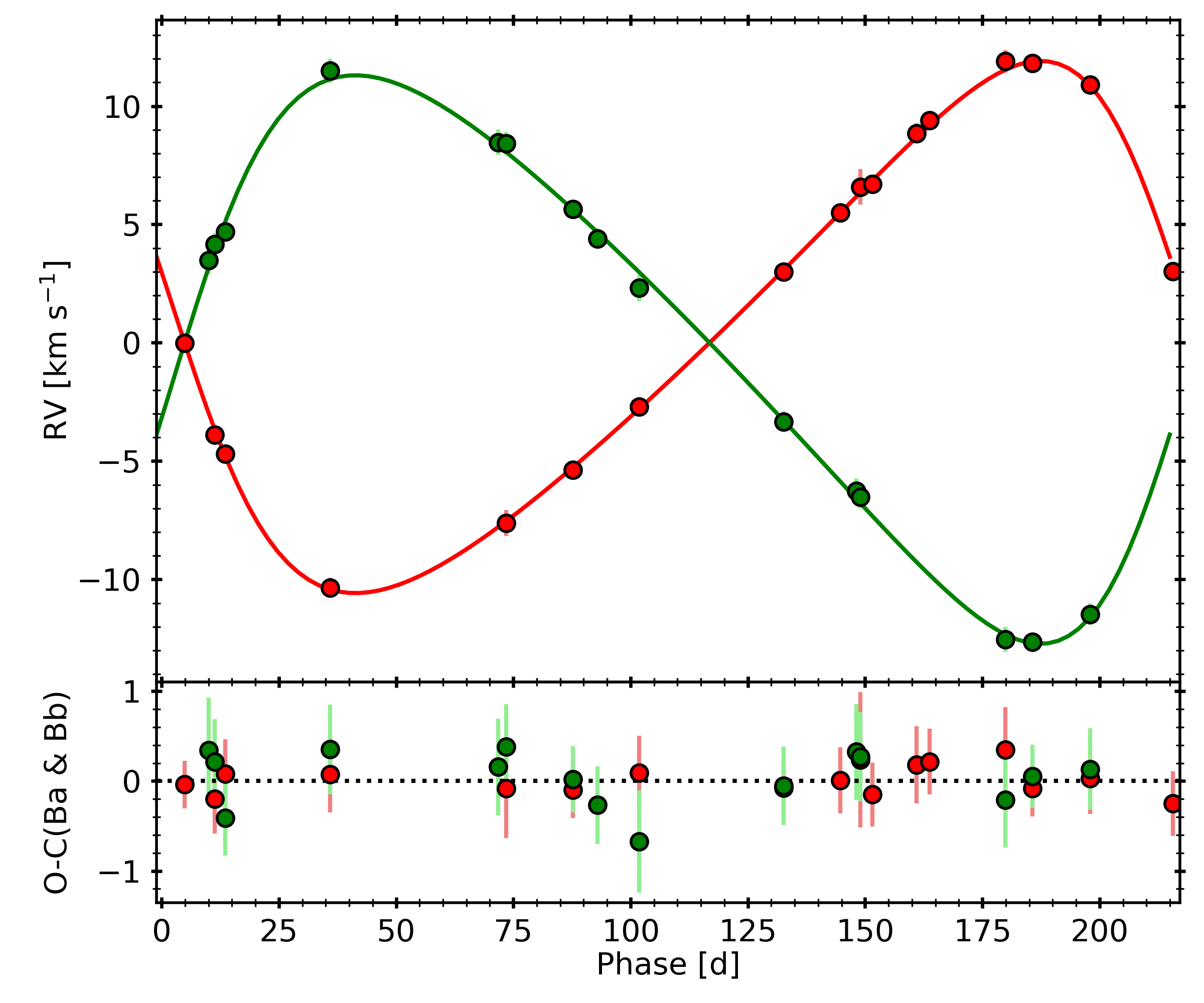}
\caption{Same as Fig.\,\ref{J15474orbit}, but for GJ\,4383.}
          \label{J23585orbit}%
\end{figure}

\begin{figure}[t]
\centering
\includegraphics[width=\columnwidth]{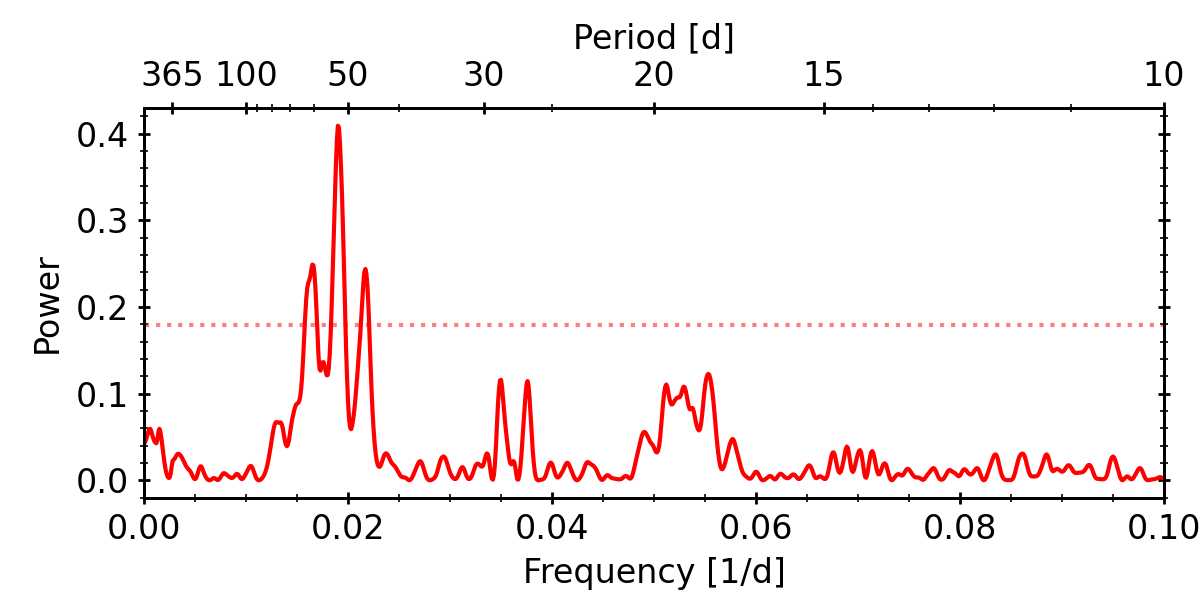}
\caption{GLS periodogram of the SWASP photometry of GJ\,4383. The horizontal dotted line indicates the corresponding 0.1\,\% FAP level.}
\label{J23585GLS}%
\end{figure}

GJ\,4383 is an M3.0\,V star located at a distance of $16.43\pm0.07$\,pc. It is an inactive star, with  pEW(H$\alpha$) = $+0.04\pm0.01$\,{\AA} in absorption, and a projected rotational velocity lower than 2\,km\,s$^{-1}$. Using the measured $R'_{\rm HK}$ index, \cite{Astudillo2017} estimated a rotation period of 67\,d, while \cite{DiezAlonso2019} did not find any significant signal in the ASAS photometry associated with rotation.
Although the {\em Hipparcos} astrometry does not show an acceleration in the astrometric measurements, \cite{Makarov2005} found discrepancies between the proper motions in the {\em Hipparcos} and Tycho-2 \citep{Tycho} catalogues and concluded that GJ\,4383 could be an astrometric binary. Likewise, \cite{Kervella2019} found differences between the proper motions of {\em Hipparcos} and {\em Gaia}, from which a lower limit to the mass of the companion, normalised to 1\,au, was set to \AS{83}{23}{9}\,M$_{\rm Jup}$\,au$^{-1/2}$.

We observed GJ\,4383 with CARMENES between July 2016 and December 2020, gathering 26 high-resolution spectra. We also retrieved nine and four spectra from the HARPS and FEROS public databases, respectively. A first inspection of the spectra with \texttt{todmor} using synthetic spectra only revealed the signal of one component, whose RVs exhibit a sinusoidal-like shape, probably produced by a companion. To attempt to resolve the suspected companion, we used as templates the coadded spectra of the CARMENES observations of GJ\,752\,A and GJ\,1253 (Sect.~\ref{sec:rvs}). With this approach, we resolved the signal of not only one but two additional components. From the 26 CARMENES spectra, we were able to detect the signal of the three components in 11 cases, two components in 10 spectra, and one component in 5. 
We did not detect the companions in any of the HARPS and FEROS spectra, mainly due to the lower exposure time of their observations with respect to those from CARMENES. We list the RVs in Table\,\ref{tab:RVST3}. 

The architecture of the RVs of this system is very similar to that of GJ\,3916: two low-mass stars orbiting each other with a relatively short period, which in turn are orbiting a heavier star in a much wider orbit. We therefore fitted the same physical model to the RVs, which yielded the parameters listed in Table\,\ref{tab:paramstriple}. We show in Fig\,\ref{J23585orbit}, from top to bottom, the orbital best fit to the RVs of GJ\,4383 and the same fit phase folded to the A--B and Ba--Bb orbital periods. %, respectively.
We measured a wide orbit between A and the pair Ba--Bb with an orbital period of $4646\pm17$\,d and an eccentricity of $0.1068\pm0.0049$. The close orbit between the Ba and Bb components has a period of \AS{216.07}{0.13}{0.14}\,d and an eccentricity of \AS{0.283}{0.012}{0.011}. From the derived semi-amplitudes, we derived an approximately 4:1:1 mass ratio between the components, with minimum masses of \AS{0.509}{0.017}{0.016}\,M$_{\odot}$, \AS{0.1206}{0.0040}{0.0040}\,M$_{\odot}$, and \AS{0.1130}{0.0048}{0.0049}\,M$_{\odot}$ for components A, Ba, and Bb, respectively. The value of the minimum mass of component A deviates only 2$\sigma$ from the value obtained from mass-luminosity calibrations of $0.455\pm0.019$\,M$_{\odot}$ by \cite{Schweitzer2019}. This fact points at an almost edge-on orbit with respect to the line of sight.

Finally we searched for significant signals in the available ASAS and SWASP photometry. No significant periods were found in the ASAS photometry. However, the periodogram of the SWASP photometry, shown in Fig.\,\ref{J23585GLS}, has a significant %and clear 
signal at $52.5\pm1.8$\,d, which we attributed to the rotation period of the most massive star, given the low expected flux of the other components.

\section{Discussion and conclusions} \label{sec:conclusion}

In this study we determined the orbital parameters of eight new multiple systems using high-precision RVs collected with CARMENES and from archival data, and refined the orbital parameters of UU\,UMi by adding additional observations with respect to the analysis presented by \cite{Baroch2018}. Adaptive optics imaging data also permitted us to measure the absolute masses of the Ursa Major moving group members GJ\,282\,Ca and GJ\,282\,Cb with uncertainties of 10\,\% and 3\,\%, respectively, making it one of the youngest systems with measured dynamical masses.

%\textbf{The multiple stars presented here, along with the SB2s in \cite{Baroch2018}, add up to a total of 17 stellar systems detected among the 349 M-dwarf stars monitored with the CARMENES spectrograph. This yields a multiplicity ratio of 4.9\%. This value is much lower than the M-dwarf multiplicity ratios reported by other studies of samples with significant sizes, ranging between 20 and 30\% \citep{Bergfors2010,Janson2012,Janson2014,Winters2019}. However, this difference arises from the selection of targets for the CARMENES survey, which is aimed at the detection of exoplanets. A substantial observational effort \citep{Alonso2015,Cortes2017,Jeffers2018} was done to determine the multiplicity of the stars in the CARMENES input catalogue \citep{Caballero2013}, in order to only include single stars. However, the resolution and/or the cadence of the observations was not enough to detect the multiple systems presented here.}

The multiple stars presented here, along with the SB2s in \cite{Baroch2018}, add up to a total of 17 stellar systems detected among the 349 M-dwarf stars monitored with the CARMENES spectrograph. The new binary systems can be used to update the M-dwarf multiplicity rate measured in other works. In a volume-limited sample of 1120 M dwarfs with distances below 25\,pc, \cite{Winters2019} counted 265 multiple systems, from which they computed a multiplicity rate of $23.7\pm1.3$\,\% uncorrected for observational biases. Their sample included all the systems in this work except for UCAC4\,355--020729, GJ\,282\,C, and LP\,427--016, and already tagged as multiple only the systems UU\,UMi and GJ\,3916, based on the works by \cite{Heintz1993} and \cite{Zechmeister2009}, respectively. 
Adding GJ\,282\,C, and LP\,427-016 to the sample (UCAC4\,355--020729 is beyond 25\,pc), and tagging as multiple the systems found in this work, we compute a new multiplicity rate of $24.2\pm1.3\%$ (271 multiple systems %out 
of 1122 stars). Furthermore, the companion to GJ\,282\,Ca, GJ\,282\,Cb, enlarges the list of 512 M-dwarf stars with measured masses between 0.1 and 0.3\,M$_{\odot}$ at distances $d<$ 15\,pc compiled by \cite{Winters2021}, 
who already compiled the mass estimate of UU\,UMi\,B reported by \cite{Baroch2018}. %which already contains UU\,UMi\,B from the mass estimate reported in \cite{Baroch2018}. 

We found that the GJ\,207.1 system is probably composed of a white dwarf and an M dwarf. The short orbital period of about 14.5\,h suggests that this system may have survived a common envelope phase due to the evolution of the massive star \citep{Nebot2011}. The two components may have started off in a much wider orbit, but the engulfment of the M dwarf by the envelope of the white-dwarf progenitor caused drag forces that transported orbital angular momentum to the envelope, which was finally ejected. This process may have ended up shrinking and tidally locking the orbit to the present state \citep{Nebot2011,Skinner2017}. The number of white-dwarf plus M-dwarf close binary systems is rapidly increasing thanks to large-scale surveys such as SDSS \citep{Rebassa2016} and LAMOST \citep{Ren2018}. However, these short-period post-common-envelope systems, in which the optical light is dominated by the M dwarf, are very uncommon, comprising an estimated fraction of 0.03--0.18\,\% of all M dwarfs \citep{Law2012}, as estimated from the detection of three such systems in a photometric survey of 45\,000 M dwarfs. Only one white dwarf (in the system GJ\,207.1, identified here) has been found among the $\sim360$ stars in the CARMENES sample \citep{Reiners2018}, a ratio larger (but compatible because of small-number statistics) than that estimated by \cite{Law2012}. The larger ratio found by us % in CARMENES 
could be a by-product of the sensitivity difference between the methods used to detect white dwarfs, since their imprint over RVs should be in principle easier to detect than using photometry.

Although RV surveys have detected an abundance of exoplanets, the number of close brown dwarf companions is small, despite the fact that brown dwarfs should be more easily detected in comparable orbits, given the larger RV signals that they produce. Several studies estimated a stellar-brown dwarf multiplicity fraction around 1\,\% for a large range of primary mass and separation \citep{Halbwachs2000,Marcy2000,McCarthy2004,Metchev2009,Kraus2011}. This is also valid for M-dwarf primaries \citep{Dieterich2012}, as suggested by the only 18 M-dwarf-brown-dwarf systems found among the 1120 M dwarfs contained in the sample of \cite{Winters2019}, or by the two system candidates found in this work among the $\sim360$ M dwarfs observed with CARMENES. Observational evidence points at a bimodal companion mass function \citep{Grether2006} with a gap, also called the brown dwarf desert, at the intersection between the planetary and stellar-companion mass functions. The position of this minimum in the companion mass function seems to increase with the mass of the host star \citep{Quirrenbach2019} and, therefore, objects with similar masses may have very different formation mechanisms \citep{Armitage2002,Matzner2005}. Although the masses of the brown dwarf candidates found in this work are not fully constrained, the low mass of the host stars and the highly eccentric orbits of the systems may be indicating that they probably belong to the ``stellar'' part of the companion mass function. If that is the case, one would expect the same metallicity as for their host stars, whereas ``planet-like'' objects might be metal-enhanced \citep{Maldonado2017,Quirrenbach2019}. GJ\,912\,B and GJ\,3626\,B, with computed minimum masses of $\sim$60\,M$_{\rm Jup}$ and $\sim$30\,M$_{\rm Jup}$, respectively, and expected angular separations between 100\,mas and 300\,mas, may be directly imaged with state-of-the-art imagers such as %for example
Gemini/GPI \citep{GPI}, Subaru/SCExAO \citep{Subaru}, Keck/OSIRIS-NIRC2 \citep{KECK,OSIRIS}, or VLT/SPHERE \citep{SPHERE}, as estimated from simulated planet injections \citep{Wagner2020}. If the brown-dwarf nature of GJ\,912\,B and GJ\,3626\,B is confirmed with future observations, they will help to better constrain the multiplicity rate of these systems and the boundaries of the brown dwarf desert, providing additional insights into the formation mechanisms of brown dwarfs.
%respectively, should have a contrast magnitude in the $K$-band with respect to their companions of $\Delta K\sim6$ and $\Delta K\sim9$, according to the 1\,Ga evolutionary models by \cite{Chabrier2000b}. With expected angular separations of 150 and 250\,mas, these systems may be directly imaged with state-of-the-art imagers such as SPHERE \citep{SPHERE}, as estimated from simulated planet injections \citep{Wagner2020}. 

Finally, we unveiled the hierarchical triple nature of GJ\,3916 and GJ\,4383, for which we were able to detect the spectroscopic signals of all three components. Interestingly enough, the masses of the close, inner pair are nearly identical in both cases, and their combined mass is either equal to the outer component or half that value. These results could help constraining formation models. Understanding the processes leading to the formation of multiple systems is one of the fundamental challenges in astrophysics and has been a matter of debate for many years \citep{Duchene2013}. One possible mechanism is the formation of these systems through the fragmentation of collapsing and rotating protostellar clouds \citep{Boss1988}. Another possible explanation is the dynamical break-up of rapidly rotating clouds, or fission, although this mechanism does not seem to lead to the production of equal-mass binary protostars \citep{Durisen1986}, and thus is not capable of explaining the mass ratio distribution for binary systems, which shows an intrinsic excess at about unity %$\simeq1$
\citep{Lucy2006,Simon2009}.
The formation of both democratic (all components with similar masses) and hierarchical systems could be explained by fragmentation. In the first case, the cloud fragments into three or more protostars, while in the other case the cloud first collapses into binary protostars, which then can undergo subsequent fragmentations. Most observed triples have hierarchical structures, consisting of an inner binary and a distant star \citep{Hut1983,Tokovinin2006}, because democratic triples tend to be unstable and short-lived \citep{Berk2007}. The presence of a third body on a wide orbit around a tight binary causes an angular momentum exchange between the two orbits, which in turn produces dynamical variations, such as periodic oscillations in the eccentricity of the inner binary and in the mutual inclination through Lidov-Kozai interactions \citep{Lidov1962,Kozai1962,Mazeh1979,Naoz2016}. Usually, the timescale of such interactions is much larger than the orbital periods and, thus, their
effect in triple systems has only been detected indirectly through unusual inner eccentricity values \citep{Kiseleva1998,Ford2000}, or measured directly in very few systems \citep{Jha2000}. However, tight orbital configurations produce dynamical variations with timescales of the order of the orbital period, which lead to the disruption of the system. 

In order to check the long-term stability of the triple systems found in this work, we used the \cite{Eggleton1995} criterion for the stability of triple stellar systems, which \cite{He2018} found to work reasonably well for a wide range of parameters in their N-body simulations. According to this criterion, the systems are stable when the orbit separation parameter $r_{\rm ap}\equiv a_{\rm AB}a_{\rm B}^{-1}(1-e_{\rm AB})(1+e_{\rm B})^{-1}$ is higher than the value $Y_{\rm EK}$ defined as:

\begin{equation}
    Y_{\rm EK} \equiv 1+\frac{3.7}{q_{\rm AB}^{1/3}}-\frac{2.2}{1+q_{\rm AB}^{1/3}}+\frac{1.4}{q_{\rm B}^{1/3}}\frac{q_{\rm AB}^{1/3}-1}{1+q_{\rm AB}^{1/3}}
\end{equation}

\noindent  \citep{Eggleton1995,He2018}.

For the cases of the triple systems GJ\,3916 and GJ\,4383, which have an orbit separation parameter $r_{\rm ap}$ of 5.03 and 7.89, we obtained values of $Y_{\rm EK}$ of 3.56 and 4.17, respectively. Therefore, both system have stable orbital configurations.

The time scale of the Lidov-Kozai interactions is also useful to assess the stability of the system. Actually, this parameter is crucial for verifying if the orbit of the systems can be approximated by a Keplerian function in a hierarchical system during the time-span of all the RV measurements, as we have done in this work. This timescale $\tau_{\rm LK}$ is defined as:

\begin{equation} \label{eq:tlk}
    \tau_{\rm LK} = \alpha \frac{P^2_{\rm AB}}{P_{\rm B}}\frac{M_{\rm A}+M_{\rm Ba}+M_{\rm Bb}}{M_{\rm A}}(1-e_{\rm AB}^2)^{3/2},
\end{equation}

\noindent where $\alpha$ is close to one, and weakly depends on the orbital parameters \citep{Kinoshita1999,Toonen2020}. Equation~\ref{eq:tlk} yields timescales of the Lidov-Kozai oscillations for GJ\,3916 and GJ\,4383 of 272\,a and 414\,a, respectively, which, just by chance, in both cases is equivalent to 33 orbital periods of the outer binary. The time-span of the RV measurements used in the fits covered just 5.6\,\% and a 2.3\,\% of $\tau_{LK}$ for GJ\,3916 and GJ\,4383, respectively. This small fraction of the Lidov-Kozai cycle does not have a significant effect on the orbital fits of the RVs of these systems. This is also confirmed by the lack of systematic deviations in the residuals shown in the top panels of Figs.\,\ref{J15474orbit} and~\ref{J23585orbit}. 

\begin{figure}[t]
\centering
\includegraphics[width=\columnwidth]{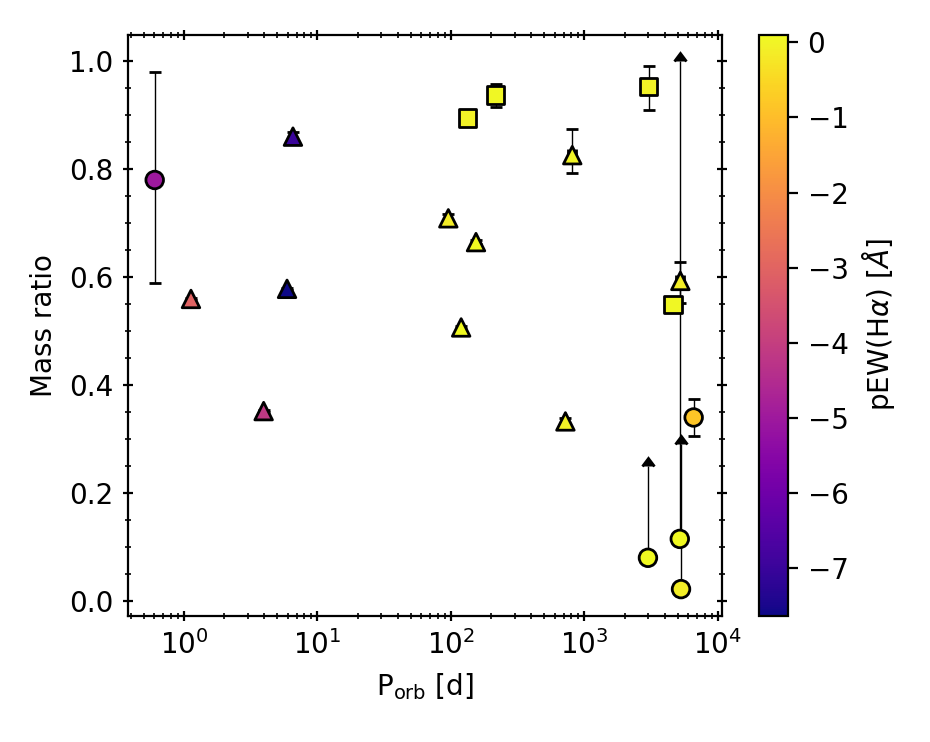}
\caption{Mass ratio as a function of the orbital period of the multiple systems characterised in this work and in \cite{Baroch2018}, color coded by the pEW(H$\alpha$). Circular, triangular, and squared symbols correspond to SB1, SB2, and ST3 systems, respectively. The upward arrows indicate the mass ratio limits for the systems for which only upper and lower bounds to the secondary masses could be given. We plot outer and inner orbits of the ST3 as individual binaries, with the same pEW(H$\alpha$).}
\label{Pvsq}%
\end{figure}

Figure\,\ref{Pvsq} shows the mass ratios of the multiple systems found in this work and by \cite{Baroch2018} as a function of their orbital period, and colored by the pEW(H$\alpha$). Apart from the lack of systems with mass ratios lower than $\sim0.3$, as expected from the flux ratio needed to discern the two components in the spectra with \texttt{todmor}, no clear dependence on the orbital period of the systems can be observed. 
We do find, however, a dependence of the stellar activity on the orbital period: systems with periods shorter than $\sim10$\,d have strong emission in the H$\alpha$ line, whereas systems with larger periods have weak emission or even, in the earliest types, show the same spectral line in absorption. This is probably caused by the synchronisation between the rotation of the stars and the orbital motion due to tidal effects, which for main-sequence stars should be achieved for orbital periods below 8--10\,d \citep{Mazeh2008}. Besides, short rotation periods are associated to increased stellar activity in M dwarfs \citep{Newton2016,Suarez2016,Astudillo2017,Jeffers2018}. 

The determination of the spectroscopic orbits and properties of the systems in this work will provide help to understand the galactic population as a whole. The new systems contribute to statistical studies of stellar multiplicity and of the frequency of singular objects, and provide additional constraints to star formation theories, stellar evolutionary and structure models, and empirical calibrations. These constraints could be enhanced by the determination of the individual masses of the systems, which will be possible with upcoming {\it Gaia} data releases containing the astrometric epoch data, and with the direct imaging of some of the systems.

\begin{acknowledgements}
Based on observations collected at the Centro Astron\'omico Hispano Alem\'an (CAHA) at Calar Alto, operated jointly by the Junta de Andaluc\'ia and the Instituto de Astrof\'isica de Andaluc\'ia (CSIC). CARMENES is funded by the German Max-Planck-Gesellschaft (MPG), the Spanish Consejo Superior de Investigaciones Cient\'ificas (CSIC), the European Union through FEDER/ERF FICTS-2011-02 funds, and the members of the CARMENES Consortium (Max-Planck-Institut f\"ur Astronomie, Instituto de Astrof\'isica de Andaluc\'ia, Landessternwarte K\"onigstuhl, Institut de Ci\`encies de l’Espai, Insitut f\"ur Astrophysik G\"ottingen, Universidad Complutense de Madrid, Th\"uringer Landessternwarte Tautenburg, Instituto de Astro\'isica de Canarias, Hamburger Sternwarte, Centro de Astrobiolog\'ia and Centro Astron\'omico Hispano-Alem\'an), with additional contributions by the Spanish Ministry of Economy, the German Science Foundation through the Major Research Instrumentation Programme and DFG Research Unit FOR2544 “Blue Planets around Red Stars”, the Klaus Tschira Stiftung, the states of Baden-W\"urttemberg and Niedersachsen, and by the Junta de Andaluc\'ia. Based on data from the CARMENES data archive at CAB (INTA-CSIC). Based on observations made with the 0.9-m telescope at the Sierra Nevada Observatory (Granada, Spain), operated by the Instituto de Astrof\'isica de Andaluc\'ia, the 0.8-m Joan Or\'o telescope (TJO) of the Montsec Astronomical Observatory (Lleida, Spain), owned by the Generalitat de Catalunya and operated by the Institut d'Estudis Espacials de Catalunya (IEEC), on observations collected at the European Southern Observatory under ESO programs 098.C-0739(A) and 192.C-0224(C) (P.I. A.\,M.\,Lagrange), 180.C-0886(A) and 183.C-0437(A) (P.I. X.\,Bonfils), 074.D-0016(A) (P.I. D.\,Montes), 078.A-9048(A) (P.I. J.\,Setiawan), 085.A-9027(A) (P.I. R.\,Gredel), 090.A-9003(A) and 091.A-9004(A) (P.I. R.\,Mundt), 173.C-0606(C) (P.I. M.\,K\"urster), 096.D-0818(A) (P.I. K.~Ward-Duong), 094.C-0625(A) and 097.C-0972(A) (P.I. J.~H.~Girard), and 081.A-9005(A), 081.A-9024(A), 083.A-9002(A), 083.A-9012(A,B), 085.A-9009(A), and 086.A-9016(A) (P.I. M.\,Zechmeister). This work has made use of data from the European Space Agency (ESA) mission {\it Gaia} (\url{https://www.cosmos.esa.int/gaia}), processed by the {\it Gaia}
Data Processing and Analysis Consortium (DPAC,
\url{https://www.cosmos.esa.int/web/gaia/dpac/consortium}). Funding for the DPAC has been provided by national institutions, in particular the institutions participating in the {\it Gaia} Multilateral Agreement. {\tt IRAF} was distributed by the National Optical Astronomy Observatories, which is operated by the Association of Universities for Research in Astronomy, Inc. (AURA) under cooperative agreement with the National Science Foundation. We acknowledge financial support from the Spanish Agencia Estatal de Investigaci\'on of the Ministerio de Ciencia e Innovaci\'on (AEI-MCINN) and the European FEDER/ERF funds through projects
  PGC2018-098153-B-C33,     % ICE
  PID2019-109522GB-C51/52/53/54,     %CAB + IAA + UCM + IAC
  PID2019-107061GB-C64,
  ESP2017-87143-R,
and the Centre of Excellence ``Severo Ochoa'' and ``Mar\'ia de Maeztu'' awards to the Instituto de Astrof\'isica de Canarias (SEV-2015-0548), Instituto de Astrof\'isica de Andaluc\'ia (SEV-2017-0709), and Centro de Astrobiolog\'ia (MDM-2017-0737), the Secretaria d'Universitats i Recerca del Departament d'Empresa i Coneixement de la Generalitat de Catalunya and the Ag\`encia de Gesti\'o d’Ajuts Universitaris i de Recerca of the Generalitat de Catalunya, with additional funding from the European FEDER/ERF funds, \emph{L'FSE inverteix en el teu futur}, and from the Generalitat de Catalunya/CERCA programme.
%This work makes use of data from the 80\,cm Telescopi Joan Or\'o (TJO) of the Montsec Astronomical Observatory (OAdM), owned by the Generalitat de Catalunya and operated by the Institut d'Estudis Espacials de Catalunya (IEEC), and includes data collected by the {\em TESS} mission. Funding for the {\em TESS} mission is provided by the NASA Explorer Program. 

\end{acknowledgements}

\bibliographystyle{aa} 
\bibliography{bibtex.bib}

\begin{appendix}

\section{Radial velocity tables}
Tables\,\ref{tab:RVSB1} to \ref{tab:RVST3} list the radial velocities used in this work. For stars with observations from different instruments, RV from each one are labelled as C (CARMENES), H (HARPS), F (FEROS), and U (UVES). For SB1 systems in Table\,\ref{tab:RVSB1}, the methods used to obtain the RV values are labelled as T (\texttt{todmor}) and S (\texttt{serval}).

\begin{table}[t]
\centering
\caption{RVs of SB1s.}
\label{tab:RVSB1}
\begin{tabular}{lccc} 
\hline\hline
\noalign{\smallskip}
 Obs. date & RV & \multirow{2}{*}{Inst.} & \multirow{2}{*}{Method} \\
\noalign{\smallskip}
 [BJD] & [m\,s$^{-1}$] & & \\
\noalign{{\smallskip}}
\hline
\noalign{{\smallskip}}
\multicolumn{4}{c}{\textit{GJ\,207.1}}\\
\noalign{{\smallskip}}
\hline
\noalign{{\smallskip}}
     2453708.8474 & $-3700\pm1200$ & F & T \\ 
     2453424.6545 & $55250\pm940$ & F & T \\ 
     2454455.6240 & $-2810\pm260$ & H & T \\ 
     2457693.6299 & $64069\pm12$ & C & S \\ 
     2457985.6872 & $11450\pm43$ & C & S \\ 
     2458007.6799 & $29835.5\pm9.6$ & C & S \\ 
     2458026.6315 & $57691.9\pm8.5$ & C & S \\ 
     2458078.5036 & $65398\pm15$ & C & S \\ 
     2458094.5798 & $-67.0\pm9.4$ & C & S \\ 
     2458160.3600 & $-2658\pm17$ & C & S \\ 
     2458167.3276 & $65932\pm19$ & C & S \\ 
     2458173.2985 & $54835\pm19$ & C & S \\ 
     2458182.3093 & $38259\pm16$ & C & S \\ 
     2458187.4240 & $31455\pm35$ & C & S \\
     2458206.3402 & $-1822\pm15$ & C & S \\ 
     2458434.5484 & $24896\pm12$ & C & S \\ 
     2458435.7075 & $42661.4\pm9.4$ & C & S \\ 
\hline
\noalign{{\smallskip}}
\multicolumn{4}{c}{\textit{GJ\,912}}\\
\noalign{{\smallskip}}
\hline
\noalign{{\smallskip}}
     2455046.8642 & $-1016.2\pm3.2$ & H & S \\ 
     2455053.8299 & $-968.7\pm1.4$ & H & S \\ 
     2455054.8321 & $-965.6\pm1.0$ & H & S \\ 
     2455056.8260 & $-931.5\pm6.0$ & H & S \\ 
     2455056.9116 & $-952.9\pm1.9$ & H & S \\ 
     2455057.8606 & $-948.9\pm2.9$ & H & S \\ 
     2455125.6477 & $-568.2\pm1.2$ & H & S \\ 
     2455464.6903 & $65.9\pm1.0$ & H & S \\ 
     2456235.5404 & $2.2\pm1.2$ & H & S \\ 
     2456502.9029 & $815\pm447$ & F & T \\ 
     2456612.6329 & $507\pm396$ & F & T \\ 
     2456617.7031 & $749\pm362$ & F & T \\ 
     2457593.6588 & $288.8\pm2.0$ & C & S \\
     2457595.6683 & $283.3\pm2.0$ & C & S \\ 
     2457604.6409 & $283.5\pm2.4$ & C & S \\ 
     2457608.6448 & $283.1\pm2.1$ & C & S \\ 
     2457625.6036 & $276.8\pm2.2$ & C & S \\ 
     2457634.5720 & $265.4\pm2.4$ & C & S \\ 
     2457644.5412 & $263.2\pm2.3$ & C & S \\ 
     2457672.5148 & $257.9\pm1.7$ & C & S \\ 
     2457753.2839 & $228.4\pm2.2$ & C & S \\ 
     2457931.6397 & $137.9\pm2.2$ & C & S \\ 
     2457951.6217 & $129.5\pm2.2$ & C & S \\ 
     2457971.5873 & $123.9\pm1.8$ & C & S \\ 
     2457997.5621 & $110.2\pm1.6$ & C & S \\ 
     2458017.4397 & $105.4\pm1.9$ & C & S \\ 
     2458065.4532 & $79.4\pm1.6$ & C & S \\ 
     2458095.3024 & $65.2\pm1.5$ & C & S \\ 
     2458134.2945 & $45.3\pm1.6$ & C & S \\ 
     2458289.6531 & $-30.1\pm1.7$ & C & S \\ 
     2458349.6060 & $-57.8\pm1.1$ & C & S \\ 
     2458416.4309 & $-95.7\pm1.2$ & C & S \\ 
     2458476.3298 & $-133.2\pm1.3$ & C & S \\
     2458663.6613 & $-242.1\pm1.9$ & C & S \\
     2458723.5831 & $-282.1\pm2.1$ & C & S \\
     2458783.4458 & $-326.4\pm1.8$ & C & S \\
     2458844.2613 & $-373.0\pm3.6$ & C & S \\
     2459033.6486 & $-528.8\pm2.2$ & C & S \\
     2459093.5803 & $-590.4\pm2.1$ & C & S \\
     2459153.4087 & $-649.2\pm2.6$ & C & S \\
\hline
\end{tabular}
\end{table}

\addtocounter{table}{-1}
\begin{table}[t]
\centering
\caption{Continued.}
\begin{tabular}{lccc} 
\hline\hline
\noalign{\smallskip}
Obs. date & RV & \multirow{2}{*}{Inst.} & \multirow{2}{*}{Method} \\
\noalign{\smallskip}
 [BJD] & [m\,s$^{-1}$] & & \\
\noalign{{\smallskip}}
\hline
\noalign{{\smallskip}}
\multicolumn{4}{c}{\textit{GJ\,3626}}\\
\noalign{{\smallskip}}
\hline
\noalign{{\smallskip}}
2457472.5069 & $-1426.4\pm1.7$ & C & S \\
2457534.3920 & $-1489.2\pm2.6$ & C & S \\
2458078.6861 & $-604.4\pm1.3$ & C & S \\
2458092.6519 & $-532.8\pm1.5$ & C & S \\
2458110.6715 & $-442.8\pm1.6$ & C & S \\
2458112.6498 & $-431.5\pm1.3$ & C & S \\
2458161.5900 & $-210.8\pm5.6$ & C & S \\
2458205.4683 & $-47.6\pm1.4$ & C & S \\
2458236.5044 & $35.5\pm1.8$ & C & S \\
2458270.4148 & $119.9\pm1.7$ & C & S \\
2458300.3645 & $166.2\pm1.5$ & C & S \\
2458413.7056 & $260.2\pm1.3$ & C & S \\
2458449.7141 & $272.8\pm3.6$ & C & S \\
2458480.6987 & $265.3\pm1.4$ & C & S \\
2458510.6934 & $268.0\pm1.6$ & C & S \\
2458540.7452 & $253.1\pm1.8$ & C & S \\
2458570.5584 & $246.2\pm1.8$ & C & S \\
2458600.4789 & $229.8\pm1.1$ & C & S \\
2458608.5203 & $222.0\pm1.5$ & C & S \\
2458615.4695 & $219.8\pm1.4$ & C & S \\
2458626.4686 & $219.4\pm1.4$ & C & S \\
2458633.4138 & $212.6\pm2.3$ & C & S \\
2458641.4210 & $210.4\pm1.6$ & C & S \\
2458649.3983 & $206.3\pm1.9$ & C & S \\
2458656.3797 & $199.9\pm1.2$ & C & S \\
2458663.3722 & $196.6\pm2.2$ & C & S \\
2458664.3700 & $193.8\pm1.4$ & C & S \\
2458804.7036 & $100.5\pm2.6$ & C & S \\
2458811.7346 & $91.4\pm2.8$ & C & S \\
2458831.7261 & $82.0\pm2.4$ & C & S \\
2458844.7072 & $67.9\pm3.1$ & C & S \\
2458851.6835 & $63.0\pm1.3$ & C & S \\
2458860.6576 & $58.2\pm1.7$ & C & S \\
2458878.6207 & $44.3\pm2.2$ & C & S \\
2458885.6051 & $34.1\pm1.6$ & C & S \\
2458893.6085 & $28.4\pm1.2$ & C & S \\
2458903.5747 & $18.4\pm1.4$ & C & S \\
2458911.6684 & $20.8\pm4.4$ & C & S \\
2458918.5292 & $9.1\pm1.9$ & C & S \\
2458978.3929 & $-37.3\pm3.1$ & C & S \\
2459149.6687 & $-174.1\pm2.3$ & C & S \\
2459212.7667 & $-231.5\pm3.8$ & C & S \\
\hline
\noalign{{\smallskip}}
\multicolumn{4}{c}{\textit{LSPM\,427-016}}\\
\noalign{{\smallskip}}
\hline
\noalign{{\smallskip}}
     2456353.6532 & $12860\pm130$ & F & T \\
     2456353.6711 & $12830\pm240$ & F & T \\
     2457449.5866 & $-202.2\pm1.8$ & C & S \\
     2457673.7141 & $-160.3\pm1.7$ & C & S \\
     2458059.6730 & $-73.3\pm1.9$ & C & S \\
     2458091.6469 & $-72.4\pm1.3$ & C & S \\
     2458110.4906 & $-66.0\pm1.5$ & C & S \\
     2458121.5078 & $-72.1\pm1.4$ & C & S \\
     2458135.6016 & $-62.1\pm1.9$ & C & S \\
     2458166.5454 & $-54.1\pm1.3$ & C & S \\
     2458209.4565 & $-47.6\pm1.9$ & C & S \\
     2458265.3527 & $-34.3\pm1.7$ & C & S \\
     2458405.6881 & $3.4\pm6.0$ & C & S \\
     2458435.7292 & $7.8\pm1.7$ & C & S \\
     2458468.6965 & $11.9\pm1.8$ & C & S \\
     2458498.6532 & $20.0\pm1.4$ & C & S \\
     2458528.5540 & $25.2\pm1.2$ & C & S \\
     2458558.4795 & $33.0\pm1.5$ & C & S \\
     2458589.3391 & $38.1\pm1.4$ & C & S \\
     2458631.3536 & $51.8\pm1.6$ & C & S \\
\hline
\end{tabular}
\end{table}

\addtocounter{table}{-1}
\begin{table}[t]
\centering
\caption{Continued.}
\begin{tabular}{lccc} 
\hline\hline
\noalign{\smallskip}
 Obs. date & RV & \multirow{2}{*}{Inst.} & \multirow{2}{*}{Method} \\
\noalign{\smallskip}
 [BJD] & [m\,s$^{-1}$] & & \\
\noalign{{\smallskip}}
\hline
\noalign{{\smallskip}}
     2458769.6981 & $76.8\pm2.1$ & C & S \\
     2458801.7426 & $94.5\pm2.3$ & C & S \\
     2459151.6447 & $180.0\pm1.3$ & C & S \\
     2459183.7294 & $187.5\pm1.4$ & C & S \\
\hline
\noalign{{\smallskip}}
\multicolumn{4}{c}{\textit{GJ\,282\,C}}\\
\noalign{{\smallskip}}
\hline
\noalign{{\smallskip}}
     2454040.7842 & $-20540\pm210$ & F & T\\
     2454040.8059 & $-20490\pm200$ & F & T \\
     2456294.6551 & $-16250\pm190$ & F & T \\
     2456982.8285 & $2255.3\pm1.8$ & H & S \\
     2456982.8190 & $2257.7\pm1.9$ & H & S \\
     2456986.8692 & $2238.7\pm1.4$ & H & S \\
     2456987.8191 & $2231.3\pm1.7$ & H & S \\
     2456987.8289 & $2233.0\pm1.5$ & H & S \\
     2456988.8492 & $2209.4\pm1.8$ & H & S \\
     2457342.8600 & $2016.3\pm1.4$ & H & S \\
     2457342.8709 & $2015.4\pm1.7$ & H & S \\
     2457344.8348 & $2010.5\pm1.7$ & H & S \\
     2457344.8453 & $2010.3\pm1.6$ & H & S \\
     2457395.5526 & $898.9\pm2.7$ & C & S \\
     2457401.5286 & $874.4\pm1.8$ & C & S \\
     2457444.4238 & $842.0\pm2.0$ & C & S \\
     2457466.4221 & $828.7\pm2.1$ & C & S \\
     2457489.3509 & $804.0\pm2.6$ & C & S \\
     2457493.4898 & $1903.3\pm1.4$ & H & S \\
     2457493.5007 & $1901.8\pm1.4$ & H & S \\
     2457494.5592 & $1885.5\pm1.4$ & H & S \\
     2457494.5699 & $1884.9\pm1.4$ & H & S \\
     2457699.7112 & $590.1\pm3.4$ & C & S \\
     2457704.6305 & $573.3\pm2.1$ & C & S \\
     2457712.7522 & $1645.8\pm1.7$ & H & S \\
     2457712.7627 & $1642.7\pm1.6$ & H & S \\
     2457760.5315 & $500.0\pm2.4$ & C & S \\
     2457769.6272 & $1587.7\pm1.0$ & H & S \\
     2457769.6380 & $1589.4\pm1.0$ & H & S \\
     2457771.7639 & $1596.6\pm1.8$ & H & S \\
     2457771.7745 & $1599.5\pm1.6$ & H & S \\
     2457788.5567 & $488.4\pm2.4$ & C & S \\
     2457791.4466 & $421.5\pm3.7$ & C & S \\
     2457793.4649 & $457.6\pm2.3$ & C & S \\
     2457800.4426 & $483.2\pm2.9$ & C & S \\
     2457821.4008 & $420.9\pm1.9$ & C & S \\
     2457853.3705 & $357.7\pm2.0$ & C & S \\
     2457857.3579 & $374.3\pm2.3$ & C & S \\
     2458025.7141 & $116.4\pm1.9$ & C & S \\
     2458029.6965 & $131.0\pm2.7$ & C & S \\
     2458034.6884 & $90.9\pm2.3$ & C & S \\
     2458040.6785 & $91.7\pm2.7$ & C & S \\
     2458048.6727 & $82.8\pm2.0$ & C & S \\
     2458052.6611 & $66.0\pm1.9$ & C & S \\
     2458054.7226 & $78.1\pm2.2$ & C & S \\
     2458056.6422 & $66.2\pm1.4$ & C & S \\
     2458059.6546 & $50.3\pm2.1$ & C & S \\
     2458065.5992 & $48.7\pm1.5$ & C & S \\
     2458074.6706 & $41.2\pm1.7$ & C & S \\
     2458081.5932 & $6.6\pm2.4$ & C & S \\
     2458091.5506 & $1.7\pm1.9$ & C & S \\
     2458094.5913 & $-20.9\pm1.8$ & C & S \\
     2458097.7185 & $-5.0\pm2.3$ & C & S \\
     2458123.4250 & $-48.3\pm1.8$ & C & S \\
     2458140.3853 & $-83.3\pm2.0$ & C & S \\
     2458166.4032 & $-147.2\pm2.5$ & C & S \\
     2458200.3147 & $-167.9\pm2.2$ & C & S \\
     2458382.7032 & $-565.3\pm2.5$ & C & S \\
\hline
\end{tabular}
\end{table}

\addtocounter{table}{-1}
\begin{table}[t]
\centering
\caption{Continued.}
\begin{tabular}{lccc} 
\hline\hline
\noalign{\smallskip}
 Obs. date & RV & \multirow{2}{*}{Inst.} & \multirow{2}{*}{Method} \\
\noalign{\smallskip}
 [BJD] & [m\,s$^{-1}$] & & \\
\noalign{{\smallskip}}
\hline
\noalign{{\smallskip}}
     2458451.6149 & $-704.8\pm2.0$ & C & S \\
     2458518.5736 & $-889.2\pm2.6$ & C & S \\
     2458524.5624 & $-900.0\pm2.2$ & C & S \\
     2458526.5304 & $211.3\pm1.5$ & H & S \\
     2458526.5410 & $212.6\pm1.5$ & H & S \\
     2458527.5310 & $226.9\pm1.5$ & H & S \\
     2458527.5574 & $224.3\pm1.2$ & H & S \\
     2458532.5496 & $152.2\pm1.1$ & H & S \\
     2458532.5769 & $152.1\pm1.2$ & H & S \\
     2458533.5557 & $181.4\pm1.0$ & H & S \\
     2458533.5807 & $180.4\pm1.0$ & H & S \\
     2458534.5343 & $198.3\pm1.1$ & H & S \\
     2458534.5614 & $198.3\pm1.1$ & H & S \\
     2458585.3259 & $-1027.5\pm2.1$ & C & S \\
     2458592.4805 & $12.6\pm1.6$ & H & S \\
     2458592.5159 & $11.3\pm1.7$ & H & S \\
     2458595.4924 & $45.6\pm1.1$ & H & S \\
     2458595.5258 & $44.6\pm1.2$ & H & S \\
     2458606.4830 & $-0.2\pm0.9$ & H & S \\
     2458606.5131 & $-1.2\pm0.8$ & H & S \\
     2458607.4725 & $7.6\pm0.9$ & H & S \\
     2458607.4982 & $9.5\pm0.9$ & H & S \\
     2458758.7044 & $-1461.6\pm2.2$ & C & S \\
     2458836.6590 & $-1687.4\pm7.6$ & C & S \\
     2458896.4318 & $-1869.8\pm2.0$ & C & S \\
     2459132.7130 & $-2539.1\pm3.4$ & C & S \\
\hline
\end{tabular}
\end{table}

\begin{table}[t]
\centering
\caption{RVs of SB2s.}
\label{tab:RVSB2}
\begin{tabular}{lcc} 
\hline\hline
\noalign{\smallskip}
Obs. date & RV$_{\rm A}$ & RV$_{\rm B}$ \\
 $[$BJD$]$  & [km\,s$^{-1}$] & [km\,s$^{-1}$] \\
\noalign{\smallskip}
\hline
\noalign{{\smallskip}}
\multicolumn{3}{c}{\textit{UCAC4\,355-020729}} \\
\noalign{{\smallskip}}
\hline
\noalign{{\smallskip}}
2457695.6848 & $13.294\pm0.062$ & $14.909\pm0.094$  \\ 
2457759.5452 & $11.756\pm0.050$ & $16.571\pm0.075$  \\ 
2457788.4644 & $15.282\pm0.049$ & $12.382\pm0.077$  \\ 
2457799.4012 & $11.619\pm0.041$ & $16.650\pm0.069$  \\ 
2457824.3610 & $12.993\pm0.047$ & $15.180\pm0.071$  \\ 
2457832.3566 & $11.779\pm0.043$ & $16.532\pm0.062$  \\ 
2458034.7061 & $12.218\pm0.047$ & $16.072\pm0.077$  \\ 
2458043.7118 & $14.697\pm0.045$ & $12.940\pm0.073$  \\ 
2458047.7093 & $12.394\pm0.039$ & $15.835\pm0.064$  \\ 
2458051.7069 & $15.888\pm0.039$ & $11.804\pm0.061$  \\ 
2458054.7050 & $11.699\pm0.042$ & $16.507\pm0.069$  \\ 
2458058.6955 & $15.850\pm0.047$ & $11.804\pm0.075$  \\ 
2458061.7292 & $11.578\pm0.084$ & $16.671\pm0.135$  \\ 
2458064.6885 & $15.795\pm0.059$ & $11.754\pm0.096$  \\ 
2458074.6442 & $11.563\pm0.041$ & $16.687\pm0.062$  \\ 
2458078.6349 & $15.687\pm0.040$ & $11.895\pm0.067$  \\ 
2458084.6432 & $15.848\pm0.051$ & $11.732\pm0.082$  \\ 
2458123.5290 & $15.697\pm0.043$ & $11.873\pm0.071$  \\ 
2458167.4181 & $12.518\pm0.043$ & $15.629\pm0.069$  \\ 
2458172.3816 & $12.309\pm0.051$ & $15.911\pm0.081$  \\ 
2458173.3962 & $11.710\pm0.042$ & $16.525\pm0.072$  \\ 
2458174.3848 & $13.157\pm0.053$ & $15.019\pm0.086$  \\ 
2458175.3863 & $15.206\pm0.049$ & $12.505\pm0.077$  \\ 
2458182.3797 & $15.587\pm0.065$ & $12.055\pm0.099$  \\ 
2458191.3431 & $13.968\pm0.031$ & $\cdots$  \\ 
2458426.6586 & $15.395\pm0.047$ & $12.300\pm0.073$  \\ 
2458433.6589 & $14.877\pm0.046$ & $12.815\pm0.076$  \\ 
\noalign{{\smallskip}}
\hline
\noalign{{\smallskip}}
\multicolumn{3}{c}{\textit{UU\,UMi}} \\
\noalign{{\smallskip}}
\hline
\noalign{{\smallskip}}
2456418.4924\tablefootmark{a} & $-41.365\pm0.162$ & $\cdots$  \\ 
2457472.6413 & $-39.007\pm0.035$ & $-45.157\pm0.141$  \\ 
2457504.5670 & $-38.904\pm0.038$ & $-45.141\pm0.147$  \\ 
2457529.4912 & $-38.968\pm0.037$ & $-45.298\pm0.141$  \\ 
2457556.4629 & $-38.860\pm0.041$ & $-45.265\pm0.161$  \\ 
2457559.5434 & $-38.858\pm0.043$ & $-45.269\pm0.174$  \\ 
2457763.6439 & $-38.781\pm0.039$ & $-45.590\pm0.148$  \\ 
2457800.7530 & $-38.690\pm0.025$ & $-45.638\pm0.111$  \\ 
2457815.5392 & $-38.692\pm0.030$ & $-45.635\pm0.135$  \\ 
2457832.5606 & $-38.701\pm0.025$ & $-45.645\pm0.122$  \\ 
2457848.6126 & $-38.663\pm0.027$ & $-45.662\pm0.115$  \\ 
2457867.5390 & $-38.681\pm0.027$ & $-45.678\pm0.133$  \\ 
2457897.4662 & $-38.693\pm0.031$ & $-45.680\pm0.135$  \\ 
2457931.5387 & $-38.677\pm0.033$ & $-45.713\pm0.152$  \\ 
2457961.3985 & $-38.612\pm0.029$ & $-45.732\pm0.142$  \\ 
2457993.3479 & $-38.677\pm0.036$ & $-45.774\pm0.166$  \\ 
2458054.3032 & $-38.614\pm0.029$ & $-45.769\pm0.122$  \\ 
2458117.7403 & $-38.643\pm0.030$ & $-45.772\pm0.122$  \\ 
2458161.7033 & $-38.647\pm0.033$ & $-45.784\pm0.131$  \\ 
2458200.5577 & $-38.666\pm0.034$ & $-45.763\pm0.143$  \\ 
2458263.4946 & $-38.681\pm0.031$ & $-45.727\pm0.133$  \\ 
2458353.3698 & $-38.665\pm0.033$ & $-45.661\pm0.145$  \\ 
2458451.7424 & $-38.766\pm0.029$ & $-45.599\pm0.127$  \\ 
2458541.6890 & $-38.788\pm0.029$ & $-45.484\pm0.120$  \\ 
2458631.5092 & $-38.895\pm0.030$ & $-45.274\pm0.138$  \\ 
2458723.3696 & $-38.985\pm0.029$ & $-45.091\pm0.136$  \\ 
2458816.7528 & $-39.113\pm0.028$ & $-44.931\pm0.129$  \\ 
2458913.6613 & $-39.255\pm0.028$ & $-44.616\pm0.121$  \\ 
2459003.4687 & $-39.474\pm0.035$ & $-44.412\pm0.144$  \\ 
2459093.3467 & $-39.692\pm0.029$ & $-44.050\pm0.144$  \\ 
\noalign{{\smallskip}}
\hline
\end{tabular}
\tablefoot{\tablefoottext{a}{RV from a CAFE spectrum.}}
\end{table}

\begin{table}[t]
\centering
\caption{RVs of ST3s.}
\label{tab:RVST3}
\begin{tabular}{lcccc} 
\hline\hline
\noalign{\smallskip}
Obs. date & RV$_{\rm A}$ & RV$_{\rm Ba}$ & RV$_{\rm Bb}$ & \multirow{2}{*}{Inst.} \\
 $[$BJD$]$  & [km\,s$^{-1}$] & [km\,s$^{-1}$] & [km\,s$^{-1}$] & \\
\noalign{\smallskip}
\hline
\noalign{{\smallskip}}
\multicolumn{5}{c}{\textit{GJ\,3916}} \\
\noalign{{\smallskip}}
\hline
\noalign{{\smallskip}}
2453099.8240 & $4.073(95)$ & $-10.21(41)$ & $12.11(33)$  & U  \\ 
2453112.6457 & $4.25(12)$ & $\cdots$ & $\cdots$  & U  \\ 
2453453.7944 & $8.291(97)$ & $\cdots$ & $\cdots$  & U  \\ 
2453480.6825 & $8.879(97)$ & $-18.55(41)$ & $\cdots$  & U  \\ 
2453488.6672 & $8.736(92)$ & $-18.63(36)$ & $\cdots$  & U  \\ 
2453505.8286 & $8.31(10)$ & $-8.90(37)$ & $2.58(25)$  & U  \\
2454167.8224 & $3.18(95)$ & $-5.0(3.3)$ & $11.50(1.84)$  & F  \\ 
2454569.6487 & $-1.675(99)$ & $\cdots$ & $14.34(30)$  & F  \\ 
2454570.8108 & $-2.037(86)$ & $\cdots$ & $12.84(27)$  & H  \\ 
2454580.8485 & $-1.24(10)$ & $\cdots$ & $\cdots$  & F  \\ 
2454583.8633 & $-2.029(84)$ & $11.57(25)$ & $3.25(26)$  & H  \\ 
2454628.5259 & $-2.43(72)$ & $16.9(2.5)$ & $\cdots$  & F  \\ 
2454639.7481 & $-1.83(70)$ & $13.2(2.5)$ & $\cdots$  & F  \\ 
2454661.5681 & $-2.248(79)$ & $\cdots$ & $14.54(29)$  & H  \\  
2454939.7950 & $-2.616(82)$ & $-9.26(21)$ & $20.77(26)$  & F  \\ 
2454943.8549 & $-2.544(96)$ & $-9.67(26)$ & $22.06(30)$  & F  \\ 
2454951.8029 & $-2.623(91)$ & $-9.71(24)$ & $22.11(30)$  & F  \\ 
2454984.6058 & $-0.876(82)$ & $13.04(31)$ & $\cdots$  & F  \\ 
2454993.7310 & $-1.771(80)$ & $17.20(32)$ & $\cdots$  & F  \\ 
2455003.5563 & $-2.36(14)$ & $20.80(48)$ & $\cdots$  & F  \\ 
2455053.5303 & $-0.846(87)$ & $\cdots$ & $9.75(35)$  & F  \\ 
2455066.4967 & $-1.995(93)$ & $\cdots$ & $17.59(28)$  & F  \\ 
2455075.4993 & $-2.10(70)$ & $\cdots$ & $20.9(2.4)$  & F  \\ 
2455338.7519 & $-0.91(10)$ & $-8.80(25)$ & $19.37(28)$  & F  \\ 
2455397.5650 & $-0.798(88)$ & $18.45(29)$ & $-7.40(17)$  & F  \\ 
2455407.5645 & $-1.127(86)$ & $19.65(32)$ & $-7.97(20)$  & F  \\ 
2455430.5447 & $-0.392(81)$ & $14.24(33)$ & $\cdots$  & F  \\ 
2455434.5660 & $-0.20(75)$ & $12.5(2.3)$ & $\cdots$  & F  \\ 
2455438.4973 & $0.14(64)$ & $\cdots$ & $\cdots$  & F  \\ 
2455615.8289 & $0.77(10)$ & $-11.16(34)$ & $19.57(26)$  & F  \\ 
2456352.8951 & $6.87(75)$ & $\cdots$ & $-10.2(2.4)$  & F  \\ 
2456500.6423 & $7.59(91)$ & $\cdots$ & $-6.0(2.1)$  & F  \\ 
2457476.6461 & $-1.098(78)$ & $-8.69(22)$ & $22.32(24)$  & C  \\ 
2457493.6138 & $-1.448(34)$ & $-1.22(29)$ & $14.76(22)$  & C  \\ 
2457504.5889 & $-1.511(43)$ & $6.67(25)$ & $7.70(18)$  & C  \\ 
2457540.4868 & $-2.167(41)$ & $21.32(27)$ & $-4.59(37)$  & C  \\ 
2457910.4391 & $-1.398(60)$ & $14.35(29)$ & $1.48(48)$  & C  \\ 
2457916.5426 & $-1.794(47)$ & $17.79(34)$ & $-1.61(36)$  & C  \\ 
2457949.3780 & $-1.865(42)$ & $18.61(32)$ & $-1.99(34)$  & C  \\ 
2457977.3697 & $-1.537(36)$ & $3.81(34)$ & $11.21(26)$  & C  \\ 
2458166.6713 & $-1.126(35)$ & $6.91(55)$ & $7.00(42)$  & C  \\ 
2458177.7400 & $-0.872(62)$ & $14.23(33)$ & $-0.63(36)$  & C  \\ 
2458186.7186 & $-1.501(40)$ & $18.85(27)$ & $-2.95(38)$  & C  \\ 
2458200.6650 & $-1.417(41)$ & $20.75(22)$ & $-5.71(23)$  & C  \\ 
2458207.6529 & $-1.469(41)$ & $19.72(23)$ & $-4.83(29)$  & C  \\ 
2458225.6859 & $-0.626(49)$ & $13.26(28)$ & $0.26(36)$  & C  \\ 
2458237.5692 & $-0.749(65)$ & $6.68(35)$ & $6.63(27)$  & C  \\ 
2458244.5965 & $-0.544(38)$ & $1.02(55)$ & $10.43(27)$  & C  \\ 
2458249.6074 & $-0.835(33)$ & $-0.65(33)$ & $13.27(24)$  & C  \\ 
2458261.4581 & $-0.905(37)$ & $-7.11(25)$ & $19.01(23)$  & C  \\ 
2458269.5321 & $-0.803(35)$ & $-9.38(25)$ & $21.28(22)$  & C  \\ 
2458275.5228 & $-0.789(56)$ & $-9.26(30)$ & $21.09(26)$  & C  \\ 
2458283.5380 & $-0.813(46)$ & $-6.85(28)$ & $18.48(24)$  & C  \\ 
2458289.5200 & $-0.921(38)$ & $-1.68(47)$ & $14.87(26)$  & C  \\ 
2458297.4920 & $-0.390(42)$ & $4.63(26)$ & $8.82(18)$  & C  \\ 
2458575.7001 & $0.331(53)$ & $12.69(25)$ & $-0.46(32)$  & C  \\ 
2458587.6266 & $0.657(37)$ & $18.01(22)$ & $-6.07(21)$  & C  \\ 
2458597.6156 & $0.725(62)$ & $19.26(29)$ & $-7.15(25)$  & C  \\ 
2458603.5602 & $0.753(38)$ & $18.55(24)$ & $-6.79(22)$  & C  \\ 
2458609.5737 & $0.732(42)$ & $17.15(23)$ & $-5.63(21)$  & C  \\ 
2458615.5395 & $0.446(38)$ & $15.19(25)$ & $-4.18(26)$  & C  \\ 
2458626.5359 & $0.830(46)$ & $10.28(31)$ & $0.60(31)$  & C  \\ 
2458636.4799 & $1.381(35)$ & $5.70(16)$ & $5.74(14)$  & C  \\ 
\hline 
\end{tabular}
\end{table}

\addtocounter{table}{-1}
\begin{table}[t]
\centering
\caption{Continued.}
\begin{tabular}{lcccc} 
\hline\hline
\noalign{\smallskip}
Obs. date & RV$_A$ & RV$_{Ba}$ & RV$_{Bb}$ & \multirow{2}{*}{Inst.} \\
 $[$BJD$]$  & [km\,s$^{-1}$] & [km\,s$^{-1}$] & [km\,s$^{-1}$] & \\
\noalign{\smallskip}
\hline
\noalign{{\smallskip}}
2458643.4509 & $0.894(35)$ & $0.64(35)$ & $9.36(26)$  & C  \\ 
2458656.4064 & $0.993(39)$ & $-7.21(25)$ & $15.64(22)$  & C  \\ 
2458662.3779 & $1.102(43)$ & $-9.81(26)$ & $17.89(25)$  & C  \\ 
2458679.3749 & $1.157(39)$ & $-9.89(25)$ & $17.94(22)$  & C  \\ 
2458690.3681 & $1.000(38)$ & $-2.14(43)$ & $11.34(28)$  & C  \\ 
2458696.3722 & $1.774(60)$ & $-0.16(31)$ & $7.22(30)$  & C  \\ 
2458708.3595 & $0.956(40)$ & $12.13(30)$ & $-0.72(44)$  & C  \\ 
\noalign{{\smallskip}}
\hline
\noalign{{\smallskip}}
\multicolumn{5}{c}{\textit{GJ\,4383}} \\
\noalign{{\smallskip}}
\hline
\noalign{{\smallskip}}
2455046.8563 & $-12.63(11)$ & $\cdots$ & $\cdots$  & H  \\ 
2455053.8419 & $-12.608(96)$ & $\cdots$ & $\cdots$  & H  \\ 
2455054.8560 & $-12.622(87)$ & $\cdots$ & $\cdots$  & H  \\ 
2455056.8369 & $-12.58(12)$ & $\cdots$ & $\cdots$  & H  \\ 
2455056.8994 & $-12.58(11)$ & $\cdots$ & $\cdots$  & H  \\ 
2455124.6171 & $-12.257(83)$ & $\cdots$ & $\cdots$  & H  \\ 
2455537.5585 & $-10.154(87)$ & $\cdots$ & $\cdots$  & H  \\ 
2455541.5532 & $-10.067(85)$ & $\cdots$ & $\cdots$  & H  \\ 
2455550.5263 & $-9.997(94)$ & $\cdots$ & $\cdots$  & H  \\ 
2456496.8926 & $-7.47(28)$ & $\cdots$ & $\cdots$  & F  \\ 
2456496.9060 & $-7.49(22)$ & $\cdots$ & $\cdots$  & F  \\ 
2456614.6481 & $-7.54(22)$ & $\cdots$ & $\cdots$  & F  \\ 
2456614.6626 & $-7.58(22)$ & $\cdots$ & $\cdots$  & F  \\ 
2457587.6611 & $-10.559(38)$ & $\cdots$ & $\cdots$  & C  \\ 
2457604.6500 & $-10.629(30)$ & $\cdots$ & $\cdots$  & C  \\ 
2457622.6367 & $-10.730(29)$ & $\cdots$ & $-18.93(53)$  & C  \\ 
2457623.5259 & $-10.57(31)$ & $\cdots$ & $\cdots$  & C  \\ 
2457623.5458 & $-10.717(33)$ & $-6.08(75)$ & $-19.17(50)$  & C  \\ 
2457654.5045 & $-10.941(31)$ & $-0.46(48)$ & $-24.88(53)$  & C  \\ 
2457672.4896 & $-10.984(29)$ & $-1.26(40)$ & $-23.64(45)$  & C  \\ 
2457695.3885 & $-11.147(33)$ & $\cdots$ & $\cdots$  & C  \\ 
2457786.2850 & $-11.615(43)$ & $\cdots$ & $\cdots$  & C  \\ 
2457916.6483 & $-12.511(25)$ & $\cdots$ & $-6.01(58)$  & C  \\ 
2457942.6396 & $-12.612(27)$ & $-19.57(42)$ & $2.28(50)$  & C  \\ 
2457999.5659 & $-12.937(24)$ & $\cdots$ & $-4.17(43)$  & C  \\ 
2458051.4269 & $-13.287(28)$ & $-2.49(37)$ & $\cdots$  & C  \\ 
2458092.3378 & $-13.540(20)$ & $4.25(31)$ & $-20.19(35)$  & C  \\ 
2458122.2522 & $-13.636(21)$ & $-4.21(36)$ & $\cdots$  & C  \\ 
2458283.6381 & $-14.622(29)$ & $3.26(43)$ & $\cdots$  & C  \\ 
2458343.6430 & $-14.896(22)$ & $-5.05(26)$ & $\cdots$  & C  \\ 
2458410.5819 & $-15.148(33)$ & $\cdots$ & $3.99(54)$  & C  \\ 
2458471.3527 & $-15.450(22)$ & $-1.05(34)$ & $-7.37(43)$  & C  \\ 
2458656.6395 & $-15.975(24)$ & $-5.82(41)$ & $-0.80(57)$  & C  \\ 
2458718.5956 & $-15.981(23)$ & $6.44(36)$ & $\cdots$  & C  \\ 
2458784.4240 & $-16.111(25)$ & $-7.59(38)$ & $1.80(42)$  & C  \\ 
2458844.3742 & $-16.019(27)$ & $-10.53(55)$ & $5.52(47)$  & C  \\ 
2459074.6456 & $-15.606(23)$ & $-9.04(31)$ & $1.97(37)$  & C  \\ 
2459138.4984 & $-15.398(26)$ & $2.64(35)$ & $\cdots$  & C  \\ 
2459214.3691 & $-15.135(25)$ & $-8.49(38)$ & $-0.45(47)$ & C \\
\noalign{{\smallskip}}
\hline 
\end{tabular}
\tablefoot{The number between parentheses indicate the uncertainty of the last digits.}
\end{table}

\end{appendix}

\end{document}